\RequirePackage{fix-cm}
\documentclass[natbib,smallextended]{svjour3}       
\smartqed  
\usepackage{graphicx}
\usepackage{longtable}
\usepackage{txfonts}
\usepackage{rotating}
\usepackage{lscape}
\usepackage{color}
\usepackage[colorlinks=true,citecolor=blue,urlcolor=blue,breaklinks]{hyperref}
\journalname{The Astronomy and Astrophysics Review}
\begin{document}

\title{What is a Globular Cluster?
}
\subtitle{An observational perspective}


\author{Raffaele Gratton$^1$         \and
        Angela Bragaglia$^2$         \and
        Eugenio Carretta$^2$         \and
        Valentina D'Orazi$^{1,3}$    \and
        Sara Lucatello$^1$           \and
        Antonio Sollima$^2$}


\institute{$^1$INAF-Osservatorio Astronomico di Padova,
              Vicolo dell'Osservatorio 5, 35122 Padova (Italy),  
              Tel.: +39-049-661303,
              \email{raffaele.gratton@inaf.it}           
           \and
           $^2$INAF-Osservatorio di Astrofisica e Scienza dello Spazio, via P. Gobetti 93/3, 40129 Bologna (Italy) 
           \and
           $^3$Monash Centre for Astrophysics, School of Physics and Astronomy, Monash University, Melbourne, Clayton 3800, Australia.
         }

\date{Received: date / Accepted: date}

\maketitle

\begin{abstract}
Globular clusters are large and dense agglomerate of stars. At variance with smaller clusters of stars, they exhibit signs of some chemical evolution. At least for this reason, they are intermediate between open clusters and massive objects such as nuclear clusters or compact galaxies. While some facts are well established, the increasing amount of observational data is revealing a complexity that has so far defied the attempts to interpret the whole data set in a simple scenario. We review this topic focusing on the main observational features of clusters in the Milky Way and its satellites. We find that most of the observational facts related to the chemical evolution in globular clusters are described as being primarily a function of the initial mass of the clusters, tuned by further dependence on the metallicity -- that mainly affects specific aspects of the nucleosynthesis processes involved -- and on the environment, that likely determines the possibility of independent chemical evolution of the fragments or satellites where the clusters form. We review the impact of multiple populations on different regions of the colour-magnitude diagram and underline the constraints related to the observed abundances of lithium, to the cluster dynamics, and to the frequency of binaries in stars of different chemical composition. We then re-consider the issues related to the mass budget and the relation between globular cluster and field stars. Any successful model of globular clusters formation should explain these facts.
\end{abstract}

\setcounter{tocdepth}{3}
\tableofcontents
\clearpage
\vspace*{0.5cm}
\hfill\begin{minipage}{\dimexpr\textwidth-3cm}

\xdef\tpd{\the\prevdepth}
 {\it ``16. Of Globular Clusters of Stars.

The objects of this collection are of a sufficient brightness to be seen with any good common telescope, in which they appear like telescopic comets, or bright nebulae, and under this disguise, we owe their discovery to many eminent astronomers; but in order to ascertain their most beautiful and artificial construction, the application of high powers, not only of penetrating into space but also of magnifying are absolutely necessary; and as they are generally but little known and are undoubtedly the most interesting objects in the heavens, I shall describe several of them, by selecting from a series of observations of 34 years some that were made with each of my instrument, that it may be a direction for those who wish to view them to know what they may expect to see with such telescopes as happen to be in their possession.''}\\
Herschel, W. 1814, Philosophical Transactions, 104, 248

\end{minipage}

\section{Introduction}
\label{Sec:1}

Globular clusters (GCs: \citealt{Herschel1814}) are usually considered as a class of stellar agglomerates characterized by being compact (half-light radius up to a few tens of pc, with more typical values of about 3 to 5 pc), bright (mean absolute visual magnitude around $M_V=-7$), old (in most cases, ages around 10 Gyr), and (within the Milky Way - MW) to be representative of the halo, thick disk and bulge, but being absent in the thin disk\footnote{Globular clusters may have been formed in-situ or have been accreted, see the classical paper by \citet{Searle1978} and the recent results coming out from the Gaia mission, as presented e.g. in \citet{GaiaCollaboration2018, Myeong2018c, Helmi2018}.}. Being abundant in the halo and thick disk, they are often metal-poor and have quite extreme kinematics. There is evidence that the peak of the formation of GCs pre-dates most of the stellar formation in galaxies and that they may have played an important role in the early formation of galaxies and even in re-ionization (see e.g. the discussion in \citealt{Renzini2017}). However, at a closer scrutiny the classical definition of GCs becomes a bit vague and there are many objects that are classified as GCs based on only a few of these criteria. About ten years ago, \citet{Carretta2010c} proposed a new definition of GCs, that is related to the chemical inhomogeneities that are characteristic of these objects, and that differentiate them from the less massive open clusters: genuine GCs are stellar systems showing anti-correlations among the abundances of light-elements, whose main and most widespread example is the Na-O anticorrelation. While classification according to this criterion is still not perfect, it has the advantage to shift the attention to a fundamental characteristics of GCs, that is their complex formation scenarios and the clear signatures of a chemical evolution within them. In this sense, GCs are objects intermediate between normal stellar clusters and the blue compact galaxies, as indicated by their location in the mass-to-light ratio versus luminosity plane \citep{Dabringhausen2008, Forbes2008}. While how a GC forms and what are its very early phases are still strongly debated topics, we have at least in theory the possibility of separating chemically, and in some case dynamically, quite pure populations, sharing very similar chemical composition reflecting single nucleosynthesis effects. In this environment, stars form with very peculiar chemical composition, that are very rarely observed in the main population of galaxies, where we generally see the combination of many different nucleosynthesis processes.

It is then not surprising that since their discovery from low resolution spectra and intermediate band photometry in the '70s, chemical ``inhomogeneities'' in individual GCs raised a considerable interest. Reviews of very early results can be found in \citet{Kraft1979} and of later progress obtained mainly thanks to echelle spectrographs on 4m telescopes in \citet{Kraft1994}. The availability of high quality spectroscopic data for large samples of stars in many individual clusters allowed by multi-fiber high-resolution spectrographs on 8m class telescopes and the exquisite photometry provided by HST have provided an important breakthrough in the first decade of the new millennium, with the acknowledgment that the ``inhomogeneities'' are due to a distinct chemical evolution within the GCs that seem to be characterized by multiple stellar populations. Reviews of the progress, more focused on the observational side, were given by \citet{Gratton2004, Piotto2010, Gratton2012}. 

In the mean time, a number of discussions were rather more focused on the scenarios that may explain the multiple populations (see e.g. \citealt{Renzini2008, Schaerer2011, Krause2013, Renzini2015, Bastian2015, D'Antona2016, Dercole2016, Bastian2015b, Prantzos2017, Bastian2018, Gieles2018}). The conclusion of the review by \citet{Bastian2018} is that there is not a unique simple scenario able to explain the variety of issues related to the multiple population problem. This suggests that GCs are likely not a homogeneous sample of objects, but rather may include different histories. This fact should be perhaps considered not too surprising for objects that represent transitions between single episodes of star formation - characteristics of open clusters - and more continuous stories of star formation - characteristics of galaxies. 

Notwithstanding these difficulties, in this review we will generally assume that the different populations are indeed different generations of stars. As already discussed in the previous reviews \citet{Gratton2004, Gratton2012}, there are various arguments supporting this assumption. A successful scenario should not only explain enrichment in some element such as He or Na - this is relatively easy to achieve also in alternative scenarios because it may be obtained by integrating in a star a small fraction of polluted material - but also the large depletion of quite robust nuclei such as O and Mg, that are observed in a significant fraction of the GC stars. This depletion can only be obtained by having stars that are mainly composed of the ejecta of previous generations, in spite of the fact that only a small fraction of the ejecta of a generation of stars are expected to be actually depleted of these elements. In fact, alternative schemes such as selective chemical enrichment during star formation and deep mixing fail completely to reproduce either the observed abundance pattern or the fact that the abundance anomalies are observed in a similar way throughout the whole colour-magnitude diagram. Finally, in many cases - though possibly not in all cases - discrete populations can be clearly discerned. However, we concur with \citet{Bastian2018} that the scenarios based on multiple generations considered so far have major difficulties and that we should be open to other possible explanations.

\begin{figure*} 
\includegraphics[width=1.0\textwidth]{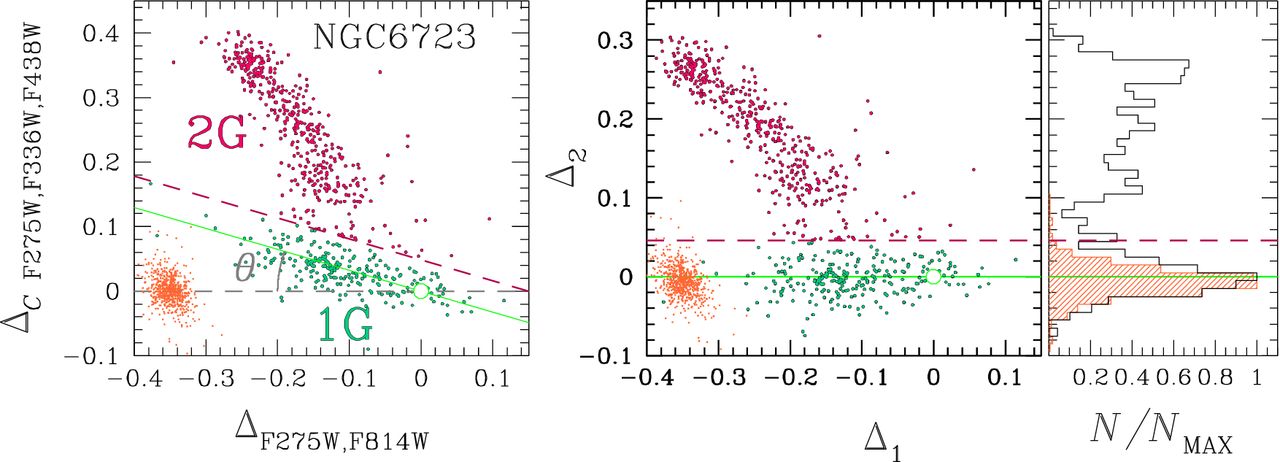}
\caption{Example of a chromosome map as taken from \citet{Milone2017}. As explained in the caption of the figure of their paper, this map is used to identify the two samples of bona-fide first-generation (indicated as 1G, FG in the present text) and second-generation (2G in the figure, SG in this text) stars in NGC 6723. The left-hand panel shows $\Delta$F275W, F336W, F438W versus $\Delta$F275W, F814W. The green line through the origin of the frame is a fit to the sequence of candidate 1G=FG stars. The middle panel shows the $\Delta_2$ versus $\Delta_1$ plot where these new coordinates have been obtained by a suitable rotation of the plot in the left-hand panel. The histogram in the right-hand panel shows the distributions of the $\Delta_2$ values. The orange points in the left-hand and middle panels show the distribution of the observational errors and their $\Delta_2$ distribution is represented by the shaded orange histogram in the right-hand panel. The dashed magenta lines separate the selected 1G=FG and 2G=SG stars, which are coloured aqua and magenta, respectively, in the left-hand and middle panels.}
\label{fig:chromosome}       
\end{figure*}

In this review we present an update of the field, exploiting the new look that is provided by additional spectroscopic data, both at high and low-resolution, that is accumulating \citep[see the review by][]{Bastian2018}, by the extensive UV photometry obtained with HST-WFC3 (e.g. \citealt[][for the MW]{Piotto2015}, \citealt[][for the MW satellites]{Niederhofer2017b}) and by the estimates of the initial cluster masses \citep[e.g.][]{Baumgardt2019} that are now possible thanks to the orbital parameters extracted from the Gaia DR2 data \citep{GaiaCollaboration2018}, as well as a large number of other important contributions. This impressive amount of data is revealing a complex scenario, with different types of clusters likely having different evolution, that left traces imprinted in the chemistry of their stars. An help can be given by the so-called ``chromosome map'', first introduced by \citet{Milone2012a, Milone2017}, for several tens of GCs (see Fig.~\ref{fig:chromosome}). In the following, we will make extensive use of their notation of Type~I and II clusters, based on this diagram (see Section 3.3), even if we alert that exact classification of some GC may be questioned.

Due to space limitations, we have to operate a selection on topics, privileging the observer's ``route'' and focusing on papers discussing large samples. In addition, we will not speak of the wide main sequence turn off (MSTO) of young and intermediate age clusters in the MCs. This had been originally considered as an evidence for multiple populations with large age differences in these clusters too \citep[e.g.][]{Bertelli2003, Milone2009, Goudfrooij2014}. However, recent developments seem rather to indicate a combination of spread in stellar rotation and possibly presence of binaries as an explanation \citep[e.g.][]{Dupree2017,Dantona2017,Milone2018c,Marino2018b, bastian_rot, lim2019}, as originally proposed by \citet{Bastian2009}. We will also not describe the evidence from variable stars (in particular, RR Lyrae) for which we refer to \citet{Catelan2009, Gratton2010b}, and \citet{Jang2014}. In addition, we will limit our analysis to the meta-Galaxy, that is the MW and its close satellites; reviews of the properties of GCs in further galaxies can be found in \citet{Brodie2006, Kruijssen2014, Grebel2016}. 

Finally and more importantly, we will not discuss the vast literature on models of cluster formation and evolution and will only touch upon some scenarios for explaining the multiple population phenomenon (for this, see \citealt{Bastian2018} and  more recent papers).

In Section~\ref{Sec:2} we will introduce the chemical anomalies usually seen in GCs. In Section~\ref{Sec:3} we will describe the main observed dependencies considering various classes of GCs. In Section~\ref{Sec:4} we will review the impact of chemical anomalies on the stellar evolution. In Section~\ref{Sec:5}, \ref{Sec:6} and \ref{Sec:7} we will discuss three important pieces of information, often neglected in the discussion of GCs: that is, the evidence that concern Lithium abundances, dynamics, and binarity. In Section~\ref{Sec:8} we will revisit the connection existing between GCs and the general field. Finally, we draw some conclusions in Section~\ref{Sec:9}. Relevant data used throughout this review are collected in the Appendices.

\section{Chemical anomalies in GCs}
\label{Sec:2}

The simplest and clearer way to define multiple stellar populations in GCs is through their opposite. A {\it simple} stellar population (SSP) is an ensemble of coeval (single) stars with the same initial chemical composition. Thus, when we see stellar systems hosting stars of different starting chemistry and with (even small) age differences we observe multiple stellar populations.

\subsection{Basics of multiple stellar populations in GCs}

Almost a century may have gone by since pioneering observations pointed out that star-to-star abundance variations existed among stars of otherwise mostly chemically homogeneous GCs (see the nice historical notes in \citealt{Smith2006}). By comparing elemental abundances in field and cluster stars (e.g. \citealt{Gratton2000, Smith2003}; see the review by \citealt{Gratton2004}) it is immediately evident what is observed for the multiple population phenomenon, in what stellar systems, and at which evolutionary phase.

Large abundance variations are mostly seen among light elements, starting from the elusive He up to Sc. The GC NGC~2808 is the ideal ``showroom'' for all involved elements in mono-metallic GCs, because it was extensively studied  with spectroscopy at different evolutionary phases: red giant branch(RGB)/red horizontal branch (RHB): \citet{Norris1983, Carretta2003}; blue hook: \citet{Moehler2004}; RGB: \citet{Carretta2004, Carretta2006a, Carretta2006b, Pasquini2011, Carretta2014b, Mucciarelli2015, D'Orazi2015, Carretta2015, Carretta2018}; blue horizontal branch (BHB): \citet{Pace2006}; main sequence (MS): \citet{Bragaglia2010b}; horizontal branch (HB): \citet{Gratton2011b, Marino2014}; RGB/asymptotic giant branch (AGB): \citet{Wang2016}; AGB: \citet{Marino2017}.
If we add also the indirect evidence provided by photometry on the MS \citep[e.g.][]{D'Antona2005, Piotto2007,Milone2019a} we see that the multiple population phenomenon concerns stars in {\it all} the stages of their lifetime, from very low-mass dwarfs to giants.

\begin{figure*} 
\includegraphics[width=1.0\textwidth]{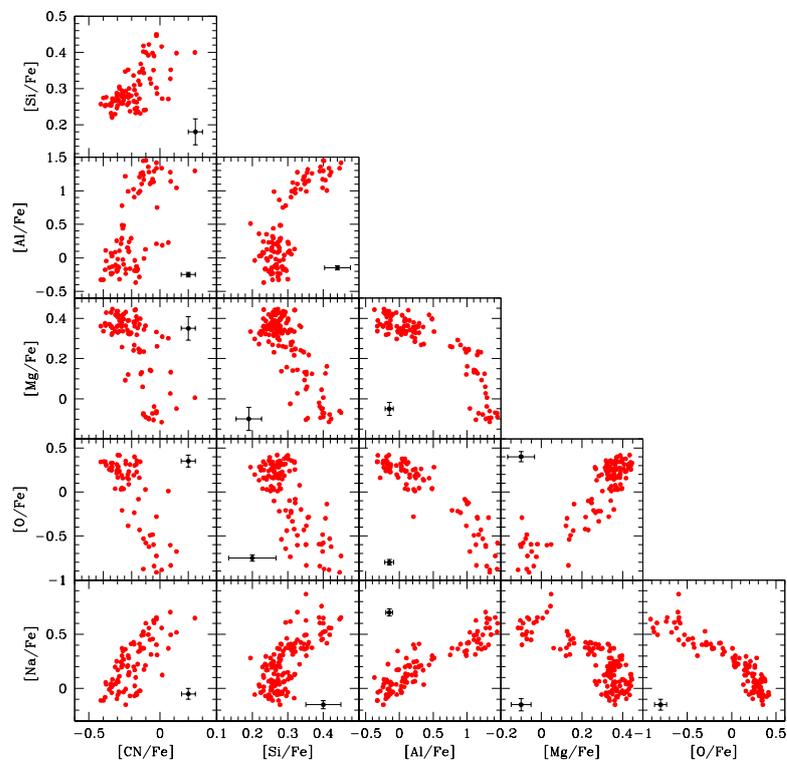}
\caption{Run of abundance ratios for light elements in RGB stars in NGC~2808. O, Na, Si, Mg are from \citet{Carretta2015b}, Al and CN abundances from \citet{Carretta2018}. The figure is adapted from the invited review by \citet{Carretta2016}.}
\label{fig:abu2808a}       
\end{figure*}

\begin{figure*} 
\includegraphics[width=0.75\textwidth]{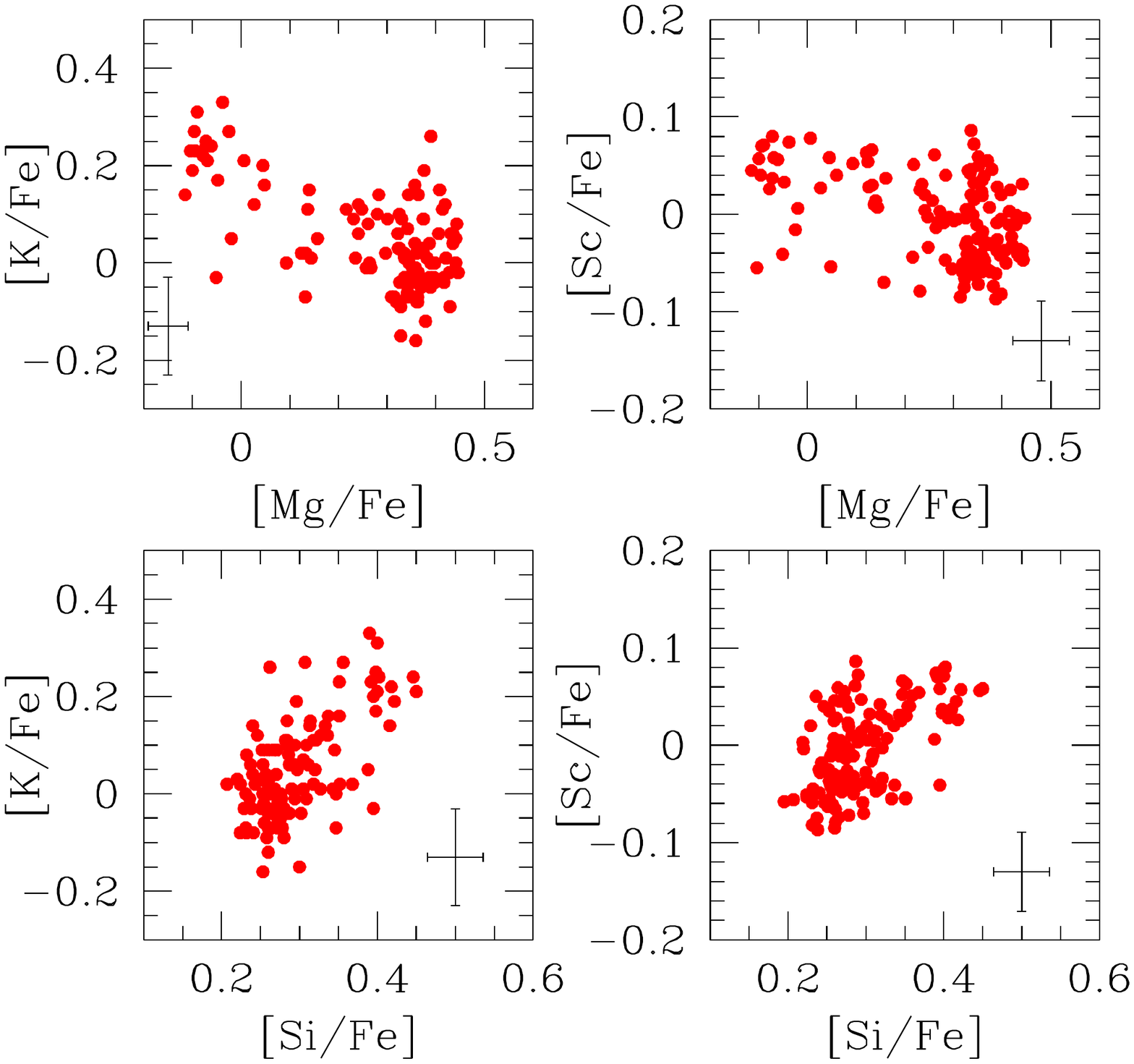}
\caption{As in previous figure for Mg, Si, Sc from \citet{Carretta2015b} and K abundances from \citet{Mucciarelli2015}.}
\label{fig:abu2808b}       
\end{figure*}

Abundance variations are not randomly distributed. By looking at the relationships summarized in Fig.~\ref{fig:abu2808a} and Fig.~\ref{fig:abu2808b} and made using the same sample of stars, we can appreciate that the star-to-star variations are all linked by (anti)correlations with each other. Moreover, the  examples of NGC~2808 and of almost all GCs studied so far show that the ``direction'' of these (anti)correlations is uniquely determined: the departures from the level imprinted by supernova nucleosynthesis all go in the same sense. Oxygen abundances may only be depleted, there is no observed star with [O/Fe] enhanced much above the plateau given by SN II ejecta, nor is Na found much below the level observed in field stars of similar metallicity, its abundance can be only enhanced above this level. This is crucial to understand the origin of the chemical pattern observed in GCs.

Multiple populations are found in almost all Galactic GCs, regardless of their Galactic parent population (halo, disk, bulge); for an updated list see \cite{Bragaglia2017} and the Appendix. Both those GCs formed {\it in situ} and those thought to have formed in external galaxies and later accreted by the Milky Way are found to host multiple populations. The phenomenon is ubiquitous in ``tiny'' GCs as well as in the most massive ones (probably former nuclear clusters of dwarf galaxies, like $\omega$~Cen=NGC~5139 and M~54=NGC~6715), in mono-metallic (intended as overall metal abundance [Fe/H]) as well as iron complex GCs. In the last, the multiple populations are simply repeated in each individual metallicity component. Multiple populations are not found among field stars of dwarf galaxies (e.g. Sgr, Fornax, Large Magellanic Cloud - LMC) but only observed in their associated old\footnote{See,however, Sect. 3.6 for the recent extension to lower ages.} GCs \citep[e.g.][]{Carretta2010, Carretta2010b, Letarte2006, Johnson2006, Mucciarelli2009, Larsen2014}. Finally, no anti-correlation among light elements was ever observed in open clusters, whose average abundances follow the pattern of other field thin disc stars \citep[see e.g.][]{deSilva2009, Bragaglia2014, Maclean2015} and Sect. 3.3.

\subsection{Observational approaches}

Since the finding by \citet{Lindblad1922} of giant stars with anomalously weak CN bands in NGC~6205=M~13, observations of abundance variations in GCs were performed either through spectroscopy or photometry. Of course, spectroscopy is the privileged method, because the essence of multiple stellar populations resides in abundance differences. However, we may think that photometry in different bands is equivalent to very low resolution spectroscopy and it allows access to large samples in a short observing time and better handling of crowded regions close to the cluster center.

The information accessible via the different techniques on the various involved species is principally dictated by the spectral resolution. At low resolution, lines are blended in spectra. This implies resorting to spectral synthesis and/or line indices to extract the information on abundances. Traditionally, this is the most followed approach to detect differences between populations based on the molecular features of CN and CH (and, less often, NH). To obtain detailed estimates of C and N, calibration through spectral synthesis is required (see the review by \citealt{Smith1987, Gratton2004, Gratton2012} for references on early and following works using this technique). The gain in observing time due to the low spectral resolution is exploited in particular for stars with severe flux limitations, either because they are intrinsically faint (as the case of unevolved stars in Galactic GCs, e.g. \citealt{Briley2004, Harbeck2003, Pancino2010}) or distant, as giants in old clusters in the LMC (e.g. \citealt{Hollyhead2019}). The chief drawback is that there is no indicator of O abundance from low resolution spectroscopy, though, interestingly, O abundance variations can be seen using UV photometry. This fact in turn generates uncertainties in the determination of C and N abundances, since their derivation from molecular CH and CN features requires the knowledge of O abundance, in particular for bright giants, where the CO formation is favoured. On the other hand, the strength of molecular bands decreases in warmer stars (such as dwarfs near the turn-off), with a bias against low metallicity clusters, where features of bi-metallic molecules like CN may become vanishingly weak. Despite these limitations, low resolution spectroscopy is widely used to detect ``first-pass'' evidence of abundance variations in C/N for moderately large sample of stars in many Galactic and extra-galactic GCs, in particular when multi-object spectroscopy is feasible. Only a few heavier elements can be measured with this technique. The atomic lines visible even at low resolution allow to compute e.g. the Lick indexes, that are essentially expressing the overall metallicity (see Fig. 9 in \citealt{Kim2016}), or to derive abundances of species like Cr, Sr, Ba, and Mg using spectrum synthesis \citep[e.g. ][]{Gratton2012c,Dias2016}. The calibration of low resolution spectroscopic observations of Ca features are also frequently used. In the near-IR the reduced equivalent width of the Ca II triplet is considered (see \citealt{Olszewski1991, Armandroff1991}; and \citealt{Rutledge1997, Dacosta2016} for extensive applications), whereas in the near-UV spectral region, the calcium index HK' \citep{Lim2015} based on the Ca II H \& K resonance lines provides metallicity estimates from a set of calibrating GCs with [Fe/H] determined from high resolution spectroscopy. In the last case, all these Ca (and Mg) lines are among the strongest lines in the stellar spectra; although the origin of multiple population is not supposed to be related primarily to polluting source altering the Fe content, the growing number of iron complex GCs makes useful to have methods for a quick first screening of the cluster overall metallicity and dispersion, in particular for distant objects.

Blending of lines is more easy to deal with high-resolution spectroscopy, and accurate equivalent widths of many transitions for numerous elements may be obtained, provided the spectral coverage is large enough. Abundances of many different species are then simultaneously derived, in particular those for elements that differ in multiple stellar populations. Abundance  ratios may be then constructed for all the relevant species\footnote{For most elements, we adopt the usual spectroscopic notation, $i.e.$  [X]=$\log{X_{\rm star}} -\log{X_\odot}$ for any abundance quantity X, and  $\log{\epsilon(X)} = \log{N_{\rm X}/N_{\rm H}} + 12.0$ for absolute number density abundances. For Helium, we use Y, that is the fraction of He in mass.}. In other words, with  high-resolution spectroscopy we are measuring the concentration of atoms of different species in the atmosphere of stars of different stellar populations. With respect to lower resolution spectra, the chief disadvantage is that, for a given magnitude, less flux is available per resolution element and thus longer integration times are required to achieve high signal-to-noise ratios (SNR). The problem is partly alleviated by using modern multi-object spectrographs mounted at 8-10 m class telescopes. Nevertheless, despite the multiplex advantage, the increased observing times forcibly reduces the sample size. Good statistics is routinely possible only for RGB and AGB stars, is more limited for HB stars (mainly for concerns related to abundance analysis in hot objects), and for dwarf stars is more or less limited to nearby GCs (such as 47~Tuc=NGC~104, M~4=NGC~6121, NGC~6397, NGC~6752, $\omega$~Cen=NGC~5139). 

A particular care must be applied for He, since the direct determination of its abundance is plagued by the lack of photospheric lines in cool stars. Therefore, only a handful of giants have He abundances derived from the near IR line at 10,830~\AA. Moreover, this line forms in the upper chromosphere, in non-LTE conditions and  the analysis requires using complex chromospheric models, and even accounting for spherical geometry of the atmosphere \citep[see][]{Pasquini2011, Dupree2013}. The only other direct He determinations are based on the weak He I triplet at 5875.6~\AA\ discovered in the spectra of late-type stars by \citet{Wilson1956}. This line is vanishingly weak in stars cooler than about 8000 K and cannot be safely used above 11500~K.  In warmer stars the measured He content is not the original one of the star at birth, but the value resulting from sedimentation caused by diffusion and element stratification (e.g.\citealt{Behr1999, Behr2000, Behr2003}). For these reasons, He is only measured directly in HB stars in this limited temperature range \citep[e.g.][]{Villanova2009, Gratton2014}, and its abundance in GCs is mainly obtained by indirect estimates based on photometry. Apart from He and related problems, high-resolution spectroscopy is a powerful tool to study multiple stellar populations because the measured indicators are directly related to the stellar composition, i.e., to the essence of multiple populations. Age differences of a few or a few tens of Myr expected in most scenarios of multiple populations are not detectable in colour-magnitude diagrams. On the other hand, the abundance differences related to these scenarios, several tenths of a dex in the abundances of elements such as C, N, Na, O, Mg, Al, are well measurable from spectra. These differences cannot be produced within the currently observed low mass stars.
 
Finally, the photometric approach consists in tracing the abundance variations through their effects (flux variations) in selected band-passes where some molecular features of CNO elements are located. As such, panoramic photometry can be considered as spectroscopy at very low resolution and with very high multiplexing gain.  Investigation of multiple stellar populations is possible with any photometric system including some filters (located in the UV/blue regions) whose band-passes cover features of interest (essentially CN, NH, OH, CH molecular bands). Broad band (Sloan: \citealt{Lardo2011}; Johnson-Cousins: e.g. \citealt{Monelli2013}; HST: e.g. \citealt{Piotto2015, Larsen2014}; intermediate band Str\"omgren: e.g. \citealt {Grundahl1998, Yong2008a, Carretta2011}, and narrow band systems such as Ca-photometry: \citealt{Anthony-Twarog1991, Lee2009a, Lee2009b, Lee2015}) have all been used to detect variations in light-elements abundances, separate different populations on colour-magnitude diagrams, and trace their radial distribution thanks to the large statistics possible with photometry. The dichotomy low/high resolution is partially reproduced also for photometric observations, since the crowded cores of GCs can be only resolved with space-based photometry. However, this is possible at the price of a limited spatial coverage, and this occurrence may give some problems when radial gradients in the distribution of the population ratios are present \citep[e.g.][]{Lee2019}, unless they are combined with ground-based data \citep[e.g.][]{Milone2012a, Savino2018}. Also, the definition of what are the multiple populations may differ somewhat according to the study and the adopted photometric system, although in most cases there is a reasonable agreement between spectroscopy and photometry classification (see the discussions in \citealt{Carretta2011, Savino2018, Lee2019} and \citealt{Marino2019}).

Due to the above limitations related to spectroscopic detection of He, photometry is the most used approach to estimate the He abundance and variations, by exploiting the prediction of stellar evolution for He-enhanced models \citep[e.g.][]{Salaris2006}. Even if the absolute He abundance cannot be given by these methods, relative estimates can be provided by differences in magnitude of RGB-bump stars and in colour of RGB stars \citep[e.g.]{Bragaglia2010a}, colour spreads on the main-sequence \citep[e.g.]{Norris2004, Piotto2007, Gratton2010b}, and comparison of multi-wavelength HST photometry with synthetic spectra \citep[e.g.][]{Milone2018}. An additional drawback of photometry is that the accessible pattern of multiple population is limited to the lightest elements (He, C, N). Oxygen can be estimated only when the HST filter F275W is available, and this introduces the same degree of uncertainty as seen for low dispersion spectroscopy of CN and CH bands. No indicator is available for heavier elements, apart from Ca (see \citealt{Lee2019} and references therein), measuring the Ca II H \& K lines through narrow band photometry \citep{Anthony-Twarog1991, Lim2015}.

Further improvements of photometric methods can come from the upcoming survey J-PAS (see \citealt{Benitez2014} and http://www.j-pas.org/ for information), which will observe about 8000 deg$^2$ of the Northern sky with 56 narrow band filters (about 150 \AA \ each, covering the $\sim 3750-8100$ \AA \ interval) which will hopefully permit to measure different light element abundances. A tentative separation of the upper RGB of NGC~7078=M~15 has already been presented by \citet{Bonatto2019} who use J-PLUS data (J-PLUS is meant to calibrate J-PAS and uses a combination of 12 wide, intermediate, and narrow band filters).

\begin{figure}[htb] 
\includegraphics[width=\textwidth]{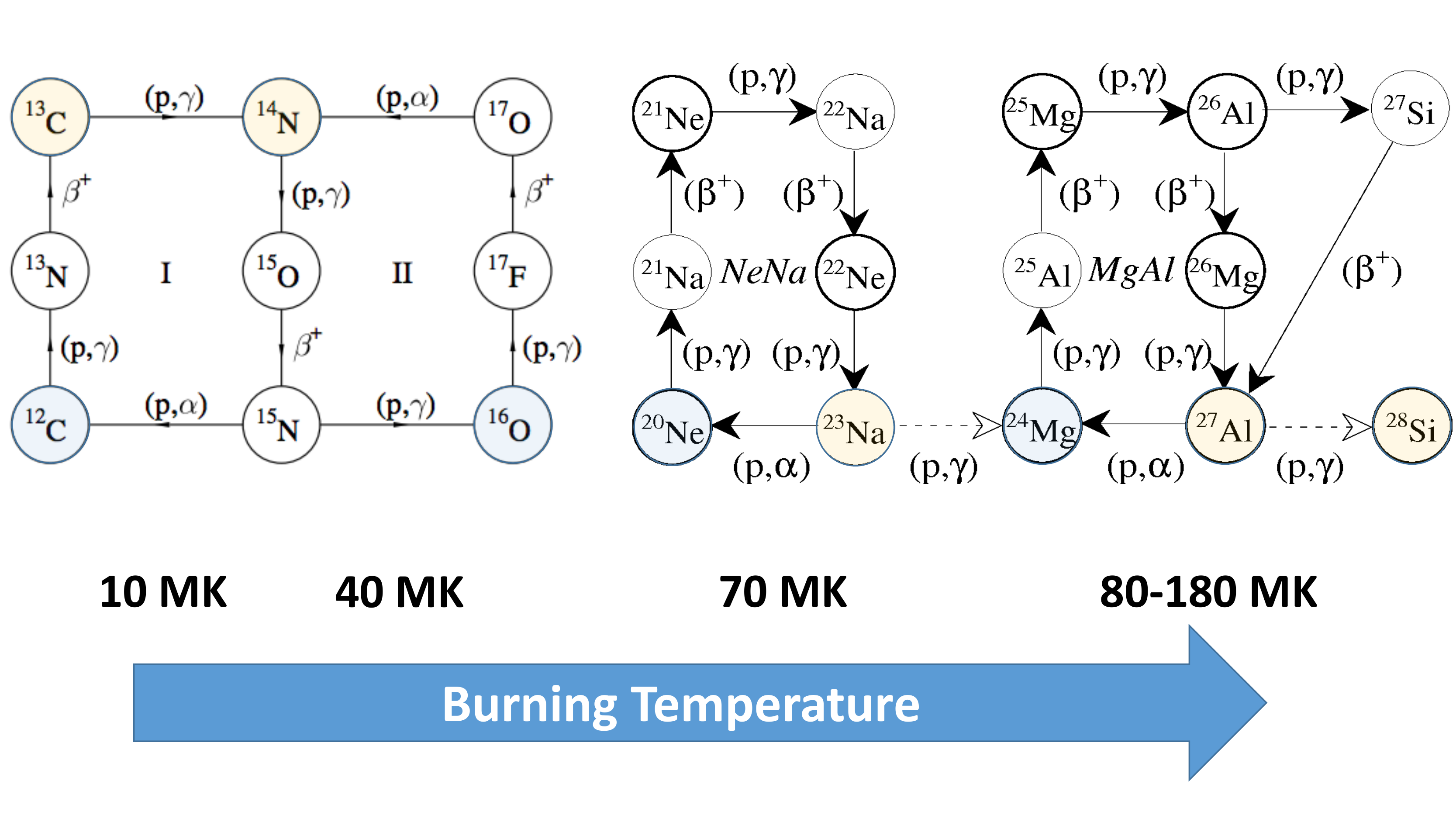}
\caption{The upper part of the figure shows the main p-capture cycles considered in this review: form left to right the CNO-cycle, the Ne-Na cycle, and the Mg-Al cycle. We marked in light orange/light blue those nuclei considered in this review that are produced/destroyed by the various cycles. The lower part of the figure shows the values of the temperature at which the various cycles become efficient. These temperatures are indicative; they depend on the evolutionary phase where the burning occurs. Adapted from \citet{Salaris2002}}
\label{fig:cycles}       
\end{figure}

\subsection{A mechanism to rule them all}

To understand the nature and origin of multiple stellar populations in GCs, the abundance ratios represent our privileged investigation tool, because they provide an unique source of {\it thermometers} and {\it chronometers} to get insights on this phenomenon. Different elements are forged/destroyed at different temperatures, hence by looking at abundance ratios we have an accurate  description of the temperature stratification of the sites where these species were altered (e.g.\citealt{Prantzos2007, Prantzos2017, D'Antona2016}). Moreover, stars of different masses are able to reach different inner temperature, and mass simply means different evolutionary times. Of course, this picture is schematic because the material from which stars of the second generation forms likely comes from stars of the first generation in a range of masses that mix together, possibly blurring and making more confuse the picture.

Simply by looking at the full set of abundance ratios, (anti-)correlated to each other, as observed e.g. in NGC~2808, we may understand several key facts. In the vast majority of GCs (those also classified as Type~I GCs, see Sect. 3.4), the main nucleosynthesis mechanism is very likely proton-capture reactions in H-burning at high temperature \citep{Denisenkov1989, Langer1993}; see Fig.~\ref{fig:cycles}. In the H-burning of main sequence stars, the conversion of C $\rightarrow$ N proceeds at fusion temperature $\geq 10 \times 10^6$ K, while the activation of the ON branch of the CNO cycle requires higher inner temperatures ($\geq 40 \times 10^6$~K). The Na-O anti-correlation, widely observed in almost all GCs, simply results from this conversion of O and the simultaneous production of Na from the NeNa chain, operating at the same temperatures.  Higher temperatures ($\geq 70 \times 10^6$) allow Al to be produced from Mg, whereas T$> 80 \times 10^6$ and $> 180 \times 10^6$ are necessary to produce Si and K, respectively (see \citealt{Prantzos2017}), that are observed in a fraction of the GCs. Temperatures are even higher when considering hot bottom burning \citep[see e.g.][]{D'Antona2016}. Armed with these basics thermometers, an ESO Large Program led \citet{Gratton2001} to change once and for all the paradigm of GCs as simple stellar population. They found a clear Mg-Al anti-correlation among unevolved stars in NGC~6752. Currently observed low mass stars cannot reach the temperature threshold required to activate the Mg $\rightarrow$ Al conversion, thus this nuclear burning must have occurred in more massive stars, already evolved and dead, of a first stellar  generation. The (anti-)correlation among light elements are simply a manifestation of the multiple stellar populations.

There are still problems for theoretical models to quantitatively reproduce the observations \citep{BastianCZS2015}. This led these authors to suggest that possibly the observed pattern cannot be attributed to nucleosynthesis and that alternative scenarios not invoking nuclear burning must be explored. Unluckily, no practical alternative scenario has been found so far. Let us recall the most relevant facts.

\begin{itemize}
\item We know since long time that C and N are anti-correlated in GC stars, but at the same time the sum C+N increases as C decreases, already in unevolved stars (e.g. \citealt{Briley2001, Briley2004}). Either the conversion of O into N must be happening or we are seeing a variable level of N. However, when the trio C, N, O is simultaneously available, the sum C+N+O is found to be  quite constant in the majority of GCs, both in dwarfs \citep{Carretta2005} and in giants (e.g. \citealt{Ivans1999, Smith2005, Yong2008b, Yong2015, Meszaros2015}). Notably, there are exceptions to this rule, that is, clusters where the sum C+N+O is not constant; we will come back to this point Section 3.4.
\item The Na-Al correlation mimics the Na-O anti-correlation, in the sense that links two chains (Ne-Na and Mg-Al) sampling different temperature ranges. Moreover, the sum Al+Mg does not vary in most GCs \citep{Meszaros2015}.
\item In other GCs, while the sum Mg+Al does not stay constant, the sum Mg+Al+Si does, as in NGC~6388 \citep{Carretta2018a}. This agrees with the findings that in very metal-poor and/or massive GCs Si is anti-correlated with Mg and/or correlated with Al abundances (e.g. \citealt{Yong2015, Carretta2009a, Meszaros2015}), an occurrence explained by leakage of Mg on Si bypassing the Al production, when interior temperatures exceed $\sim 65 \times 10^6$ K \citep{Karacas2003}.
\item Star-to-star abundance variations can be traced up to the heavier elements like K and Sc. Abnormal abundances of these elements are actually observed only in about 10-15\% of stars in peculiar GCs such as NGC~2419 and NGC~2808 \citep{Mucciarelli2012, Mucciarelli2015, Carretta2015}; they are however  robustly documented and explained as the output of the Ar-K chain, that bypasses Al production for temperatures in excess of 150 MK \citep{Ventura2012, Prantzos2017}.
\item Finally, there is no correspondence between the observed anti-correlations and either the temperature of condensation on grains or the sensitivity to radiation pressure and sedimentation, the only other selective effects that are known to affect the chemical composition of main sequence stars. For instance, in both cases we should expect that the abundances of CNO are correlated with each other, and Li with Na \citep{Melendez2009, Behr1999}; these patterns are at odds with what is required to explain abundances in GC stars.
\end{itemize}

All these observations show not only the effects of complete CNO cycling, but also that all possible chains of proton-capture reactions (Ne-Na, Mg-Al, Ar-K) were active: the resulting abundances all go in the sense predicted by stellar nucleosynthesis (some being depleted, some produced) in the proton-capture reactions occurring during H-burning, with a few exceptions. Other nucleosynthesis processes such as triple-$\alpha$, $s-$process, and explosive nucleosynthesis should be considered for a minority of GCs (also classified as Type~II, see Sect. 3.4). In addition, we know that the gas polluted by these ejecta was not simply accreted to the surface of forming stars \citep{Gratton2001,Cohen2002,Briley2004}, but went into the formation of the whole star. This conclusion stems from the observation that the same chemical patterns (e.g. the Na-O anti-correlation), are found with very similar  extent in both dwarf and giant stars (e.g. \citealt{Dobrovolskas2014} and \citealt{Cordero2014} for 47~Tuc=NGC~104), despite very different stellar structures (negligible convective envelopes in MS {\it versus} convection extended to more than half of star on the RGB) and H-burning mode (p-p in core burning on the MS and CNO-shell burning along the RGB).

Differences in the multiple stellar populations are due to the ashes of nuclear burning. However, this is not the whole story. Both theory and observations strongly suggest that a majority of the stellar populations following the first burst of star formation must be composed by a mix of nuclearly processed ejecta and gas with pristine composition. First, since star formation is far from being a 100\% efficient process (e.g. \citealt{Lada2003}) there should be gas leftover from the first star formation burst, available for new episodes of stellar formation until pushed out from the cluster. At the same time, if GCs form in high pressure discs of high redshift galaxies (e.g. \citealt{Kruijssen2015}), some fresh gas may be also re-collected from the surrounding ambient medium (see \citealt{Dercole2016}). More importantly, most of the proposed candidate polluters of first generation (FG) are not able to provide more than a few percent of their mass in the chemical feedback process, not enough to reproduce the observed change in chemical composition of second generation (SG) stars (e.g. \citealt{deMink2009}). Dilution with unprocessed gas is mandatory for some kind of polluters, like the AGB stars, fast rotating massive stars and supermassive stars, to turn their correlated O and Na yields into the observed anti-correlation (e.g. \citealt{Dercole2010,Dercole2011,Dercole2012}). Finally, the observations show the presence of Li in second generation stars (e.g. \citealt{Pasquini2005, D'Orazi2015}). Since Li is destroyed at much lower temperature than those experienced in hot-H-burning, either fresh Li must be provided by some class of polluters (see \citealt{Ventura2009}) or dilution with unprocessed matter must be taken into account (as first suggested by \citealt{Prantzos2007}),  or both. We will come back on Lithium in Section~\ref{Sec:5}.

A dilution model where processed material is mixed with variable amounts of gas with primordial composition may easily explain the run of the observed negative and  positive correlations among light-elements (see the discussions in \citealt{Carretta2009a, Carretta2009b}). In this case, two other complications must be considered. First, there is the difficulty to distinguish between SG stars formed by pure processed ejecta and those including a minimum amount of unprocessed matter, as well as to reliably separate FG stars from SG stars affected by large amount of dilution. In turn, the translation of the observed correlations into a temporal sequence may be not trivial or unambiguous (see e.g. the complex sequence devised by \citealt{D'Antona2016} to explain NGC~2808). Second, when large samples of polluted, second (and possibly further) generation stars are scrutinized in details with spectroscopy, discrete (as  opposed to continuous) distributions are detected along the anti-correlations in a growing number of GCs (e.g. M~4=NGC~6121, 47~Tuc=NGC~104, M~28=NGC~6626, NGC~6752, NGC~5986, NGC~2808, NGC~6388, NGC~6402). This pairs with the observations of very common multiple sequences in the colour-magnitude diagram (see e.g. \citealt{Bedin2004, Piotto2007}) and multiple groups in the photometric chromosome map \citep{Milone2017}. Note however that these different groups may be not exactly homogeneous and that the match between groups selected from spectroscopy and photometry is not always exact (see \citealt{Carretta2015} for the exemplary case of NGC~2808). For several GCs a single dilution model is not adequate to reproduce simultaneously the components with extreme and intermediate composition, leading to the conclusion that the operation of different classes  of FG polluters is likely required to fit all the observations \citep{Carretta2012, Johnson2017a, Carretta2018, Carretta2018a,Johnson2019}. Alternatively, we might perhaps consider the possibility that the whole scenario of multiple generations is incorrect, or at least, complicated by the presence of other mechanisms too.

Strictly connected to the above issues of yields from FG polluters and dilution another observation at present seems very difficult to reconcile with model prediction. Even if large amounts of uncontaminated gas were available at early times in the lifetime of GCs the most popular scenarios for self-enrichment are seen to clash with the observed number ratios of SG stars. Since these stars must include in their composition  matter ejected by only a small fraction of FG stars, and their fraction represent the majority of current cluster stars (as tagged from both spectroscopy and photometry) most scenarios are confronted with a sort of paradox. This problem is known as the mass-budget problem, discussed below (see Section 8.1). The most common way out has been to assume that current GCs are only a small remnant of their starting mass, and were able to get rid of most of their FG components in early times, in particular during the phase of expansion driven by the loss of SNe II ejecta (e.g. \citealt{D'Ercole2008}).

Closing this discussion, we note that according to most models, the extra-He observed in the SG stars is produced during the main sequence phase and brought to the surface of the polluter by the second dredge-up; on the other hand, if the massive AGB stars are considered as polluters, the anti-correlations described above are produced by hot bottom burning during the early AGB phase, that is a later evolutionary phase. The correlation between the production of He and other elements then depends on details of the models, such as the efficiency of convection and mass-loss.

\subsection{Neutron-capture elements}

Trans-iron elements are produced via two main mechanisms, involving neutron capture processes because of the high Coulomb barriers: the {\it slow} neutron capture process (the $s-$process) and the {\it rapid} neutron capture process (the $r-$process), where slow and rapid are defined with respect to the $\beta$ decay timescale. The exact site of the $r-$process is still controversial, however due to the necessary conditions of high neutron density and high temperature, core collapse supernovae and neutron star mergers are the most likely candidates (see e.g., \citealt{qian96}, \citealt{freiburghaus99}, and the recent review by \citealt{cowan19}).

The majority of the $s-$process elements between Fe and Sr (60 $<A<$ 90) are produced in massive stars (with initial mass M $>$ 8 M$_\odot$), defining the {\it weak} s component \citep{kappeler99}. In these stars, the main neutron source is provided by the $^{22}$Ne($\alpha,n$)$^{25}$Mg reaction, activated at the end of the convective He-burning core \citep{prantzos1990}  and in the following convective C-burning shell (e.g., \citealt{raiteri1991}). The $^{22}$Ne abundance available in the He core is produced from the initial CNO isotopes, which are converted to $^{14}$N during the H-burning phase, and then to $^{22}$Ne by two $\alpha-$captures (see \citealt{pignatari2010} and references therein). 

For $A>$ 90, the $s-$process elements are produced in AGB stars ($\approx$ 1.3--8 M$_\odot$ ) forming the {\it main} $s-$component (see \citealt{karakas2014} for an updated review on the nucleosynthesis of low-mass and intermediate single stars). In AGB stars, carbon and $s$-process elements are produced during the thermal pulse (TP) phase and brought to the surface by mixing episodes known as third dredge-up events. The $s-$process elements are produced via the capture of neutrons on Fe seeds, with neutrons released both during the H-burning phases, within a so-called $^{13}$C ``pocket'' by the $^{13}$C($\alpha$,n)$^{16}$O reaction at temperatures in the order of 100 MK, and during the TPs by the $^{22}$Ne($\alpha$,n)$^{25}$Mg reaction, if the temperature reaches above 300 MK \citep{cristallo2009}. In AGB stars of relatively low mass ($\lesssim 4M_\odot$) the temperature barely reaches 300 MK, and the $^{13}$C nuclei are the predominant neutron source, at variance with higher masses for which $^{22}$Ne neutron is the main mechanism. The $^{13}$C and $^{22}$Ne neutron source produce different $s-$process paths, implying that isotopes beyond the first s-process peak can be reached (up to Pb at low metallicity) at lower burning temperatures (smaller masses) while a significant Rb production is only predicted with high temperature burning (larger masses).

In general, GCs are homogeneous as far as n-capture elements are concerned (e.g., \citealt{armosky1994}; \citealt{james2004}; \citealt{D'Orazi2010c}), with the exception of ``anomalous'' (or Type~II) clusters (e.g., \citealt{Yong2008c}; \citealt{Marino2015}), where variations in the s-process content are detected in conjunction with iron, CNO and p-capture element abundances (see dedicated discussion of these GCs in Section~\ref{sec:types}). An interesting case in this framework is represented by the metal-poor GC NGC~7078=M~15, where \cite{Sneden1997} first detected a Ba variation simultaneous with Eu, suggesting an $r$-process enrichment within this cluster. This trend was later confirmed by \cite{Otsuki2006}. As for the primordial composition, GC stars exhibit a typical $r$-process rich pattern, which reflects pollution episodes (before the cluster formation) related to massive star nucleosynthesis. Interesting enough, in some cases the n-capture primordial abundance is conversely $s$-process rich: e.g. NGC~6121=M~4 displays an average $s$-process element content significantly larger than other GCs (including its $twin$ NGC~5904=M~5), with an enrichment more than a factor of two larger than typical values for field stars of similar metallicity (\citealt{Ivans1999}, \citealt{D'Orazi2010c}, \citeyear{D'Orazi2013}).

\subsection{Some caveats}

When considering the various abundance anomalies observed in GCs, a few facts should be considered. First, the spread is typically expressed as a logarithm of the variation, essentially because this is how observational errors scale, and the O-Na anti-correlation that we observe in virtually all GCs is due to the transformation of previously existing (from cluster formation) O and Ne into N and Na, respectively. Similarly, the Mg-Al anti-correlation is due to the transformation of already existing Mg into Al. Destruction and production are then proportional to the initial chemical composition, and the spread (expressed in the logarithm) is quite independent of the original metal abundance of the cluster, save for the possible dependence of the burning temperature on the chemical composition. The situation is different for the variation of the total content of CNO and of Fe, observed in a fraction of the clusters (called Type~II clusters by \citealt{Milone2017} or iron-complex clusters by \citealt{Johnson2015}; see Section 3.4), where newly produced metals sum up to the existing ones. In this case, much more production is required to obtain the same spread in the logarithm of the abundance in metal-rich clusters than in metal-poor ones. For instance, while the ejecta of a few SNe are enough to cause detectable star-to-star variations in the Fe abundance of a cluster with [Fe/H]=-1.7 (such as M~2=NGC~7089), about 15 times more SNe are required to produce the same (logarithmic) change in a cluster with [Fe/H]=-0.5 (such as NGC~6388), even if the two clusters likely had a similar original mass. The case for He is simpler, first because the He abundances are not measured in logarithmic units, and second because all GCs have similar original values of the Y abundance.

An additional important point concerns the sensitivity of observations to abundance variations. We note here that a variation of $dY=0.01$, that is detectable on the colour-magnitude diagram e.g. considering HB stars, implies an He abundance variation of only 4\%, that is $\sim 0.017$~dex that is not detectable through spectroscopy, neither for He nor for other elements. This should be considered in particular when considering the correlation of a spread in He abundances derived from photometry with spreads for other elements obtained from spectroscopy (see e.g. \citealt{Cabrera-Ziri2019}), especially in a context where we expect a strong dilution effect is present.


\section{What is a GC}
\label{Sec:3}

The aim of this section is to provide a basic classification of GCs, useful to understand the relative roles of origin vs environment/evolution.

\begin{figure}[hbt]
\includegraphics[width=\textwidth]{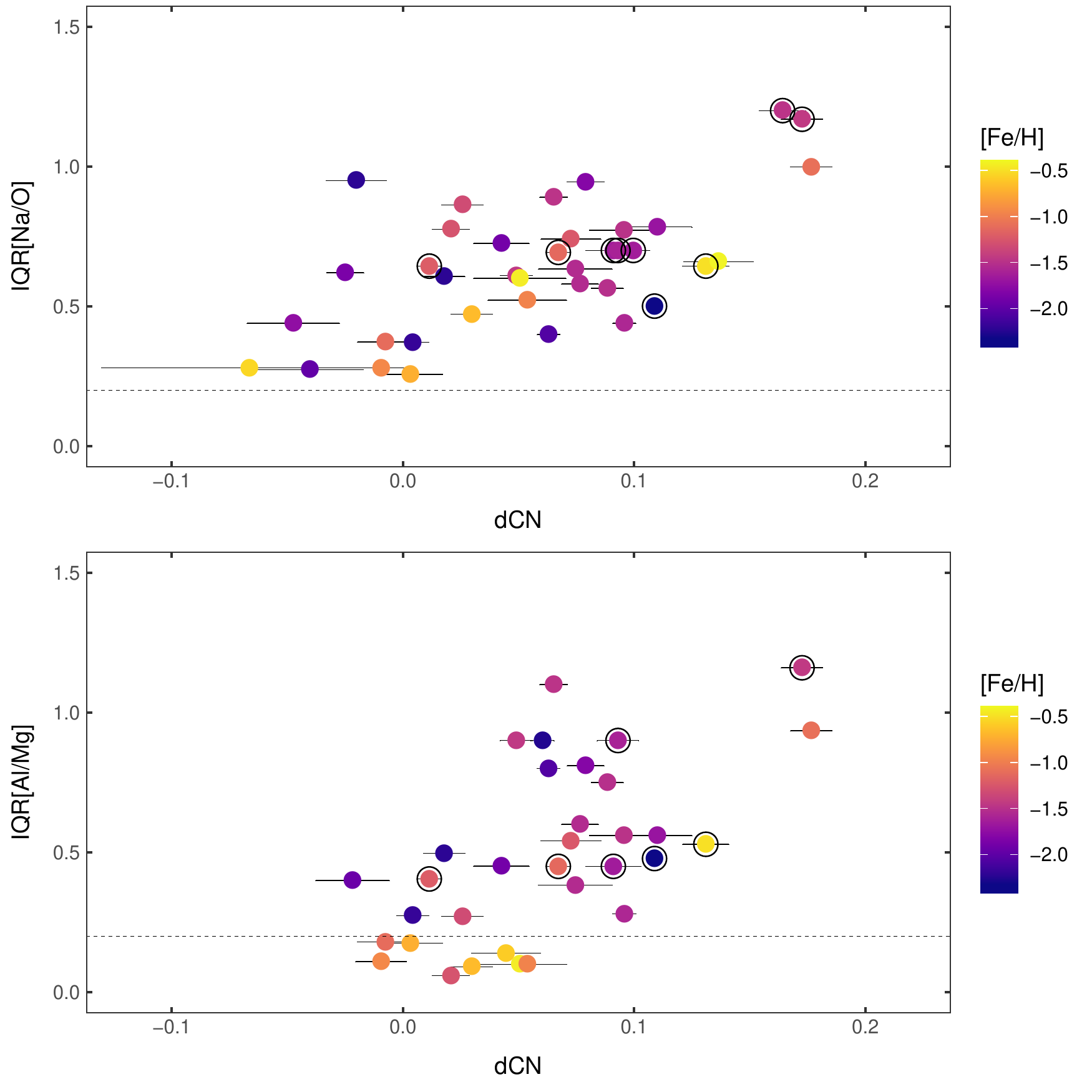}
\caption{Upper panel: run of the interquartile IQR[Na/O] of the distribution of [Na/O] abundance ratios within a cluster (see compilation in Appendix 1) and the spread in CN as derived from the spread in UV colours along the RGB $\Delta_{\rm F275W,F336W,F438W}$\ from \citet{Milone2017}. Lower panel: the same, but for the interquartile IQR[Al/Mg] of [Al/Mg] abundance ratios. Circled dots are for Type~II clusters according to the classification by \citet{Milone2017}. In both panels, points below the dashed lines are actually consistent with no spread at all because they corresponds to the observational errors. Colours code metallicity (see scale on the right of the plot)}
\label{fig:anti_anti}       
\end{figure}

\subsection{Relation between anti-correlations}

While representative of a unique broad phenomenon (see e.g. the good correlations between different element distributions in Fig.~\ref{fig:abu2808a}, and the extensive discussion in \citealt{Marino2019}), the various anti-correlations are not strictly identical with each other. In Fig.~\ref{fig:anti_anti} we compare the index of the spread of the N abundances from HST photometry \citep{Milone2017} with the interquartile of the [Na/O] and [Al/Mg]\footnote{The interquartile of a distribution is the range of values including the middle 50\% of the distribution, leaving out the highest and lowest quartiles.} distribution from the literature (see Appendix 1). The index of spread of N abundances is the value of $\Delta_{\rm F275W,F336W,F438W}$, subtracted by the best fit straight line with metallicity for the clusters with absolute visual magnitude $M_V>-7.3$, as given by the same paper. We first notice that, due to observational errors (typically of the order of 0.1-0.2 dex), the interquartiles are always positive, even when no real spread exists. We will then hereinafter assume that an interquartile value smaller than 0.2 dex is compatible with no spread at all. Second, there is no reason to think that $\Delta_{\rm F275W,F336W,F438W}=0$\ implies no spread in N abundances. Rather, a comparison with the internal spread in N abundances considered in \citet{Milone2018} indicates that the clusters with the smaller values of $\Delta_{\rm F275W,F336W,F438W}$ still have a spread in N abundances as large as $\sim 0.2$~dex, while those with the larger values have a spread as large as $\sim 1.2$~dex. In addition, the removal of the metallicity dependence using a simple offset from a reference line may be simplistic, so that $\Delta_{\rm F275W,F336W,F438W}$\ should only be considered as a proxy for the N abundances.

Once this is taken into account, there is a roughly linear relation between the spread in N abundances and the interquartile of the [Na/O] ratio. On the other hand, the Al/Mg relation is offset with respect to the two others: only clusters with positive values of $\Delta_{\rm F275W,F336W,F438W}$ - that implies a spread in N abundances larger than 0.5 dex - have a significant spread in the Al/Mg ratio. In addition, only a small fraction of the metal-rich clusters ([Fe/H]$>-0.8$) exhibits a spread in Al/Mg. We will come back on this point later, but the different behaviour of the O-Na and Mg-Al anti-correlations was first noted by \citet{Carretta2009b} and later confirmed by other studies \citep[see e.g.][]{Nataf2019}.

The variation of $\Delta_{\rm F275W,F814W}$\ at nearly constant $\Delta_{\rm F275W,F336W,F438W}$ in FG stars observed in many clusters is discussed at length in \citet{Marino2019} (and references therein). While this might in principle be an indication of a spread in He abundances without a corresponding spread in N, they argue that the most probable explanation is rather a small spread in the Fe abundances that should likely be primordial. On the other hand, while star-to-star differential reddening is corrected while deriving the chromosome map, it is also possible that some residuals may remain: this also may contribute to this spread.

\begin{figure}[htb]
\includegraphics[width=\textwidth]{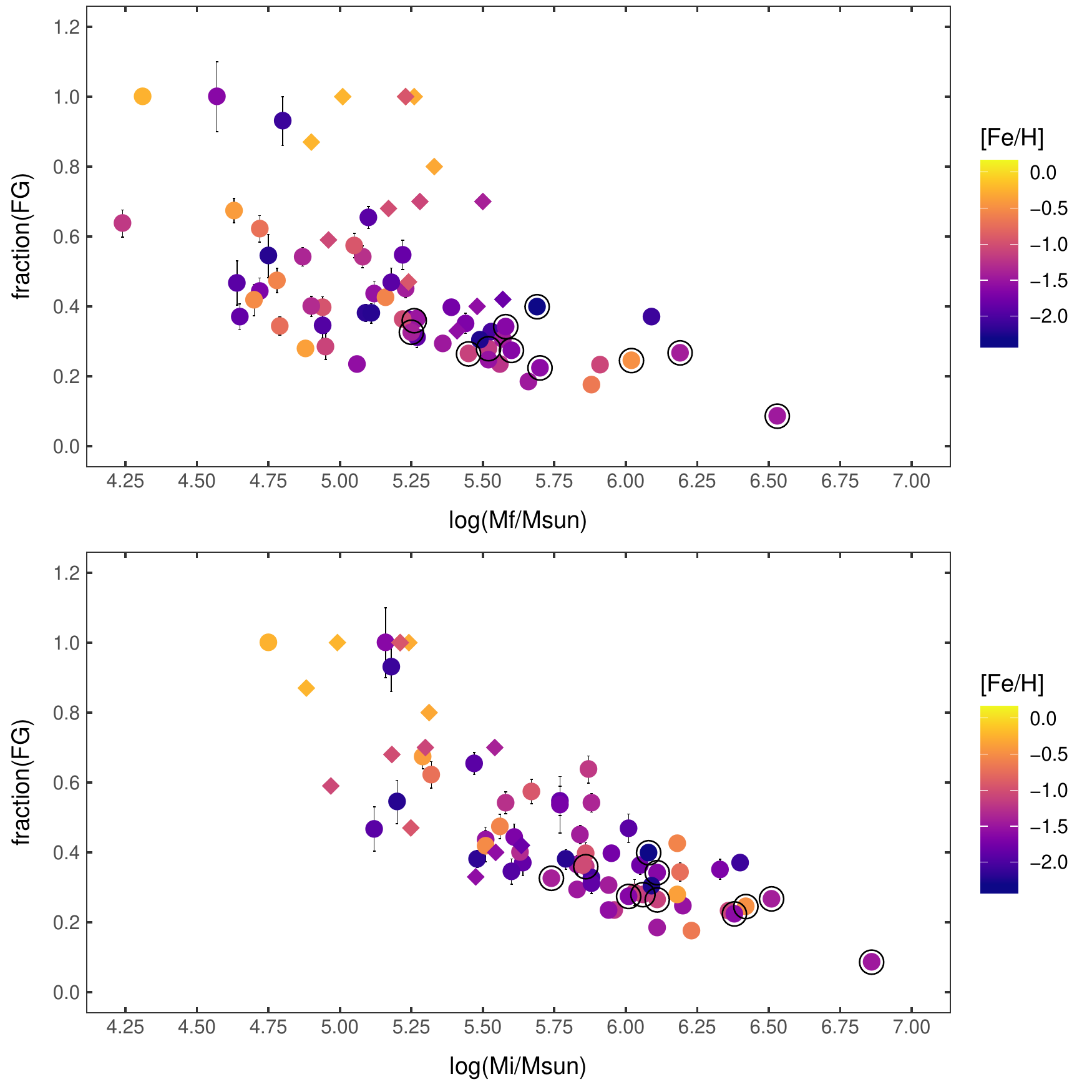}
\caption{Upper panel: run of the fraction of first generation stars within a cluster \citep{Milone2017} and the cluster mass from \citet{Baumgardt2019}. Lower panel: the same, but using the initial mass from \citet{Baumgardt2019}. Circled dots are for Type~II clusters according to the classification by \citet{Milone2017}. Diamonds are clusters in the MCs (see Appendix for references). Note that for the latter in case of unknown metallicity [Fe/H]=$-1$ was adopted for the sake of the figure.  Colours code metallicity (see scale on the right of the plot)}
\label{fig:mass_f1}       
\end{figure}

\subsection{Cluster masses}

Arguably, the most important parameter determining the chemical evolution within a cluster is its original mass. This may be largely different from the current mass because the clusters have a substantial dynamical evolution, and several attempts have been done in the past to better estimate this quantity. Very recently, \citet{Baumgardt2019} published new estimates for the current and original mass of MW GCs based on extensive comparisons between observational data for the surface luminosity and internal velocity distribution and N-body computations, and exploiting the Gaia DR2 data to better estimate distances and 3-d motion of the clusters. The original masses reported by these authors neglect many processes possibly occurring during the early stage of evolution \citep[like e.g. gas expulsion, remnant retention, collisions with giant molecular clouds, etc.][]{D'Ercole2008,Giersz2019} which could determine important amount of mass loss. Nevertheless, they represent the first attempt to account for the slow mass loss process occurred during the last Gyrs of dynamical evolution. We will then use them to estimate the masses at the end of the very complex formation phase of GCs. We will come back on this  point in Section 8.1. In Fig.~\ref{fig:mass_f1} we compare the run of the fraction of first generation stars from \citet{Milone2017} with the values for the present and original mass of the clusters from \citet{Baumgardt2019}. The scatter of points is substantially reduced when we use the initial rather than the present masses. This suggests that while uncertainties are still not negligible, the initial masses are now reliable enough that we may use them in the present discussion. We notice that using the initial rather than present mass implies to take into consideration the environment where the GC formed, at least at first order.

In addition to the results for the MW, we considered the case of clusters in the Magellanic Clouds (MCs), to enlarge the sample ranges both in mass and age. Estimates of the present mass have been provided by \citet{Mackey2003a} and \citet{Mackey2003b} for a large number of clusters in the Large and SMCs\footnote{Alternative estimates of the current masses for MC clusters are provided by other studies, e.g. by \citet{McLaughlin2005}; while these last authors did not list values for all the clusters considered here, whenever available the masses agree very well with those given by \citet{Mackey2003a} and \citet{Mackey2003b}, but for the single case of NGC~2257.}. Of course, when comparing the properties of the MC clusters with those of the MW ones, we should consider the mass loss by the MC clusters as was done by \citet{Baumgardt2019} for the Galactic clusters. Unluckily, we are not aware of a systematic determination of initial masses for clusters in the MCs similar to that described above. In general, mass loss from MC clusters is considered to be small but it is likely not entirely negligible (see e.g. the discussion in \citealt{Baumgardt2013} and \citealt{Piatti2018}). Hereinafter, we only considered the minimum mass loss that is due to the combination of  stellar evolution \citep{Lamers2010} and of the evaporation related to the two-body relaxation (neglecting then tidal effects, disk shocks, and encounters with giant molecular clouds). Both of them are function of age, and the second one also of the cluster relaxation time (that is presently of the order of 500 Myr for most of the clusters of interest here: see \citealt{Piatti2018}). With these assumptions, the oldest clusters in the MCs have lost at least 30\% of their original mass, while those about 2 Gyr old only about 10\%. We corrected the points relative to the MC clusters for these effects on the lower panel of Fig.~\ref{fig:mass_f1}. However, it is very possible that the initial masses determined in this way are underestimated for some of the oldest clusters (see \citealt{Baumgardt2013}).

We can compare these masses with the fraction of FG stars. Initial values of these quantities were derived by \citet{Carretta2010c} from spectroscopic surveys, that call Primordial (or P) the FG stars. \citet{Carretta2010c} found a quite uniform value of the fraction of FG stars in GCs of  30\%, although with cluster-to-cluster variations within quite large error bars. These values, with some addition from later papers, were used e.g. in the discussion by \citet{Bastian2015}. However, the separation between FG and SG stars according to \citet{Carretta2009a} might be affected by small number statistics and the presence of outliers. The locus of FG stars was individuated by comparing the P components in GCs to the field halo stars of similar metallicity, as shown e.g. in Fig. 10 of \citet{Carretta2016}, where it is evident that the estimate of a typical value of about 30\% of FG stars in GCs is likely correct. However, due to the adopted methodology, a fraction of the SG stars with moderate excess of Na may be disguised as FG stars, an effect that may depend on the actual shape of the distribution of Na abundances. A better statistics to derive the frequency of stars in the different populations in individual GCs is now possible thanks to the HST photometry, see \citet{Milone2017}: these are not directly abundance determinations but rather qualitative labeling based on indices that can be determined accurately for large samples of stars. Inspection of the figures in their paper shows that while in the majority of cases the distinction is clear and the measured fractions yield a clear statistical meaning, there are a few cases where the separation of stars in different populations might be questioned (e.g. NGC~5272=M~3 and a few more): care should then be taken in order not to over-interpret data. As a further note of caution, in some cases slightly different filters and procedures may result in large variations of the estimated fraction of FG and SG stars. The case of NGC~2419 (admittedly the most distant Galactic GC) is a good example. Using the same approach as \citet{Milone2017}, \citet{Zennaro2019} estimated a fraction of $37 \pm 1\%$ of FG stars, as listed in our Table 8, while \citet{Larsen2019} obtained a much higher value of 55\% for this component in the same GC. Both studies are based on HST magnitudes through filters sensitive to CNO absorption features.

Fig.~\ref{fig:mass_f1} indicates that there is a close (anti-)correlation between the initial masses of the cluster and the fraction of FG stars as found using the HST photometry by \citet{Milone2017}, in disagreement of a uniform value. The lower panel indicates that an initial mass in the range between $8\times 10^4$\ and $\sim 2\times 10^5$~M$_\odot$\ is likely required for the multiple population phenomenon\footnote{The sample of clusters in \citet{Milone2017} may suffer from a selection bias, because only rather nearby and massive GCs have been targeted (the selection is essentially that of the ACS Survey by \citealt{Sarajedini2007}). On the other hand, these are also those GCs for which more precise data can be obtained. A similar bias can of course be present in case spectroscopy is used to define the populations fractions. It would be interesting to extend the same kind of studies to a sample fully representative of all MW GCs.}. In this mass range, there is actually a considerable cluster-to-cluster scatter in the fraction of first/second generation stars. This might possibly be simply an effect of the uncertainties existing in the determination of the masses and of the fraction of FG stars in small clusters, where also the samples available from photometry become limited in size; however, we cannot exclude that some parameter(s) other than mass is (are) also important. As a matter of fact, it is not even exactly clear what we mean for initial mass in the framework of a multiple population scenario. We also notice that the fraction of FG stars is not strongly dependent on cluster metallicity.

Finally, \citet{Baumgardt2018} also found a quite close (anti-)correlation between the fraction of FG stars from \citet{Milone2017} and the escape velocity from the cluster, supporting an early suggestion by \citet{Georgiev2009}. This suggests that the correlation with cluster mass may actually be due to a higher capability of massive cluster to retain a larger fraction of the ejecta from FG stars, that may be used to produce next generations \citep{Vesperini2010}. However, this issue may be more complex, as we will see in Section 8.1.

\begin{figure}[htb]
\includegraphics[width=\textwidth]{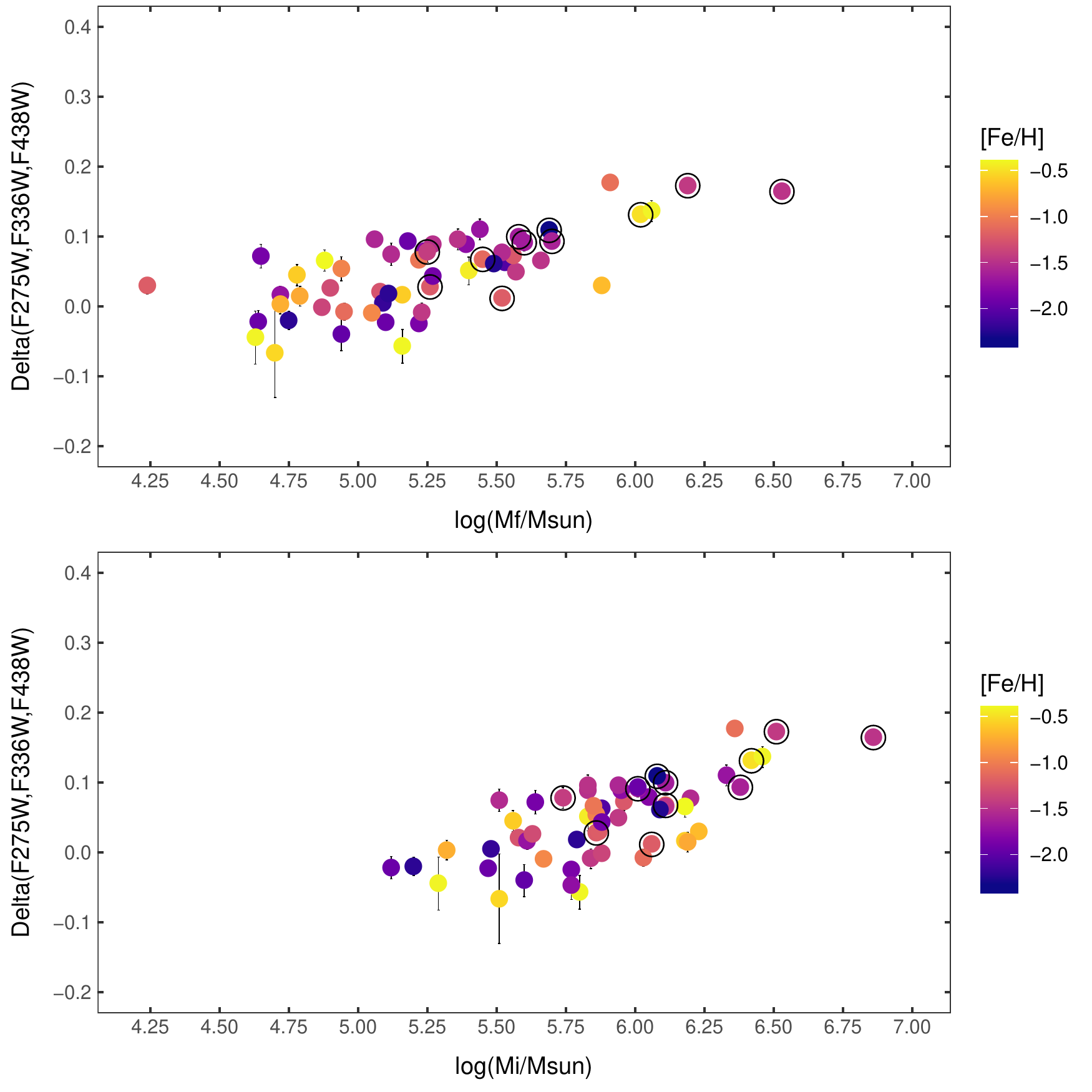}
\caption{Run of the spread of UV colours $\Delta_{\rm F275W,F336W,F438W}$ within a cluster \citep{Milone2017} with final and initial cluster mass from \citet{Baumgardt2019}. Circled symbols are for Type~II clusters according to the classification by \citet{Milone2017}.  Colours code metallicity (see scale on the right of the plot)}
\label{fig:min_cn}       
\end{figure}

\subsection{Small clusters}

\begin{figure}[htb]
\includegraphics[width=\textwidth]{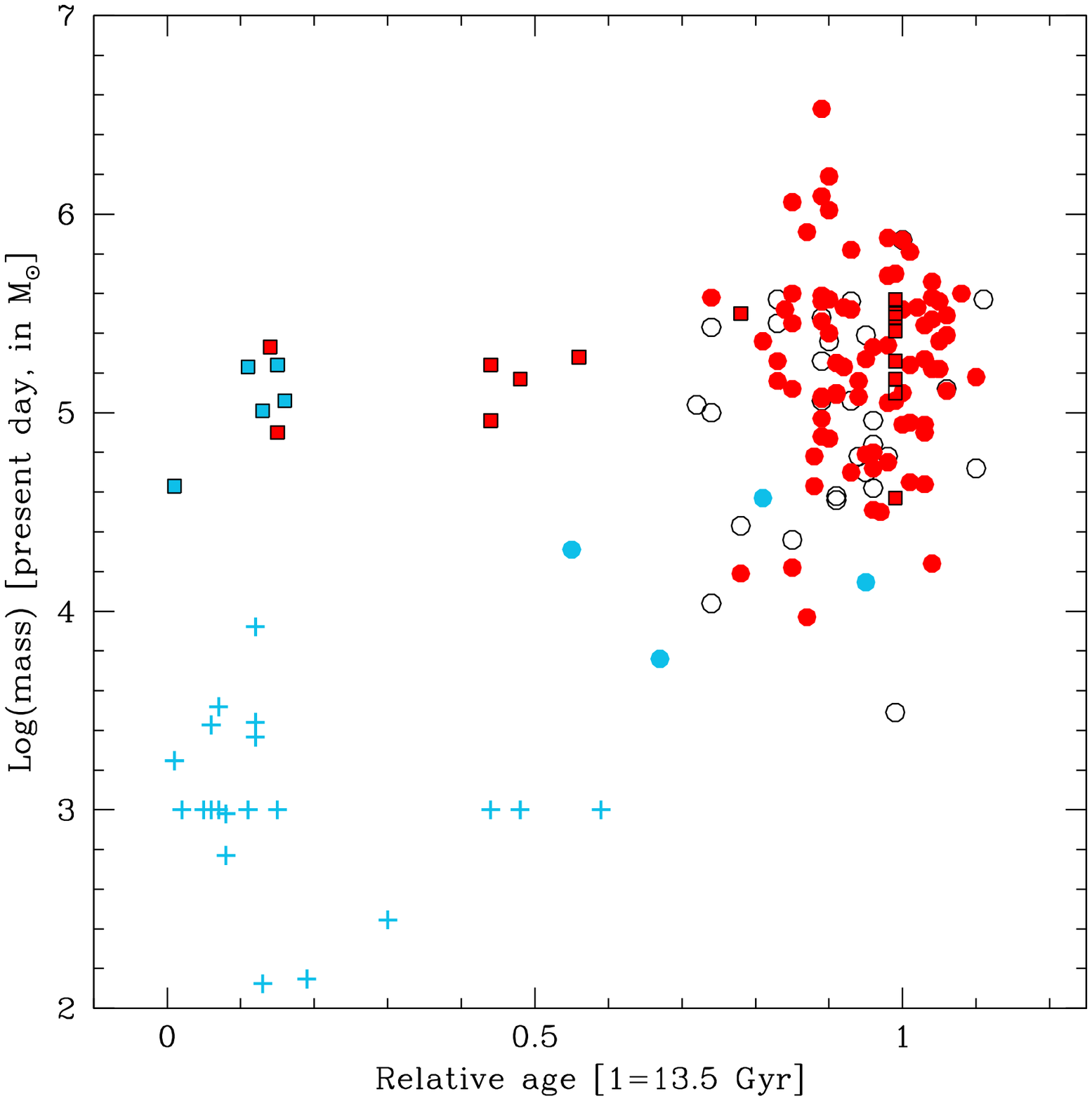}
\caption{Relative age (where 1=13.5 Gyr) versus the logarithm of the present-day cluster mass for MW GCs (open and filled circles), MW satellites GCs (SMC, LMC, Fornax; filled squares), and open clusters (plus signs; whenever the mass was not available from Piskunov et al. 2008, we adopted Log(mass)=3). Open symbols indicate no information on MP, red and light blue colour indicates the presence or absence of MPs, respectively. All open clusters are single populations, while  GCs in the MCs may show multiple populations even at comparable ages.}
\label{fig:age_mass}       
\end{figure}

Establishing that a cluster does not have MPs is not easy, because ``lack of evidence" does not necessarily means ``evidence of lack". Clusters originally thought to be homogeneous were lately shown to possess MPs, though with only a small fraction of SG stars or small spread in abundances (see e.g. the case of IC~4449, \citealt{Dalessandro2018a}). On the other hand, when the observed spread in abundance is small, MPs can be claimed where there may be not present. For instance, the original claim of MP in NGC~6791 \citep{Geisler2012} has not been confirmed by later more extensive and accurate data sets \citep{Bragaglia2014, Villanova2018}. With these caveats, a census of clusters for which MPs were observed has been presented in \citet{Carretta2010c, Gratton2012}, while \citet{Maclean2015} studied the possible presence of Na and O variations compiling data in 20 open clusters (finding none). The census has been updated in \citet{Krause2016} and in \cite {Bragaglia2017}, where more extra-galactic clusters and open clusters were added and results from photometry were considered. Presently, we have information on more than half the known MW GCs. Fig.~\ref{fig:age_mass} shows the mass-age plot for GCs in the MW, Fornax, the MCs, and for some MW open clusters  (masses are generally from \citealt{Baumgardt2019} for MW GCs, \citealt{Mackey2003a, Mackey2003b} for the Magellanic Clouds (MCs) and Fornax, and \citealt{Piskunov2008} for open clusters (these last are highly uncertain see Tables~\ref{tab:A2}, \ref{tab:extra}, and \ref{tab:oc} for references. Had we used M$_V$ as a proxy for mass, as done e.g. in \citet[][see their Fig.~9]{Bragaglia2017}, we would have seen more low-mass clusters but without information on presence or absence of MPs. Furthermore, many new low-mass clusters are being found combining large scale photometric surveys and Gaia \citep[see e.g.][]{Koposov2017, Ryu2018, Torrealba2019}. They are not easy objects to study, either with photometry or spectroscopy, given their faintness and low number of stars but it would be interesting to observe them. 

There are 5 MW GCs for which no MP has been detected to date (E~3, Pal~12, Ter~7, Pal~1, and Rup~106, see \citealt{Salinas2015, Cohen2004, Sbordone2005, Sakari2011, Villanova2013}) and they have generally a low mass, with the exception of the last one. There are lower-mass clusters showing MPs, but we show here the present-day mass, while as discussed previously the original one would be a better choice. However, we do not have the latter for the open clusters and the extra-galactic clusters. We are then forced to use the present-day mass if we want to compare the different cluster families, keeping in mind the strong and differential mass loss affecting clusters during their lifetimes. Anyway, we also note that \citet{Carretta2019} shows that present-day masses of GCs essentially preserve the ranking provided by initial masses, apart from a few exceptions located near the central regions of the Galaxy. This conclusion is here implicitly supported by the two panels of Fig.~\ref{fig:mass_f1}.

The only two open clusters where a large number of stars were observed on purpose to detect variations in Na, O are Berkeley 39 \citep{Bragaglia2012} and NGC~6791 \citep{Geisler2012, Bragaglia2014, Cunha2015, Villanova2018}. No indication of spread in these elements was found (see e.g. the different conclusions in \citealt{Geisler2012} and \citealt{Villanova2018}). After the compilation in \citet{Maclean2015}, data for  significant samples of stars in many other clusters are being acquired by studies or surveys such as the Gaia-ESO or APOGEE (see some examples and references in Table~\ref{tab:oc}). Also for those clusters the variations in O and Na never exceed the errors. Furthermore, high-resolution spectroscopic studies consistently found that open clusters have very homogeneous abundances, once evolutionary effects are taken into account. 

From Fig.~\ref{fig:age_mass}, mass seems to be the fundamental player for the presence of MPs. In fact, even if MPs are present in MCs clusters of ages comparable to those of the old open clusters, they are more massive. The possible dependence on age will be discussed later in the paper (Sect. 3.6).

\begin{figure}[htb]
\includegraphics[width=\textwidth]{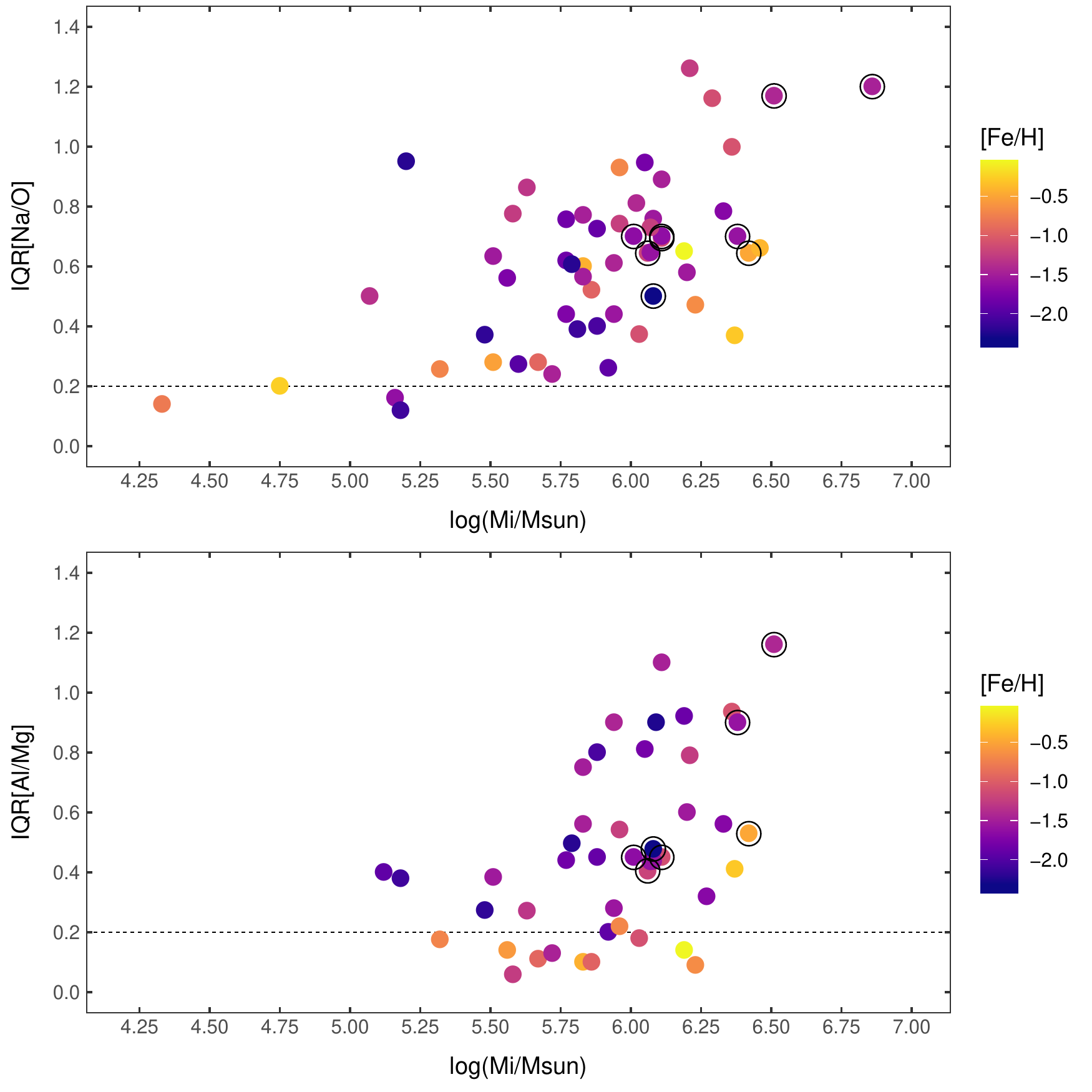}
\caption{Upper panel: run of the interquartile IQR[Na/O] of the distribution of [Na/O] abundance ratios within a cluster (see compilation in the Appendix) and the initial cluster mass from \citet{Baumgardt2019}. Lower panel: the same, but for the interquartile IQR[Al/Mg] of [Al/Mg] abundance ratios. Circled symbols are for Type~II clusters according to the classification by \citet{Milone2017}. Colours code metallicity (see scale on the right of the plot)}. Points below the dashed lines are actually consistent with no spread at all.
\label{fig:min_anti}       
\end{figure}

\begin{figure}[htb]
\includegraphics[width=\textwidth]{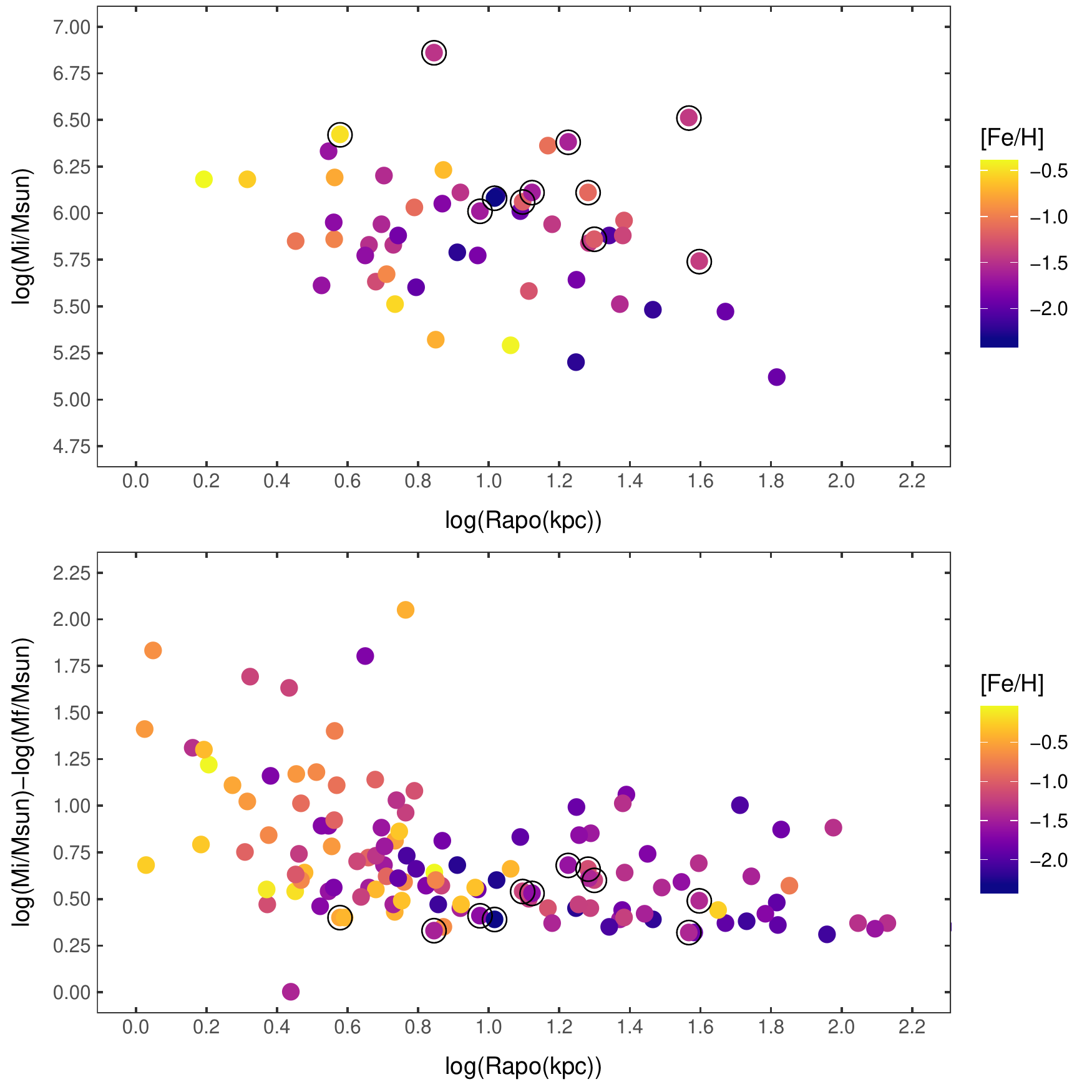}
\caption{Upper panel: run of the initial mass of GCs \citep{Baumgardt2019} with the apocenter of their orbit \citep{Baumgardt2019}; lower panel: same for the mass lost by the GCs. Circled symbols are for Type~II clusters according to the classification by \citet{Milone2017}. Colours code metallicity (see scale on the right of the plot)}
\label{fig:rapo_massin}       
\end{figure}


\subsection{Type~I and Type~II clusters}\label{sec:types}

To interpret the HST multi-colour photometry, \citet{Milone2017} introduced the concept of the chromosome map. This is the distribution of points for individual stars in the $\Delta_{\rm F275W,F336W,F438W}$ versus $\Delta_{\rm F275W,F814W}$\ pseudo-two-colour diagram within an individual GC (see Fig.~\ref{fig:chromosome}). The vertical axis in this diagram is proportional to the spread in the N abundance (higher values of $\Delta_{\rm F275W,F336W,F438W}$ corresponding to higher N abundance), while the horizontal one is proportional to the He content (lower values of  $\Delta_{\rm F275W,F814W}$\ indicate stars with higher He abundances), though it is also sensitive to metallicity and differential reddening (see discussion in \citealt{Marino2019}). In the majority of clusters, stars distribute in two main groups, one characterized by low N and He (FG stars), and the other by higher values of the abundances of these elements (SG stars). \citet{Milone2017} called these clusters of Type~I. However, in a small number of clusters the situation is more complex, with at least a third group of stars occupying a region of high N but low He abundances. \citet{Milone2017} classified them in a Type~II class, and put in this class also $\omega$~Cen=NGC~5139 and M~54=NGC~6715. However, the chromosome maps for these two clusters are more complex than those of the remaining Type~II ones, and they may well be a separate class of objects (see also \citealt{Marino2019}). A more extensive discussion of the properties of Type~II clusters is given by \citet{Marino2018}, who proposed to distinguish these clusters according to the type of chemical anomalies they show (spread in CNO, Fe, or s-process elements). While this kind of separation is not novel (e.g. the Type~II GCs are essentially the iron-complex GCs considered by \citealt{Johnson2015}), the application to a homogeneous set of GCs makes the use of Type~I/II classification a useful working tool.

Since the chromosome map is likely reflecting the imprinting of the early evolution of a cluster, Type~II clusters clearly had a more complex evolution than Type~I clusters. They are usually massive clusters, but there is not a one-to-one correspondence between the cluster mass and their classification according to the chromosome map. However, a clue may be provided by Fig.~\ref{fig:rapo_massin}, where we plot the run of the initial mass of GCs \citep{Baumgardt2019} with the apocenter of their orbit $R_{\rm apo}$\ \citep{Baumgardt2019}, with different symbols for clusters belonging to the different classes according to \citet{Milone2017}. Here, we assumed that NGC~7078=M~15 is a Type~II cluster following \citet{Nardiello2018a}. In this diagram, Type~II clusters occupy the upper envelope of the distribution, that is, they are systematically among the largest clusters for a given apocenter distance. Part of this segregation might be due to selection effects, because there are many GCs that are not plotted in this diagram because they were not included in the survey by \citet{Milone2017}, which is biased against outer halo GCs. Since they adopted the sample of GCs of the ACS Survey by \citet{Sarajedini2007}, their selection include 25 bulge/disk GCs, 25 inner halo clusters and only 6 outer halo GCs, where the classification is the one defined in \citet{Carretta2010c}. Also, there are massive GCs that are not of Type~II, such as 47 Tuc=NGC~104, NGC~2808, or NGC~2419. The two last GCs have however a complex chromosome map, even if not clearly displaying the characteristic shape of Type~II clusters\footnote{NGC~2808 has at least five different populations \citep{Milone2015c,Carretta2015}.  NGC~2419, a very massive cluster with a very large apocenter distance, also shares many characteristics of the chromosome map with NGC~2808, as suggested by the very recent study by \citet{Zennaro2019}. However, it is not plotted in Fig.~\ref{fig:rapo_massin} because it actually lacks an explicit classification in Type~I/II classes.}. We may consider $R_{\rm apo}$\ as a proxy for the distance where clusters formed; actually it should be better considered as a lower limit for this quantity, because the orbit of a GC within a satellite of the galaxy should decay with the orbit of the host, likely dominated by dynamical friction, as e.g. it was likely the case for M~54=NGC~6715 in the Sagittarius galaxy. However, statistically we may consider that a large value of $R_{\rm apo}$\ implies that the GC formed farther out in the meta-galaxy. We may then interpret the correlation between complexity of the chromosome map, $M_{\rm in}$\ and $R_{\rm apo}$\ as an indication that the environment actually played a role in the multiple population phenomenon, in the sense that very massive clusters that formed in the very outer regions of the meta-galaxy had the possibility of a more extended and complex evolution than clusters that formed closer to the center. 

Finally, there are clusters missing a clear classification but suspected to display an Fe abundance spread, such as e.g. NGC~5824 \citep{DaCosta2014, Roederer2016}; however this particular claim has been recently dismissed by \citet{Mucciarelli2018}, who rather suggested that this object has a very extreme Mg-Al anti-correlation. Even more recently, \citet{Yuan2019} proposed that NGC~5824 is the nuclear star cluster of a galaxy that originates the Cetus stream.

Type~II clusters have several other systematic differences with respect to Type~I. They in fact show a split subgiant branch, that may be interpreted as due to a variation in the total CNO content \citep{Yong2009, Yong2014b, Marino2011b, Marino2012, Marino2015, Carretta2013a, Villanova2014, Yong2009, Ventura2009, Yong2015}, and indication of some definite spread in the abundances of s-process elements \citep{Marino2011b, D'Orazi2011, Carretta2013b, Johnson2015, Marino2015, Yong2016, Marino2018}. The various features observed for these clusters likely require that in addition to H-burning at high temperatures (within supermassive stars, fast rotating massive stars or massive AGB stars) there should also be a contribution by triple-$\alpha$ reactions occurring during thermal pulses (in AGB stars), at least if the scenario of multiple generations is correct.  The timescale required for the evolution of stars that produce this nucleosynthesis is $\gtrsim 0.2-0.5$~Gyr \citep[see e.g.][]{Cristallo2015} that is much longer than that required for those stars where H-burning at high temperature occurs. In this group of clusters there are typically at least four different populations (Na-poor, CNO-poor; Na-rich, CNO-poor; Na-poor, CNO-rich; Na-rich CNO rich: see e.g. the cases of M~22=NGC~6656, \citealt{Marino2012}; NGC~1851, \citealt{Gratton2012c,Lardo2012}; NGC~5286, \citealt{Marino2015}), but there are clearly more complex cases, such as the seven populations of M~2=NGC~7089 \citep{Yong2014b, Milone2015b}. In addition, there is clear evidence for a significant spread in the Fe-peak elements, suggestive of a deep potential well able to keep at least a (very small, see Sect. 8.1.2) fraction of the supernova ejecta, for M~54=NGC~6715 \citep{Carretta2010, Carretta2010b}, $\omega$~Cen=NGC~5139 \citep{Norris1995, Suntzeff1996, Stanford2006, Johnson2010, Marino2011, D'Orazi2011, Gratton2011b, Villanova2014, Bellini2017}, and NGC~6273 \citep{Johnson2015, Johnson2017a}. More limited spread in Fe abundances ($\leq 0.2$~dex) have been claimed also for M~22=NGC~6656 \citep{Marino2012}, NGC~1851 \citep{Carretta2011b,Gratton2012c}, NGC~5286 \citep{Marino2015}, M~2=NGC~7089 \citep{Yong2014b, Milone2015b}, and NGC~6934 \citep{Marino2018}, though some of these results are controversial (see e.g. \citealt[][for M~22=NGC~6656 and M~2=NGC~7089, respectively]{Mucciarelli2015a, Lardo2016}). Finally there is the case of M~15=NGC~7078, which, unique among Type~II shows a spread in the content of the $r-$process elements \citep{Sobeck2011,Worley2013}, while its Fe content is uniform \citep{Carretta2009a}. No obvious Fe abundance variation was instead found in the Type~II clusters NGC~362 and NGC~6388 \citep{Carretta2009a, Carretta2009b, Carretta2013b, Carretta2018a}, while there are not yet published high-resolution spectroscopic data for NGC~1261. As mentioned in Section 2.5, detection of star-to-star variations in the Fe and total CNO abundance is expected to be more difficult in a metal-rich cluster. This may be the case of NGC~6388 (but neither of NGC~362 nor M~15=NGC~7078), so that lack of detection of these variations might not mean that there is a systematic difference between this cluster and the remaining Type~II clusters.

In addition, Type~II clusters tend to have a larger initial mass than Type~I with the same value of IQR[Na/O] and IQR[Al/Mg], or conversely tend to have a smaller value of IQR[Na/O] and IQR[Al/Mg] for the same mass (see Fig.~\ref{fig:min_anti}). On the other hand, there is no clear offset for the N abundance indicators ($\Delta_{\rm F275W,F336W,F438W}$\ or the fraction of first generation stars). This indicates that while the multiple population phenomenon is present as in Type~I clusters, in Type~II clusters the role played by H-burning at very high temperature is smaller, perhaps because its effect is diluted by other contributions to nucleosynthesis.

On the whole, the emerging pattern from chemistry is of significant age spreads and complex chemical evolution within individual Type~II clusters. There is clear indication that they had a quite long history within isolated fragments before the internal evolution of the fragment itself or interaction with our own Galaxy caused the final dispersal of any residual gas. This is obviously the case of M~54=NGC~6715, that is the nucleus of the Sagittarius dwarf galaxy \citep{Ibata1994, Bellazzini2008}. It has long been suggested that $\omega$~Cen=NGC~5139 also is the stripped nucleus of a dispersed dwarf galaxy (see e.g. \citealt{Bekki2003}, though direct evidence is still elusive (see e.g. \citealt{Navarrete2015}; see however also \citealt{Myeong2018a, Myeong2018b} for possible hints that the ashes of this galaxy might actually be dispersed in the Galactic halo, and the very recent result by \citealt{Ibata2019} that identifies a stellar stream found on Gaia DR2 data as the possible tidal tails). Searches for extended halos of stars around other Type~II clusters, possible remnants of the galaxies originally surrounding them and to be separated from narrow tidal tails along the orbit that may arise also for isolated clusters, have been performed  by e.g. \citet{Olszewski2009, Carballo-Bello2014}. The evidence is now established for NGC~1851 \citep{Olszewski2009, Kuzma2018}, M~2=NGC~7089 \citep{Kuzma2016}, NGC~6779=M~56 \citep{Piatti2019} while it is not clear and possibly absent for others (see e.g. NGC~1261: \citealt{Kuzma2018}). On the other hand, the chemical evolution of these objects is surely far from being well described by a closed box model, as already showed by \citet{Suntzeff1996} for the case of $\omega$~Cen=NGC~5139, where these authors estimated that some 90\% of the original mass should have been lost. We will come back on this point in Sect. 8.1.2. More in general, we notice that the typical apocenter distance of these clusters ($\sim 5-50$~kpc) likely underestimates the distance at the epoch of formation because the orbit of a large mass satellite is expected to become closer to the Galactic center with time due to the effect of dynamical friction. It is then not unreasonable to think that the fragments where Type~II clusters formed could have had a prolonged star formation phase before they interacted with the Galaxy and dispersed.

\subsection{Metallicity dependence}

While mass is the leading parameter determining the fraction of first/second generation stars, metallicity clearly plays an important role in the actual nucleosynthesis causing the pattern observed within GCs. This is shown by a comparison of Fig.~\ref{fig:min_cn} and Fig.~\ref{fig:min_anti}. In the first one we plotted the run of the spread of UV colours along the RGB $\Delta_{\rm F275W,F336W,F438W}$, a proxy for the spread in N abundances, with the initial cluster mass \citep{Baumgardt2019}. In the second one we plot the run of the interquartiles of the distribution of the [Na/O] (upper panel) and [Al/Mg] (lower panel) within a cluster with the initial mass of the clusters. These figures shows the expected correlation with cluster mass. The interquartiles are larger than the observational errors (here we assumed 0.2~dex, that is about three times the normal uncertainty in the relevant abundances) only for clusters with initial masses above $3\times 10^5$~M$_\odot$\ for the [Na/O] anti-correlation, and for even a larger mass for the [Al/Mg] one. In addition to this higher threshold with respect to what is observed for the N abundance variations, there is a clear trend for steeper relations for the metal-poor clusters than for the most metal-rich ones. This is very obvious for the [Al/Mg] anti-correlation: only extremely massive (initial mass above $3\times 10^6$~M$_\odot$) metal-rich clusters show a (limited) spread in the Al abundances. This confirms what was originally found by \citet{Carretta2009b} and seen also in APOGEE and Gaia-ESO data (see \citealt{Meszaros2015,Masseron2019, Pancino2017}). An even more extreme effect is exhibited by the Ca-K anti-correlation, that has been actually found only in the very massive and metal-poor cluster NGC~2419 \citep{Mucciarelli2012, Carretta2013b} and in NGC~2808 \citep{Mucciarelli2015}.

In Section~\ref{Sec:2} we have seen that the various anti-correlations seen in GCs can be interpreted as due to H-burning occurring at different temperatures. The trends existing with mass and metallicity in Type~I clusters can then be interpreted as due to polluters where this burning occurs at increasing temperature. This likely signals a drift toward more massive polluters with increasing cluster mass. In the scenario where the polluters are massive AGB stars, the trend with metallicity may be explained by the fact that the temperature of hot bottom burning is expected to be higher in lower metallicity stars \citep{Lattanzio2000,Ventura2009}.

In general, it is probable that even within a single GC we must consider different classes of polluters. This is quite obvious for Type~II clusters, that cannot be described by a simple mono-parametric dilution distribution. However, there is evidence that a single dilution distribution cannot reproduce simultaneously the [Na/O] and [Al/Mg] anti-correlations even in Type~I clusters such as NGC~6752 \citep{Carretta2012} and NGC~2808 \citep{Carretta2014b, Carretta2015b}, or NGC~6402 \citep{Johnson2019} for which no type is available. This suggests that often the GC formation cannot be described by a simple scenario with only two episodes of star formation and that the multiple population phenomenon is possibly more complex.

\subsection{Is there an age dependence?}

There have been several attempts to assess if mass is indeed the leading parameter determining the presence of multiple populations in massive clusters and to better understand the timescale of the multiple population process, that would be crucial to understand the nature of the polluters. The (possible) evidence for a dependence on age refers to studies of GC analogs both very young and more mature. 

A potentially attractive road is in fact to look for multiple populations in young very populous clusters in nearby galaxies -- there are not such clusters in the MW \citep{Portegies2010}. Of course, this faces with the difficulties of observing far and very dense clusters, but it can be attempted and a summary of results is given in the recent review by \citet{Bastian2018}. 

Among the important results of this line of investigation is the lack of evidence for interstellar matter in clusters older than a few Myr \citep{Longmore2015, Cabrera-Ziri2015, Bastian2014, Hollyhead2015}. If this result would also apply to young GCs, it would be a clear difficulty for scenarios where MPs are formed in various episodes of star formation, though this should be considered carefully for the different timescales involved. On the other hand, it is  not obvious that the dynamical conditions of young populous clusters are actually the same met by objects that are now GCs. Alternatively, \citet{Lardo2017} examined the case of three super-star clusters in the Antennae with ages 7--40~Myr and masses in the range $4\times 10^5$--$1.1\times 10^6$~M$_\odot$. Given the large distance, they could only observe integrated spectra of the clusters and in order to enhance the contribution by red supergiants they considered near-IR spectra where they could only measure Al lines among the possible indices of multiple populations. They did not find evidence for Al enrichment and concluded against the presence of multiple populations in these clusters. However, this is likely not conclusive because within Galactic GCs, large Al spreads are generally limited to clusters with a metallicity below 1/10th of solar (see the lower panel of Fig.~\ref{fig:min_anti}), while the clusters in the Antennae likely have near-solar metallicity \citep{Lardo2015}.  No spread in the Al abundances is actually observed e.g. in 47~Tuc=NGC~104 or NGC~6528, that when young were likely much more massive than the clusters observed by \citet{Lardo2017}.

The second line of evidence of the influence of age concerns surveys of populous clusters in the MCs, that host a population of relatively massive clusters both as old as the MW GCs and young, accessible to more traditional spectroscopic and photometric studies. \citet{Mucciarelli2009} found evidence of multiple populations (based on Na, O anti-correlation) in three very old clusters  (NGC~1786, NGC~2210, NGC~2257, see also Table~\ref{tab:extra}), while no evidence was instead found in the younger clusters NGC~1866 \citep[about 100 Myr ]{Mucciarelli2011} and NGC~1806 \citep[about 1.7  Gyr old]{Mucciarelli2014c}. The only old cluster in the Small MC (SMC), NGC~121, was found to host multiple populations by \citet{Dalessandro2016} and \citet{Niederhofer2017b}, based on HST photometry showing the effects of CN variations. The few stars for which high resolution spectra were obtained by the former work did not show spreads in Na or O. This agrees with the small fraction of second generation stars in the cluster, compared to MW analogs. 

More recently, the Bologna and Liverpool groups joined forces to search for signatures of large star-to-star variation in the N abundances, based mostly on HST photometry (i.e. based on C, N, and possibly O variations without information on Na, Mg, and Al). No cluster younger than 2 Gyr showed sign of the presence of multiple populations (see Table~\ref{tab:extra}). Interestingly, these studies resulted in the discovery of two cases of multiple population in clusters 2~Gyr old (NGC~1978: \citealt{Martocchia2018b}; Hodge~6: \citealt{Hollyhead2019}). According to the first study, there is no evidence that the SG is much younger than the first one in those particular clusters, with an upper limit at about 20~Myr. Although the results for NGC~1978 and Hodge~6 should be considered with some caution because the SG in those clusters only constitutes $\leq 20$\% of the total mass in the cluster, making them the clusters with the smallest fraction of SG stars known, these are important constraints for models of multiple populations. They possibly limit the temporal scale where the phenomenon can occur (at least in Type~I clusters) and rule out the possibility that some cosmological effects makes the multiple population phenomenon possible only at high redshift. The absence of any clear evidence for multiple populations in clusters younger than about 2 Gyr prompted the authors to argue that there is an age dependence of this phenomenon, with a threshold between 1.7 and 2 Gyr (yet unexplained by stellar evolution). These data are included in Fig.~\ref{fig:mass_f1}. However, this same figure also indicates that the young clusters considered in the survey are barely massive enough to show the multiple population phenomenon and lie in a region of the initial cluster mass - fraction of FG stars diagram where there are large cluster-to-cluster variations in the fraction of FG stars within a limited mass range. This suggests that clusters in this range of masses might have different histories, and that there may not be a single mass threshold value but rather a range of masses where the transition between single and MP cluster occurs. This is further complicated by the fact that the original cluster masses are not accurately known. We think that disentangling the effect of individual histories from an age effect needs then a large sample of clusters; fortunately,  more clusters are being added by different groups. The most massive young cluster not showing any significant spread in N abundances is NGC~419 \citep{Martocchia2017}, that likely had an original mass of $2\times 10^5$~M$_\odot$. On the other hand,  NGC~1978 \citep{Martocchia2018b} and Hodge~6 \citep{Hollyhead2019} are only slightly older than NGC~419, and of similar mass ($\sim 1-2\times 10^5$~M$_\odot$). We further notice that there are much older clusters of similar mass that also do no not show evidence of multiple populations or have a very minor fraction of SG stars, such as Ruprecht 106 \citep{Villanova2013, Dotter2018} or Terzan~8 \citep{Carretta2014}. We conclude that the role played by cluster age is not yet soundly determined. At variance with \citet{Bastian2018}, we then think it is premature to consider age as a basic parameter in the multiple population phenomenon and that more observations are required to settle this issue.


\section{Impact of chemical anomalies on stellar evolution}
\label{Sec:4}

The multiple population phenomenon impacts nearly all phases of stellar evolution, leading to significant deviations of the colour-magnitude diagrams of GCs with respect to simple stellar isochrones. Actually, the first clear evidence of the multiple populations can be traced back to the second half of the '60s, with the discovery of the so-called second parameter effect on the horizontal branch \citep{vandenBergh1967} and of the wide red giant branch of $\omega$~Cen=NGC~5139 \citep{Woolley1966}. In this section, we will briefly review these points examining some of the main features.

\subsection{The main sequence}


Multiple populations impact the main sequences of GCs. Ultra-violet CMDs show effects related to the spread of He, light elements and n-capture elements. In the optical, the effect is detectable only for what concerns the spread in Helium and heavy element abundances. The latter is actually limited to those few clusters (such as $\omega$~Cen=NGC~5139: \citealt{Piotto2005, Stanford2006, Villanova2014, Milone2017b}), where the different populations differ in the Fe content. The first realization that large variations in the He abundances can be responsible for a split of the main sequence of the mono-metallic cluster NGC~2808 came from the studies of \citet{D'Antona2005, Piotto2007}. On the same timescale, a similar conclusion was reached for the multi-metallic cluster $\omega$~Cen=NGC~5139 by \citet{Norris2004} from star counts, and by \citet{Piotto2005} from the determination of the metal abundance for stars of different colours. Splitting of sequences is however generally tiny and the high quality of HST photometry is usually needed to separate them: for instance, strict upper limits on He abundance variations were obtained from the very narrow MS of NGC~6397 by \citet{diCriscienzo2010}, consistent with the very tiny difference of $\Delta$Y=0.01 later found by \citet{Milone2012b} for this cluster. In addition, He abundance variations should be separated from other effects, such as differential reddening, contamination by field stars, binarity, and variations in heavy element abundances (for this last point, see \citealt{Marino2019}). This is best achieved using multicolour photometry, as first demonstrated by \citet{Milone2012a} for the case of 47~Tuc=NGC~104. Milone and coworkers then applied this method to several other GCs:  NGC~6397 \citep{Milone2012b}, NGC~6752 \citep{Milone2013}, M~62=NGC~6266 \citep{Milone2014}, M~2=NGC~7089 \citep{Milone2015b}, NGC~6352 \citet{Nardiello2015}, and again NGC~2808 \citep{Milone2015c}. Finally, exploiting the extensive HST survey with the WFC3 by \citet{Piotto2015, Nardiello2018b}, \citet{Milone2018} determined He abundances for a wide sample of GCs (for a recent discussion of the aspects related to stellar models, see \citealt{Cassisi2017}). These are used in the discussion in Section 4.3.

A second important piece of information that can be obtained using the main sequence is the luminosity function of the different populations. This first shows that the spread in the proton-capture elements is not limited to the external layers of the stars, as suggested by the fact that the extension of the anti-correlations is similar in main sequence and red giant branch stars that have largely different depth of the outer convective envelope \citep{Gratton2001, Cohen2002}. The determination of the luminosity function is more complex, because dynamical effects might cause selective losses of the first/second generation stars, if they have a systematically different distribution within a cluster, as e.g. observed in the case of 47~Tuc=NGC~104 \citep{Milone2018b} or $\omega$~Cen=NGC~5139 \citep{Bellini2017}. After consideration of these effects, it has been found that the fraction of stars in different generations in the lower main sequence are actually quite similar to those measured along the red giant branch, at least in the cases of M~4=NGC~6121 \citep{Milone2014} and NGC~6752 \citep{Milone2019b}, and possibly even NGC~2808 \citep{Milone2012a}. On the other hand, the only available determination of the mass function of different populations in NGC~2808 \citep{Milone2012d} revealed significant differences. This represents an important constraint for models of the multiple populations (see Section 6).

\begin{figure}[htb]
\includegraphics[width=\textwidth]{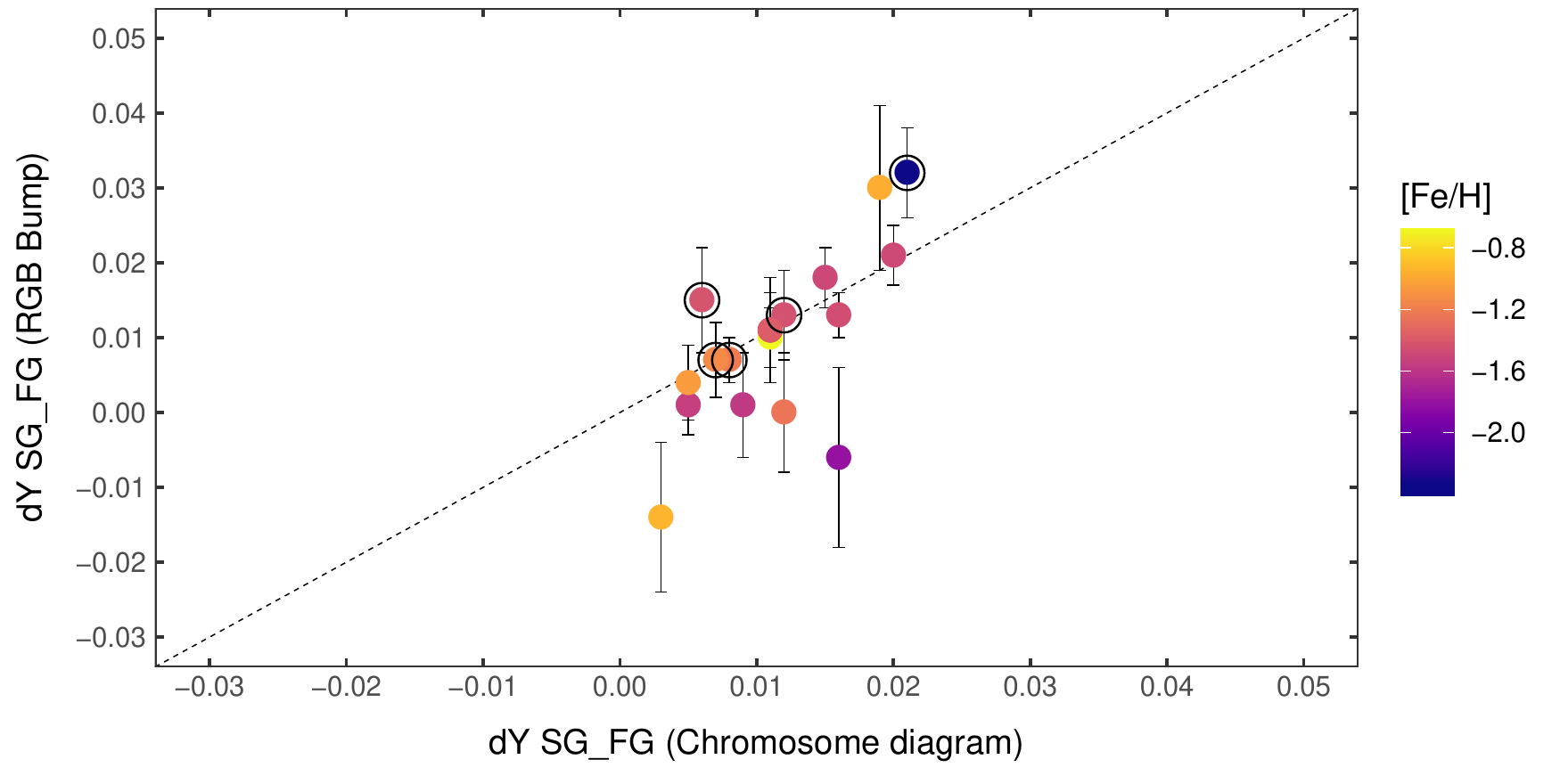}
\caption{Comparison between the difference in He abundances between second and first generation stars obtained from the chromosome map \citep{Milone2018} and from the luminosity of the RGB bump \citep{Lagioia2018}. Circled symbols are for Type~II clusters. Colours code metallicity (see scale on the right of the plot). The overimposed dashed line represents equality.}
\label{fig:rgb_bump}       
\end{figure}

\subsection{RGB bump}

Standard stellar evolutionary models \citep{Iben1969} predict that the maximum penetration of the outer convective envelope at the base of the RGB (first dredge-up) leaves behind a discontinuity in the hydrogen content. When the star evolves along the RGB, the H-burning shell is moving outward in mass and when it reaches a region that has a higher H content (i.e. the inner edge of the previous first dredge-up) a hydrostatic readjustment occurs which causes the star to expand and move down the RGB temporarily, before it continues ascending the RGB. This causes a bump in the luminosity function: this is indeed found in the observational data \citep{King1985}. The luminosity of this bump depends on age, metallicity, and helium content \citep[e.g.][]{Sweigart1979, Cassisi1997}. In particular, stars with higher He are expected to have a brighter RGB bump.\footnote{It is worth noting that current models do not reproduce the correct zero-point \citet{Cassisi2011}, however observational studies have been concentrating on differential effects, and we will limit are discussion to those in this text.}

After the pioneering work by \citet{Sollima2005} on $\omega$~Cen=NGC~5139, \citet{Bragaglia2010a} tried to see if stars of FG and SG, as defined by Na abundances, separated at the bump for 14 GCs in the \citet{Carretta2009a, Carretta2009b} sample. Indeed, a difference in the average position of the two populations showed up, implying a difference in He of about 0.05. However, due to the limited statistic, the result was obtained combining all GCs together\footnote{NGC~2808, the cluster showing the largest He differences, was not in the calculation, since no star below the RGB bump was observed for this cluster in that survey.} and differences in RGB bump luminosity and the connected He difference were scarcely significant, even if in line with other tracers and expectations.

Later on, the luminosity of the bump of different populations has been traced using photometry, thus circumventing the scarcity of stars observed spectroscopically. \cite{Nataf2011} found a gradient in RGB bump properties in 47~Tuc=NGC~104, with the bump becoming less luminous and populated moving toward the external regions. They interpreted this as a variation in He abundance, under the idea that He-enriched SG stars are more centrally concentrated. \citet{Milone2015c} detected the RGB bump in four of the five populations they identified in NGC~2808 using a combination of optical and UV HST filters. By comparing the resulting differences in magnitudes with stellar models they were able to measure the implied differences in Y, which were in line with that found with other methods employing main sequence, HB and RGB stars. More recently, \cite{Lee2017, Lee2018} employed wide field Str\"omgren photometry coupled with a narrow-band filter sensitive to CN variations to separate CN-strong (SG) and weak (FG) stars. They investigated the relative abundance in He using the bump magnitude and derived a difference of $\Delta$Y about 0.028 and 0.016 in M~5=NGC~5904 and NGC~6752, respectively, in agreement with other methods.

The most recent and exhaustive work is by \citet{Lagioia2018}, who studied the RGB bump of the 56 GCs in the UV Legacy Survey \citep{Piotto2015}, excluding only $\omega$~Cen=NGC~5139 because of its complexity and including NGC~2808 which had already been published by \citet{Milone2015c}. \cite{Lagioia2018} separated the different populations using the chromosome map and determined the RGB bump level for FG and SG stars in each of the bands. They retained for further consideration only those clusters with enough statistics (at least 15 stars in each RGB bump) and with a difference in magnitude between FG, SG significant at better than 90\%. This means they were left with 26 GCs (plus NGC~2808). Considering only clusters for which they later derived also $\Delta$Y, the difference in magnitude is $-0.033\pm 0.008$~mag (for comparison, \citealt{Bragaglia2010b} found $-0.045\pm 0.042$). They then compared the differences in magnitude with the magnitude of synthetic RGB bump stars, derived from synthetic spectra with atmospheric parameters (in particular metallicity and temperature) and light-element abundances (C, N, O) appropriate for FG, SG stars and considering two He abundances (standard and enhanced). Difference in He mostly affect the optical bands and \citet{Lagioia2018} determined the $\Delta$Y implied by the $\Delta$mag using synthetic CMDs for the RGB based on the BaSTI tracks \footnote{See http://basti.oa-abbruzzo.inaf.it/ \citep{Pietrinferni2006,Pietrinferni2004}}; they were able to derive the information for 18 clusters, finding that $\Delta{\rm Y}\leq 0.035$ for 14 of them, and about zero for 4 of them. For these GCs (not including NGC~2808), they found $<\Delta{\rm Y}>=0.011\pm 0.002$ between FG and SG stars, without correlation with [Fe/H] or M$_V$. A comparison with values derived from the main sequence \citep{Milone2018} is presented in Fig.~\ref{fig:rgb_bump}, where we use the difference in Y between FG and SG. The two methods give a consistent ranking, even if the slope is not 1.

\subsection{The horizontal branch}


\begin{figure}[htb]
\includegraphics[width=\textwidth]{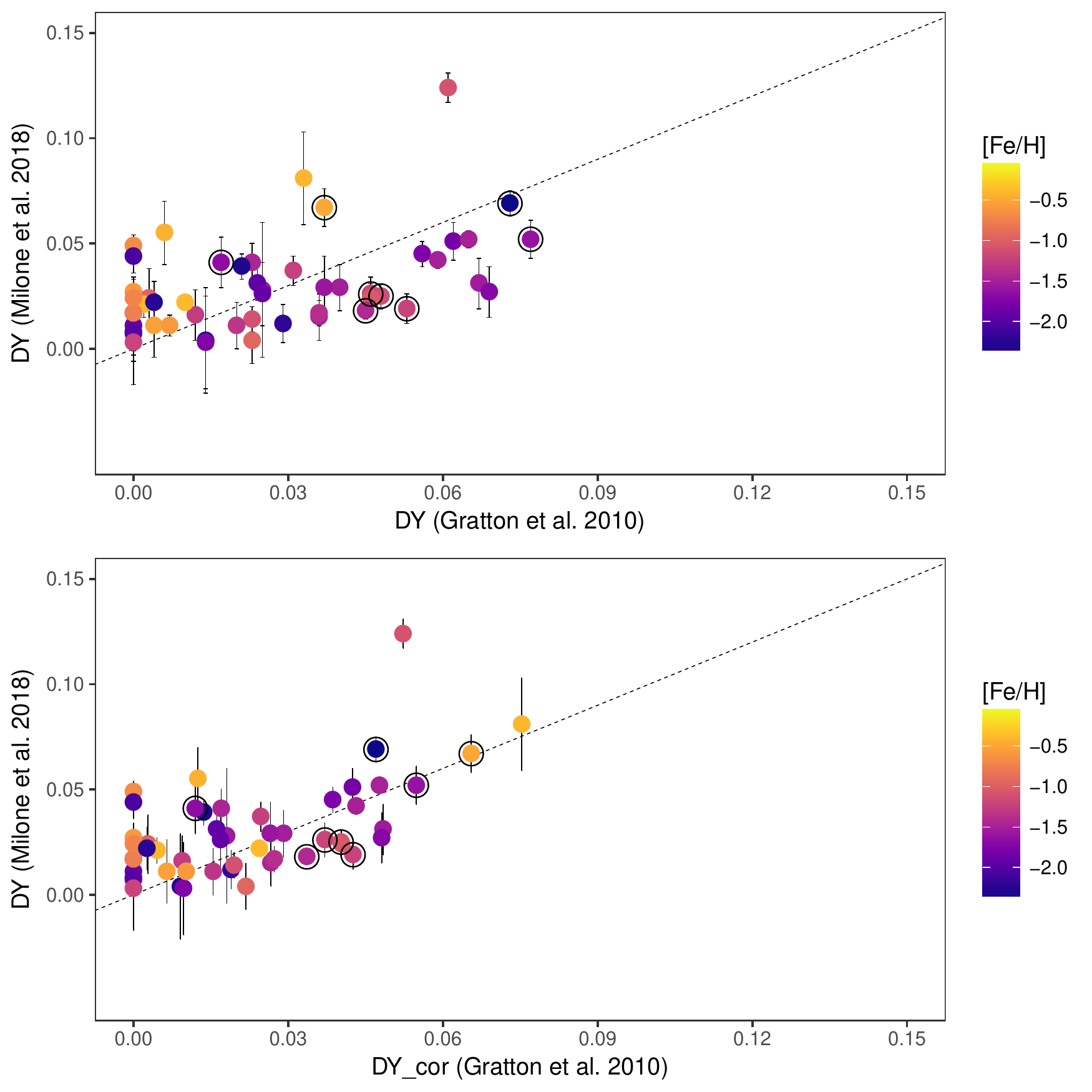}
\caption{Upper panel: comparison between the spread in He abundances within a cluster as obtained from the analysis of the horizontal branch \citep{Gratton2010b} and from the main sequence stars \citep{Milone2018}. Circled symbols are for Type~II clusters according to the classification by \citet{Milone2017}.  Colours code metallicity (see scale on the right of the plot). Lower panel: the same, after the systematic correction to the He abundance spread suggested in the text.}
\label{fig:dy}       
\end{figure}

\begin{figure}[htb]
\includegraphics[width=\textwidth]{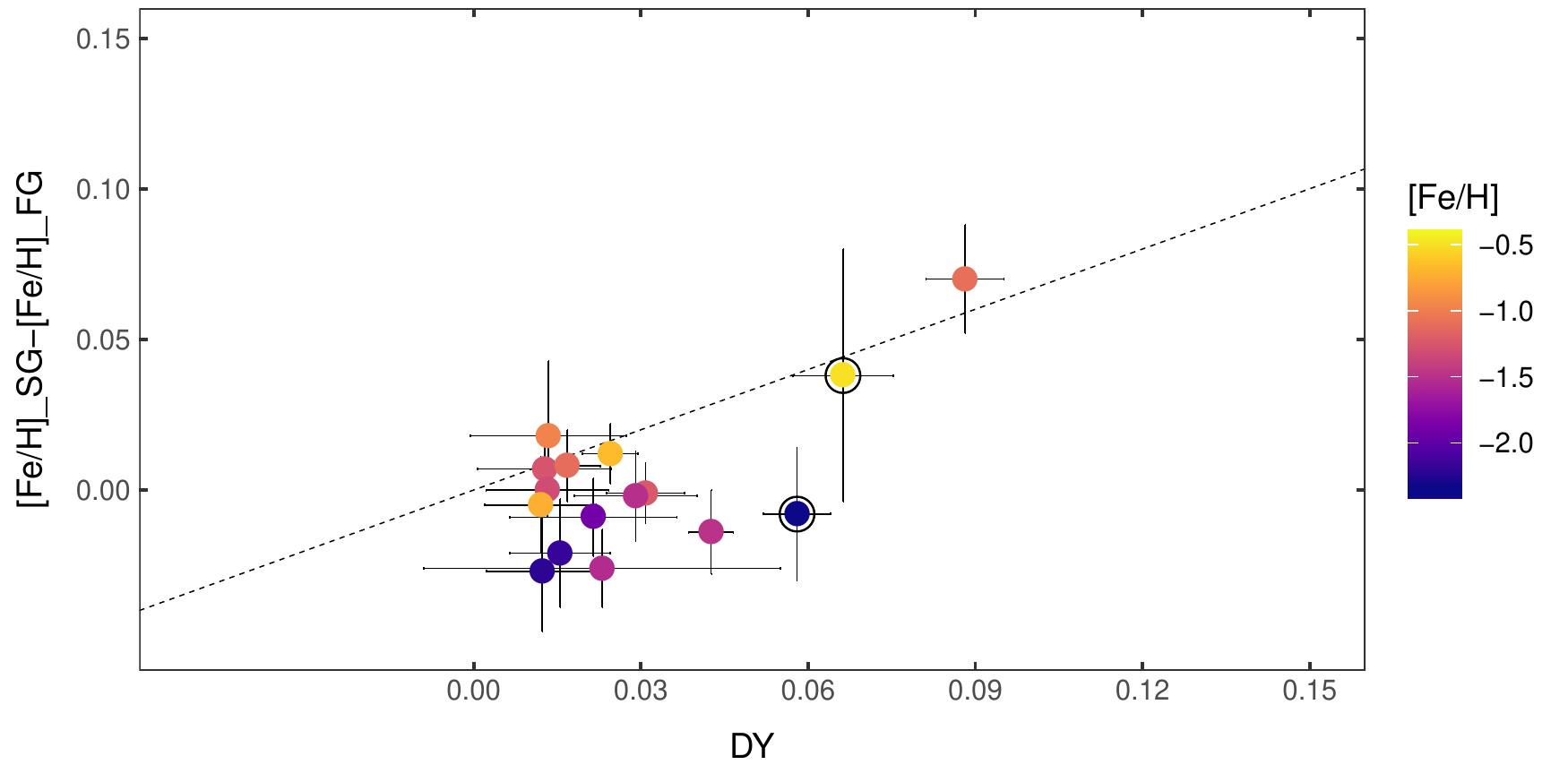}
\caption{Comparison between the spread in He abundances within a cluster (see text) and difference in [Fe/H] values between First and Second generation stars determined by \citet{Bragaglia2010a}.  Colours code metallicity (see scale on the right of the plot). The line is the expected correlation between these two quantities.}
\label{fig:dfe}       
\end{figure}

Variations in the He abundances have been called to explain the complex morphology of the horizontal branch since the '70s and early '80s (see e.g. \citealt{Norris1981}). The basic idea is that He-rich stars evolve faster than He-normal ones while on the main sequence. So He-rich horizontal branch stars are expected to be less massive than He-normal ones, and then to be bluer. After a couple of decades, this argument has been resurrected by \citet{D'Antona2002}, who used it to explain the horizontal branches of M~13=NGC~6205 and NGC~6752. In \citet{D'Antona2004} a large variation of the He content was called for the case of NGC~2808, soon brilliantly confirmed by the discovery of the splitting of the main sequence \citep{D'Antona2005, Piotto2007}. A direct link between the colours of stars along the horizontal branch and the multiple population phenomenon was obtained by in-situ observation of the expected abundance variations in NGC~6752 by \citet{Villanova2009}, then confirmed by those in other clusters: M~4=NGC~6121 \citep{Marino2011c}, NGC~2808 \citep{Gratton2011, Marino2014}, NGC~1851 \citep{Gratton2012b}, 47~Tuc=NGC~104 and M~5=NGC~5904 \citep{Gratton2013}, M~22=NGC~6656 \citep{Gratton2014}, M~30=NGC~7099 and NGC~6397 \citep{Mucciarelli2014c}, and NGC~6723 \citep{Gratton2015}. A complication in these comparisons is that the atmospheres of the hottest stars along the horizontal branch (warmer/bluer than the so-called Grundahl jump: \citealt{Grundahl1999}) are in radiative equilibrium, leading to large effects related to diffusion and radiation pressure resulting in a very odd abundance pattern (see \citealt{Behr1999}). This makes it impossible to sample the whole horizontal branch in many clusters for the abundances of the elements that are diagnostics of the multiple populations.

Another possible approach is to reconstruct the He abundances from the location of the stars on the horizontal branch by comparisons with evolutionary models, as pioneered by \citet{D'Antona2002}, and later applied by many others, e.g. \citet{Dalessandro2011,Dalessandro2013} who used a combination of ground-based and UV HST bands, where different He translates (also) into different magnitudes; they derived the He content and dispersion of NGC~2808, where they recovered the distinct populations, and M~3=NGC~5272, M~13=NGC~6205, and M~79=NGC~1904. This is not an easy task, because the horizontal branch is also shaped by the metal abundance, the age, the total CNO content, and the mass-loss along the red giant branch. Luckily, metal abundances are well established (see \citealt{Carretta2009c}) and ages are also now quite well settled, thanks to the progress allowed by HST photometry (see e.g. \citealt{Marin-Franch2009, Dotter2011, Vandenberg2013}). The main issue concerns mass-loss, that depends on metal abundance and likely contains a small random term, variable from star-to-star; it is even possible that mass-loss is different for He-normal and He-rich stars, as discussed in \citet{Salaris2016} and very recently by \citet{Tailo2019}. Since different assumptions are used by each author, comparison between results is not easy. We will then consider here only extensive data sets, such as those of \citet{Dotter2010, Gratton2010b}, and \citet{Milone2014b}. The upper panel of Fig.~\ref{fig:dy} compares the spread of the He abundances within individual GCs derived from an analysis of the distribution of stars along the horizontal branch \citep{Gratton2010b} with those from the colours of main sequence stars \citep{Milone2018}. While there is some spread exceeding the error bars, there is an overall good correlation between the two estimates, with a linear correlation coefficient of r=0.51 over 51 clusters. The significance of this correlation is very high. Residuals around the identity lines are correlated with the metal abundance of the cluster; namely, the spread in He abundance derived from the main sequence is systematically larger than that derived from the horizontal branch for the most metal-rich clusters ([Fe/H]$>-0.8$), while typically the opposite holds for more metal-poor clusters. On the whole, the existence of such differences is not surprising, given the complexity of deriving He abundances in clusters. The most uncertain case is for the derivation of the He abundance from the horizontal branch of clusters with red horizontal branches because in this case a quite large variation in the mass (that is, on the He abundance) has only a very minor effect on the horizontal branch\footnote{An example of the difficulties in deriving He abundance variations from clusters with red horizontal branch is given by a comparison of the spread in He abundances for the SMCs clusters NGC~121, NGC~339, NGC~416, and Lindsay~1 as determined from the horizontal branch by \citet{Chanterau2019}, and from a pseudo-chromosome map by \citet{Lagioia2018}. While the first study found variations in the He abundances as large as $\Delta Y=0.08$, the second one only found very tiny spreads, with the highest value being $\Delta Y=0.010\pm 0.003$. \citet{Chanterau2019} noticed this difference, and attributed it to the different meaning of $\Delta Y$\ in the two studies - maximum excursion with respect to mean difference between first and second generation stars, although it seems quite difficult to justify a factor of almost ten difference between the two results this way. We then think that the spread in He abundances derived for red horizontal branch clusters should be taken with caution.}. A considerable reduction of the scatter between the two data sets is obtained if we correct the spread in He abundances determined from the HB using the empirical formula $\Delta Y_{\rm cor}=\Delta Y_{\rm HB}~(0.58+0.36/{\rm [Fe/H]}^2)$\ (see lower panel of Fig.~\ref{fig:dy}). The only discrepant case is NGC~2808; we notice that for this cluster the spread in mass along the HB \citep{Gratton2010b}, and then in helium abundance, is likely underestimated. Hereinafter, we will adopt the average of the helium spread from \citet{Milone2018} and these corrected values for the HB as the best current estimate of the spread in He abundances within a cluster.

As discussed by \citet{Bragaglia2010a}, if we may assume that the metal abundance is the same for all stars within a cluster, it is possible to derive the difference in He abundance between second and first generation stars also from the difference in the value of [Fe/H], because a change in the He abundance is anti-correlated with the change in the H abundance, and it is then correlated with that in [Fe/H]. Since the expected variation in H abundances are small (at most, $\sim$20\%), quite large samples of accurate and uniform [Fe/H] values must be considered. Fig.~\ref{fig:dfe} compares the spread in He abundances within a cluster with the offset in [Fe/H] values obtained in \citet{Bragaglia2010a}. The correlation is good, with a linear correlation coefficient of 0.65 that is highly significant.

\begin{figure}[htb]
\includegraphics[width=\textwidth]{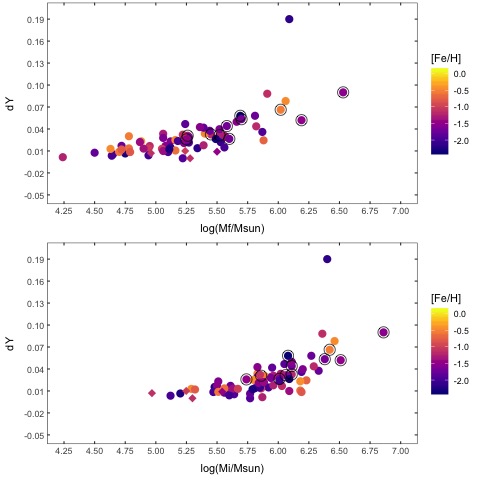}
\caption{Upper panel: run of the spread in He abundances within a cluster $\Delta$Y and the cluster mass from \citet{Baumgardt2019}. Lower panel: the same, but using the initial mass from \citet{Baumgardt2019}. The most discrepant case is NGC~2419. Diamonds and circled symbols indicate respectively MC clusters and Type~II clusters. Colours code metallicity (see scale on the right of the plot)}
\label{fig:mass_dy}       
\end{figure}

If we combine the two sets of determinations of the He spread within a cluster, it is clear that these spreads are rather small for the majority of the GCs. Also, this quantity is very well correlated with the cluster mass, as derived by \citet{Baumgardt2019} (see Fig.~\ref{fig:mass_dy}). The most discrepant case is that of NGC~2419, that has a spread in He larger than expected for its mass. The very good correlation existing between the spread in the He abundance and the cluster mass is a clear indication that cluster mass is indeed the main parameter determining the spread in He abundances. In particular, we find that only clusters with a mass larger than $3\times 10^5$~M$_\odot$ have a spread in He abundances larger than 0.01.

\subsection{Asymptotic giant branch stars vs red giant stars}

\begin{figure}[htb]
    \centering
\includegraphics[width=0.9\textwidth]{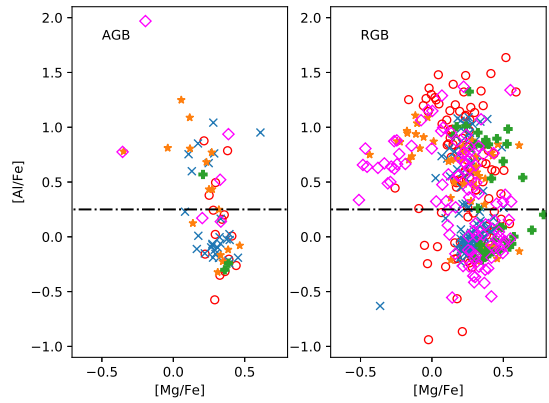}
\caption{[Al/Fe] vs [Mg/Fe] for early AGB (left-hand panel) and RGB stars (right-hand panel) by \cite{Masseron2019}. Symbols are as follows: empty circles (M~13=NGC~6205), crosses (M~3=NGC~5272), starred symbols (M~92=NGC~6341), plus (M~53=NGC~5024) and diamonds (M~15=NGC~7078). The horizontal, dot-dashed line marks our definition of Al-rich/Al-poor population at [Al/Fe]=+0.25 dex. }
\label{fig:masseron2019}
\end{figure}

\begin{table}[htb]
\centering
\caption{Number of RGB and AGB stars belonging to first and second-generation stars in GCs. The corresponding fractions of SG stars are given in Cols. 6 and 7 for RGB and AGB stars, respectively. Data references are: \citealt{Campbell2013a}; \citealt{Maclean2016}, \citeyear{MacLean2018b}, \citealt{Wang2016}; \citeyear{Wang2017}, \citealt{Lapenna2015}, \citealt{Mucciarelli2019}, and \citealt{Masseron2019}.}

\begin{tabular}{lccccccr}
\hline    
NGC & Other & RGB$_{FG}$ & RGB$_{SG}$ &  AGB$_{FG}$ & AGB$_{SG}$ & f(RGB$_{SG}$) & f(AGB$_{SG}$)\\
\hline
\multicolumn{8}{c}{MacLean/Campbell et al.}\\
NGC~6121 & M~4     & 48 &  58 &  15 &	 0 & 0.55$\pm$0.05 & $<0.06$       \\
NGC~6397 &         & 18 &  30 &   3 &    5 & 0.63$\pm$0.07 & 0.63$\pm$0.17 \\
NGC~6752 &         &  7 &  17 &  20 &	 0 & 0.71$\pm$0.09 & $<0.05$       \\
\hline
\multicolumn{8}{c}{Lapenna/Mucciarelli et al.}\\
NGC~6266 & M~62    &  7 &   6 &   5 &	 0 & 0.46$\pm$0.14 & $<0.20$       \\	
NGC~6752 &         &  6  &  8  &  11 &   8 & 0.57$\pm$0.13 & 0.42$\pm$0.11 \\
         
\hline
\multicolumn{8}{c}{Wang et al.}\\
NGC~104  & 47~Tuc  &  9 &  18  &  24 &  16 &  0.67$\pm$0.10 & 0.40$\pm$0.08\\
NGC~2808 &         & 23 &  24  &  14 &	17 &  0.51$\pm$0.07 & 0.55$\pm$0.09 \\
NGC~5904 &  M5     & 11 &  24  &   5 &  10 &  0.69$\pm$0.08 & 0.67$\pm$0.12\\
NGC~5986 &         &  5 &   8  &   3 &   4 &  0.72$\pm$0.11 & 0.57$\pm$0.19\\
NGC~6121 &  M4     & 15 &  48  &   9 &  10 &  0.76$\pm$0.06 & 0.53$\pm$0.11\\
NGC~6205 &  M13    & 23 &  73  &   2 &  14 &  0.76$\pm$0.04 & 0.88$\pm$0.08\\
NGC~6266 &  M62    &  5 &   8  &   5 &   0 &  0.62$\pm$0.13 & $<0.20$ \\
NGC~6752 &         &  6 &  18  &  17 &   3 &  0.75$\pm$0.09 & 0.15$\pm$0.08\\
NGC~6809 &  M55    & 26 &  51  &   8 &  15 &  0.66$\pm$0.05 & 0.65$\pm$0.11\\
\hline
\multicolumn{8}{c}{Masseron et al.}\\
NGC~5024 & M~53    & 20 &  17 &   2 &	 1 & 0.46$\pm$0.08 & 0.33$\pm$0.27 \\
NGC~5272 & M~3     & 66 &  40 &  21 &	 7 & 0.38$\pm$0.05 & 0.25$\pm$0.08 \\
NGC~6205 & M~13    & 28 &  68 &  11 &    4 & 0.71$\pm$0.05 & 0.27$\pm$0.11 \\
NGC~6341 & M~92    & 11 &  38 &   7 &    9 & 0.78$\pm$0.06 & 0.56$\pm$0.12 \\
NGC~7078 & M~15    & 39 &  62 &   2 &	 4 & 0.61$\pm$0.05 & 0.67$\pm$0.19 \\

\hline
\end{tabular}
 \label{tab:agb_counts}
\end{table}



In 1981, \citeauthor{Norris1981} found indication for a lack of CN-strong stars within their AGB sample in the cluster NGC~6752, as compared to stars on the first-ascent giant branch, where CN-strong stars are the dominant population. One of the possible explanations suggested by \cite{Norris1981} is that those we now recognize as SG stars and which are enriched in He and thus less massive relative to the FG stars, will fail in ascending the giant branch for a second time, evolving directly to white dwarfs from the HB phase (as AGB manqu\'e stars, e.g., \citealt{Greggio1990}). This pioneering work brought to light the presence of what is currently called the ``AGB problem'', which has received special attention in recent years. Despite considerable work, a clear picture and a comprehensive understanding is still not in hand. 

\cite{Sneden2000} compared the CN distributions along the AGB and RGB in different clusters by considering extant literature studies of NGC~6752  (\citealt{Norris1981}), M~13$=$NGC~6205 (\citealt{Suntzeff1981}), and M~4$=$NGC~6121 (\citealt{Norris1981}; \citealt{SuntzeffSmith1991}). The conclusion of their study pointed to a general lack of SG, CN-strong AGB stars, although to differing extents for each of the three clusters (with NGC~6752 the obvious case of a total dearth of CN-strong AGB stars). 

\cite{Campbell2006} also reviewed the status of the field at that time, collecting literature data and stressing the need for further investigations based on larger samples of AGB stars in clusters. In their Table 1, the authors showed GCs classified according to the presence of CN weak and CN strong stars within the AGB population for M~3$=$NGC~5272, M~4$=$NGC~6121, M~5$=$NGC~5904, M~13$=$NGC~6205, M~15$=$NGC~7078, M~53$=$NGC~5024, NGC~6752 and 47~Tuc=NGC~104 (see that paper for details and references). Out of 8 GCs, only M~5$=$NGC~5904 and 47~Tuc$=$NGC~104 displayed a significant population of CN-strong AGB stars. 

In qualitative agreement with theory \citep{Greggio1990}, these studies suggested that the stars with the smallest mass along the HB either cannot reach the AGB or they leave it earlier than more massive stars. This should appear as a correlation between HB morphology and counts of stars on the AGB. \cite{Gratton2010a} explored this relation and indeed found that there is a good correlation between the mass of the hottest 10\% stars on the HB and the count ratio between AGB and RGB stars ($f_{AGB}=n(AGB)/n(RGB)$). On the other hand, these counts suggest that even in those clusters that have the smallest ratio between the number of AGB and RGB stars, still there are more AGB stars than expected in the case that all SG stars avoid the AGB. Actually, if we compare the counts of AGB stars by \cite{Gratton2010a} with the fraction of FG/SG stars determined by \cite{Milone2017}, we find that even in extreme cases at least half of the SG stars should reach the AGB, and in clusters that do not have an extended blue HB, virtually all of them should go through this phase. We should then expect that the fraction of SG stars along the AGB is never below $\sim$50\%, in many clusters should be of the order of two/thirds, and that this fraction should depend on the morphology of the HB.

In the last few years, various groups tried to determine the ratio between FG and SG stars along the AGB using spectroscopy, mainly exploiting the Na/O anticorrelation, but in some cases also the Al/Mg distribution. The main results are collected in Table~\ref{tab:agb_counts}. Taken at face value, the picture is complex and quite at odds with expectations, with sometimes conflicting results obtained for the same cluster. For instance, in the case of NGC~6752 (that has an extended blue HB),  \citet{Campbell2013a} found that all AGB stars belong to the FG of the cluster (see their Figure 2), corroborating previous evidence by \cite{Norris1981}. A low fraction of SG stars of this GC along the AGB is also found by \citet{Wang2017}; this contrasts with the theoretical expectation that 50\% of the AGB stars are predicted to be SG \citep{Cassisi2014}. These results were questioned by \citet{Lapenna2016}, who instead detected a significant population of moderately Na-enhanced AGB stars in NGC~6752, while confirming that no extreme Na-rich stars are present within their AGB sample, which is in line with \cite{Cassisi2014} [but see also \citealt{Campbell2017}]. An extreme case is that of M~4=NGC~6121: for this cluster, \citet{Maclean2016} found that, despite the lack of an extended blue HB,  all the AGBs are found to be consistent with FG. This finding has been disputed by \citet{Wang2017}, \citet{Lardo2017b} and \citet{Marino2017}. A new discussion by \citet{MacLean2018a} concluded that a significant disagreement between theory and observations still remains for this cluster. \citet{MacLean2018b} investigated many possibilities for this hypothetical offset between AGB and RGB stars. On the other hand, \citet{Wang2016} analysed a quite numerous sample of RGB and AGB stars in NGC~2808 and found almost identical distributions. This is quite unexpected given the extended blue HB of NGC~2808. Based on only 6 AGB stars, \cite{Marino2017} concluded that multiple populations are present both in the RGB and AGB of NGC~2808; however, as it appears from their Figure 3, the Na and O distributions for AGBs are skewed towards higher Oxygen and lower Sodium abundances relative to the RGBs by \cite{Carretta2009a}, who found RGB stars in NGC~2808 exhibit extreme patterns in terms of [Na/Fe] and [O/Fe] ratios, up to $+1$ and $-1$ dex, respectively.

These studies mainly used optical spectra, from which Na and O abundances can be obtained. Near infrared spectra obtained with APOGEE have also been used. In this case, [Al/Fe] are generally used to assign stars to the different populations \citep{Garcia2015, Majewski2017}. Quite extensive samples of early AGB stars along with a control sample of RGB stars have been obtained by \citet{Masseron2019} for M~3=NGC~5272, M~13=NGC~6205, M~92=NGC~6341, M~53=NGC~5024, and M~15=NGC~7078. They have detected Al-rich AGB stars in all of them. A significant Mg-Al anticorrelation is emerging for cluster NGC~6341=M~92, and less evident for NGC~5272=M~3 and NGC~6205=M~13, while for NGC~7078=M~15 and NGC~5024=M~53 low statistics prevent from drawing any conclusion. In order to estimate if there is a lack of Al-rich stars (here defined as those with [Al/Fe] $>$ 0.25 dex) on the AGB in these clusters, we may count the number of Al-rich and Al-poor stars along the RGB and the AGB (see Figure \ref{fig:masseron2019}). In the RGB we obtain 164 Al-poor stars and 225 Al-rich ones; the same numbers in the AGB are 43 Al-poor and 25 Al-rich stars\footnote{These calculations are restricted to those stars that have Al abundances, which comprises more than 90\% of the total sample.}. This indicates a significant (at 4$\sigma$ level) lack of Al-rich stars along the AGB: the latter have $\sim 40\%$ of Al-rich (i.e. SG) stars with respect to what is expected according to the population ratios observed on the RGB.

Overall, there is a general trend for a lack of SG AGB stars. If we sum up results obtained for the whole samples observed by individual goups, \citet{Maclean2016} found 57$\pm$2\% of SG RGB stars and 32$\pm 4\%$ of SG AGB stars. The difference is less pronounced, but still clear, if we consider the results obtained by Wang and coworkers: in this case, SG stars make up 69$\pm$2\% of the RGB stars and 51$\pm$4\% of the AGB stars. Considering the total sample analysed by \cite{Masseron2019}, the fractions of SG stars along the RGB and the AGB are 58$\pm$3\% and 37$\pm$6 \%, respectively. On the other hand, the fraction of SG stars along the AGB is almost identical if we restrict to GCs with extreme HB morphology (i.e., M~13=NGC~6205, M~62=NGC~6266, M~2=NGC~7089, NGC~6752, NGC~2808, and M~15=NGC~7078); the same is true if we restrict the sample to those with red HBs. Thus the relationship with the HB morphology seems not so straightforward and other causes might be underneath. Moreover, in several cases there are no SG stars detected, and this happens also for GCs that do not have an extended blue HB. 

A more careful inspection of this complex pattern shows that we are facing several issues:
\begin{itemize}
    \item often there is the suspicion that there are small but significant offsets between the abundance ratios determined from RGB and AGB stars. This is e.g. exemplified by the study of NGC~6121=M4 by \citet{Maclean2016}, where the Na abundances determined for the AGB stars appear to be typically $\sim 0.15$~dex below those obtained for the RGB stars, or by the different results depending on the way abundances are obtained for the stars in NGC~2808 by \citet{Wang2016}. The existence and possible origin of these offsets is debated \citep{Lardo2017b, Marino2017, Campbell2017, MacLean2018a}, and the issue is not yet settled;
    
    \item stars are assigned to FG or SG by comparing their abundances with a given threshold. Often distributions are rather continuous, so that the value of the threshold is quite arbitrary and the exact choice has a large impact on the counts\footnote{Note that \citet{Maclean2016} rather use the observed minima in the [Na/H] distribution to separate FG and SG stars along the AGB and the RGB comparison dataset}. As an example of this kind of problem, we may consider the case of NGC~6752 \citep{Campbell2013a, Wang2016, Lapenna2016}: the three distribution of Na abundances in the different studies all look quite similar with each other, but the conclusions on the counts are very discrepant (see \citealt{Campbell2017}, where for NGC 6752 all the studies agree very precisely in A(Na) -their Fig 13).

    \item AGB is a rather fast evolutionary phase: as a consequence, samples of AGB stars are generally not numerous and random errors are large when compared with the expected variation on the frequency of FG/SG stars;
    
    \item finally, as we have seen in Section 3, the various anti-correlations might be telling somewhat different stories: for instance, the Mg-Al anticorrelation (considered by \citealt{Masseron2019}) is rarely present among metal-rich GCs, and there may be stars that are clearly classified as SG stars from the strength of N bands that may look quite similar to FG stars according to the Na-O or Mg-Al abundances or even the CN bands.
\end{itemize}

We may then re-examine the results collected in Table~\ref{tab:agb_counts} taking into account these issues. We find that there are facts that look solidly established and others that are still open. In the first group we may place (i) the lower overall frequency of SG stars in the AGB, that is found in all studies though with significant cluster-to-cluster differences; and (ii) the virtual absence of AGB stars with extremely large Na/Al excesses in clusters with very extended horizontal branches (e.g. NGC~6752: \citealt{Campbell2013a, Lapenna2016, Wang2016}; M~13=NGC~6205: \citealt{Masseron2019}; NGC6266: \citealt{Lapenna2015}), though there might be some stars with extreme composition in the AGB of NGC~2808 \citep{Wang2016, Marino2017}. We note however that the most extremely O-poor stars in this cluster are not very Na-rich, as shown by \citet{Carretta2007}, so it is not clear that they can be found using Na abundances alone. In general, the lack of stars with large Na/Al excess might contribute to explaining the apparent dearth of SG AGB stars in these GCs, even though there might also be a number of AGB SG stars with small or moderate excesses of these elements that are disguised in the FG group. On the other hand, the lack of SG stars in GCs that do not have a very extended HB (such as NGC~6121=M~4) is still to be established firmly.

Future observations of some critical pairs such as e.g., NGC~288 and NGC~362 (one of the most famous second-parameter {\it pairs}) or the determination of the Al abundances along the AGB of NGC~2808 might help in shedding light on this still poorly understood field of research. In these kind of investigations a careful selection of genuine AGB stars, avoiding RGB contamination, along with reliable parameter determination (with possible offset and/or NLTE effects that have to be taken into account), and the use of RGB control sample is of vital importance.

\section{Lithium}
\label{Sec:5}

\subsection{Lithium and mixing}

\begin{figure}[htb]
    \centering
    \includegraphics[width=0.95\textwidth]{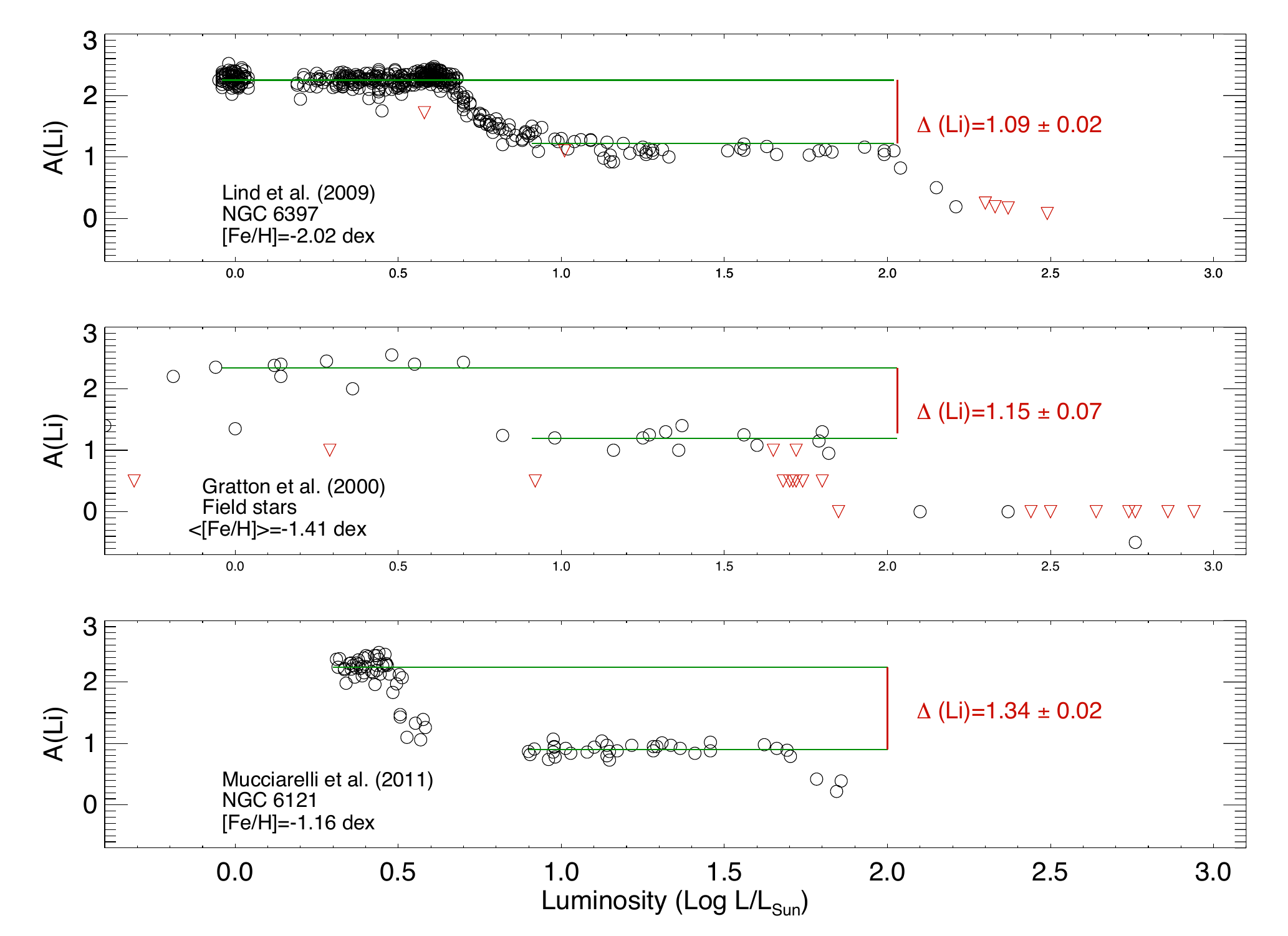}
    \caption{Lithium evolution for field stars (middle panel) analysed by \cite{Gratton2000}, NGC~6397 (upper panel) by \cite{Lind2009}, and NGC~6121 (lower panel) by \cite{Mucciarelli2011}. The $\Delta$(Li) represents the difference in abundances between main-sequence stars and  sub giants because of the {\it first dredge-up}: this values depends on metallicity, being [Fe/H]$=-1.41$ dex the average for the field stars, [Fe/H]$-2.02$ dex and [Fe/H]$=-1.16$  for NGC~6397 and NGC~6121, respectively. }
    \label{fig:li_mixing}
\end{figure}

The predicted difference in depth of the convective envelope between main sequence and lower red giants \citep{Iben1964} is confirmed by the dilution effect of lithium when stars evolve from the main sequence to the lower red giant branch (luminosity below the RGB bump, that we use here as a dividing line because Li is severely depleted after this evolutionary phase). The effect is similarly present in both field \citep{Gratton2000} and GC stars (NGC~6397, \citealt{Lind2009}; NGC~6121=M~4; $\omega$~Cen=NGC~5139 \citet{Monaco2010}; \citealt{Mucciarelli2011}; NGC~104=47 Tuc \citep{Dobrovolskas2014}, M~30=NGC~7099: \citealt{Gruyters2016}) (see Fig.~\ref{fig:li_mixing}). 

The actual Li dilution depends on the ratio of the mass contained in the region where Li is not burnt during the main sequence and in the convective envelope at the base of the RGB; this quantity is (weakly) dependent on metallicity and age. \cite{Mucciarelli2012b} examined the run of the Li abundance in field low-luminosity giants and found a quite constant value of $\log{n({\rm Li})}=0.94$ over a wide range of Fe abundances ($-3.4<$[Fe/H]$<-1.3$). Then, they considered the dependence expected from stellar evolutionary models. The quantity they considered is the expected depletion between the pre-main sequence and the RGB phases; they hence did not consider the possible surface depletion due to diffusion on the main-sequence. Even so, the result depends on the treatment of convection. Independently of this, the models show a weak dependence on metallicity, with values changing by some 0.11-0.16 dex when metallicity ranges from [Fe/H]=-2.14 to [Fe/H]=-1.01. The slope is a bit steeper when diffusion is taken into account, because diffusion is expected to have a larger impact in lower metallicity stars that have a thinner outer convective envelope.

On the whole, there is good agreement between observations and theoretical prediction. This strongly argues against the possibility that abnormal mixing (that would cause destruction of large amounts of Li) is the cause of the abundance anomalies related to multiple populations \citep{Pancino2018}.

\subsection{Lithium and multiple populations: observations}

\begin{table}[htb]
    \centering
    \caption{Fraction of stars with extreme composition along the [Na/O] anticorrelation (E-stars: \citealp{Carretta2009a}) and of Li-poor stars in various clusters}
    \begin{tabular}{lccccccc}
\hline
Cluster  & [Fe/H] & $\log{M_{rm in}}$ & Type &  E-Fraction & Ref & Li-poor & Ref \\ \hline
NGC~362   & -1.26  &  6.06 & 2 & $0.03\pm 0.02$ & 3 & $0.04 \pm 0.03 $ & 7 \\
NGC~1904  & -1.60  &  6.08 &   & $0.10\pm 0.04$ & 1 & $0.14 \pm 0.08 $ & 7 \\
NGC~2808  & -1.14  &  6.36 & 1 & $0.14\pm 0.03$ & 9 & $0.14 \pm 0.05 $ & 7 \\
NGC~5904  & -1.29  &  5.96 & 1 & $0.07\pm 0.02$ & 1 & $0.074\pm 0.030$ & 6 \\
NGC~6121  & -1.16  &  6.03 & 1 & $0.00\pm 0.01$ & 1 & $0.00 \pm 0.26 $ & 8 \\
NGC~6218  & -1.37  &  5.63 & 1 & $0.03\pm 0.02$ & 1 & $0.00 \pm 0.02 $ & 6 \\ 
NGC~6397  & -2.02  &  5.60 & 1 & $0.00\pm 0.01$ & 1 & $0.020\pm 0.008$ & 4 \\
NGC~6752  & -1.54  &  5.83 & 1 & $0.40\pm 0.06$ & 2 & $0.30 \pm 0.05 $ & 5 \\
NGC~7099  & -2.27  &  5.79 & 1 & $0.03\pm 0.02$ & 2 & $0.11 \pm 0.08 $ & 10 \\
\hline
\end{tabular}
References: 1. \citet{Carretta2010c}; 2. \citet{Carretta2012}; 3. \citet{Carretta2013a}; 4. \citet{Lind2009}; 5. \citet{Shen2010}; 6. \citet{D'Orazi2014}; 7. \citet{D'Orazi2015}; 8. \citet{D'Orazi2010b}; 9.\citet{Carretta2015b}; 10 \citet{Gruyters2016} 
\label{tab:lithium}
\end{table}

\begin{table}[htb]
    \centering
    \caption{Dilution as obtained from Na and O abundances published in Carretta et al's papers (Col. 2) along with lithium differences between intermediate SG stars and primordial stars (Col. 3), the amount of Li produced with respect to the original value (Col. 4) and the difference in maximum Na and minimum O content (Col. 5).}
    \begin{tabular}{lccccr}
\hline    
GC & I Dilution & Li(I-P) &    Li(Prod-Original)  &  [Na/Fe]$_{\max}$-[O/Fe]$_{\min}$\\
\hline	 							
NGC~362  &	 0.53 &	 -0.03$\pm$0.03	& -0.07$\pm$0.08 &	    0.90\\
NGC~1904 &	 0.47 &	 -0.14$\pm$0.04	& -0.32$\pm$0.19 &	    1.32\\
NGC~2808 &	 0.50 &	 -0.05$\pm$0.03	& -0.11$\pm$0.09 &	    1.56\\
NGC~5904 &	 0.54 &	 -0.02$\pm$0.03	& -0.04$\pm$0.08 &	    1.30\\
NGC~6121 &	 0.62 &	 ~0.04$\pm$0.05	&  0.10$\pm$0.11 &	    0.94\\
NGC~6218 &	 0.51 &	 -0.02$\pm$0.05	& -0.04$\pm$0.12 &	    0.96\\
NGC~6397 &	 0.73 &	 -0.02$\pm$0.03	& -0.08$\pm$0.17 &	    0.71\\
NGC~6752 &	 0.40 &	 -0.13$\pm$0.07	& -0.25$\pm$0.21 &	    1.05\\
\hline
\end{tabular}
\end{table}

\begin{figure}[htb]
    \centering
    \includegraphics[width=0.9\textwidth]{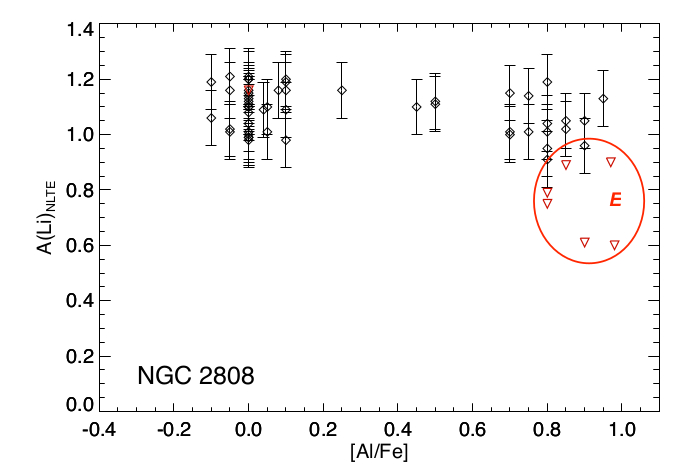}
    \caption{Lithium as a function of aluminum abundances in RGB stars (below the bump) in NGC~2808 (re-adapted from \citealt{D'Orazi2015}). Upper limits in Li abundances are marked as upside-down triangles. The orange circle marks some stars belonging to the E-group, that is, have extreme composition along the [Na/O] anticorrelation (for an exact definition see \citealt{Carretta2009a}) that are Li-poor. }
    \label{fig:li_2808}
    \label{fig:my_label}
\end{figure}    

In the multiple population framework, a number of studies have been conducted in order to ascertain the run of lithium with proton-capture element abundances in GC stars\footnote{The investigation of the Li discrepancy as measured in Pop~{\sc ii} stars with respect to the standard Big Bang nucleosynthesis is not discussed in this review, since our main focus is the multiple population scenarios. We refer the reader to \cite{Sbordone2010},  \cite{Mucciarelli2014b}, \cite{Fu2015}, and references therein for a specific discussion on this topic}. Lithium offers critical diagnostics for the investigations of the internal pollution source. In fact, because of its very fragile nature (the Li burning temperature is $T\approx 2.5 \times 10^6$K), it is expected that material processed at the much higher temperatures of the hot H-burning via CNO re-cycled in the formation of the subsequent generation(s) of stars within a GC, is free of Li. Thus,  while FG stars should exhibit a Li abundance pattern compatible with their field counterparts (A(Li)$\sim$ 2.0 - 2.2 dex), SG stars should be depleted in Li unless Li is produced in the polluters. This implies a positive correlation between Li and O (and/or C, Mg) and an anti-correlation between Li and Na (and/or N, Al).

Most interesting, while massive stars (supermassive stars, fast rotating and/or binaries) can only destroy Li, intermediate mass AGB stars may activate the {\it Cameron-Fowler} mechanism: the $^3$He($\alpha, \gamma$)$^7$Be reaction takes place in the stellar interiors and then convective processes bring the material outwards, where the temperature is much lower. By capturing one electron, the reaction $^7$Be($e^-,\nu$)$^7$Li could produce lithium, under the condition that it is not rapidly destroyed by thermonuclear reactions \citep{cameron71}. As a consequence, any Li production would tend to erase the above-mentioned Li-Na-O (anti)correlations, by furnishing compelling evidence for intermediate mass AGBs as polluters in GCs.

With this background in mind, several GCs have been investigated with the simultaneous determination of Li/Na/O/Al for samples of dwarf and giant stars. We considered here only giants fainter than the RGB {\it bump}, because Li is completely destroyed after this evolutionary phase. In 2005, \citeauthor{Pasquini2005} analysed 9 TO stars in NGC~6752 and found a statistically meaningful anti-correlation between Li and Na and a positive correlation Li-O; this result was later confirmed, upon a larger sample of 112 stars, by \cite{Shen2010}. Crucially, both studies revealed that the slope in the Li-O plane is not 1, as expected in the simple pollution scenario of ejected material that is Li free. Conversely, SG stars (Na-rich, O-poor) still exhibit a significant amount of Li, which cannot be explained by invoking a dilution of processed material with pristine matter. The presence of a Li-Na anti-correlation is also evident from the main-sequence stars analysed by \cite{gruyters2014}.

A weak hint of Li-Na anti-correlation has been also detected by \cite{Lind2009} in the cluster NGC~6397 ([Fe/H]$\approx -2$), whereas \cite{D'Orazi2010a} found a very peculiar pattern for the metal-rich GC 47~Tuc=NGC~104 ([Fe/H]$\approx -0.7$ dex): the Li content does not show an anti-correlation with Na, and only a weak correlation appears with O, with a large scatter in the Li abundance distribution of FG stars. This suggests a primordial Li dispersion that is probably related to the high-metallicity nature of this GC (i.e., a Pop~{\sc ii} analogue of M~67, see \citealt{Pace2012} and references therein), and not connected to the MP scenario.

M~4=NGC~6121 is certainly the most thoroughly examined GC in terms of Li abundances. \cite{D'Orazi2010b} determined Li and p-capture elements for 109 RGB stars (below the bump) and found that FG and SG stars share the same Li content, with lack of any (anti)correlations. At present, the only explanation we have for such a trend is that Li has been produced within the polluters, suggesting that intermediate massive AGBs are at work, at least in this cluster. This result has been later corroborated by \cite{Mucciarelli2011} and \cite{Monaco2012}, while a very weak trend has been found by \citet{Spite2016}. Similarly to M~4=NGC~6121, M~12=NGC~6218 ([Fe/H]$\approx -1.3$) and NGC~362 ([Fe/H]$\approx -1.3$) have been found to host Li-rich SG stars, with no significant Li dispersion in contrast to the large variations in Al and Na abundances (\citealt{D'Orazi2014}, \citeyear{D'Orazi2015}). In both these GCs, FG and SG stars exhibit the same Li abundance, making Li production across the different stellar generations unavoidable to explain the observed pattern. It may appears quite implausible of obtaining a constant Li content when the initial and enriched materials are mixed in different amounts because it requires the enriched material to have a similar Li abundance as the original material. This clearly requires some more insight; we will come back on this point on the next subsection.

On the other hand, the more massive clusters M~5=NGC~5904, NGC~1904 and NGC~2808 behave differently: while they are still comprised of a dominant population of Li-rich SG stars, these GCs also host an extreme population that reveals large enhancement in Al/Na accompanied by a Li depletion (\citealt{D'Orazi2014}, \citeyear{D'Orazi2015}). In Fig.~\ref{fig:li_2808} we show the run of Al abundances in RGB stars (below the bump luminosity) for NGC~2808 (originally from \citealt{D'Orazi2015}): this GC hosts a large fraction of SG stars (including extreme stars in terms of their Al abundances) that are Li normal, that is with a Li content in agreement with its original value. Most interesting, a handful of extreme stars exhibit a substantial Li depletion (red upside-down triangles in Fig.~\ref{fig:li_2808}). The fraction of Li-poor stars in NGC~2808 ($0.14\pm 0.08$) is actually consistent with the fraction of E-stars found by \citet{Carretta2015b} and of stars belonging to E-population defined by \citet{Milone2015c}. These are stars with extreme overabundance of He, Al and Si and underabundance of O and Mg. These are also likely the stars with anomalous K and Sc abundances \citep{Carretta2015b, Mucciarelli2015}. Unluckily, a star-to-star correspondence between these abundance patterns cannot be obtained because the samples of stars analyzed do not overlap. For this group of stars, we are not able to discriminate whether polluters are either massive stars or a sub-class of intermediate mass AGB stars that are not able to produce Li (because for example of their mass, metallicity, or a combination of both). The observational evidence is that in this kind of GCs more than one polluter class has to be involved in order to reproduce the complex chemical pattern, as also recently found by \citet{Carretta2018a} in NGC~2808 using O, Na, Al abundances in a large sample of brighter RGB stars and by \citet{Johnson2019} in NGC~6402.

\begin{figure}[htb]
\centering
\includegraphics[width=0.9\textwidth]{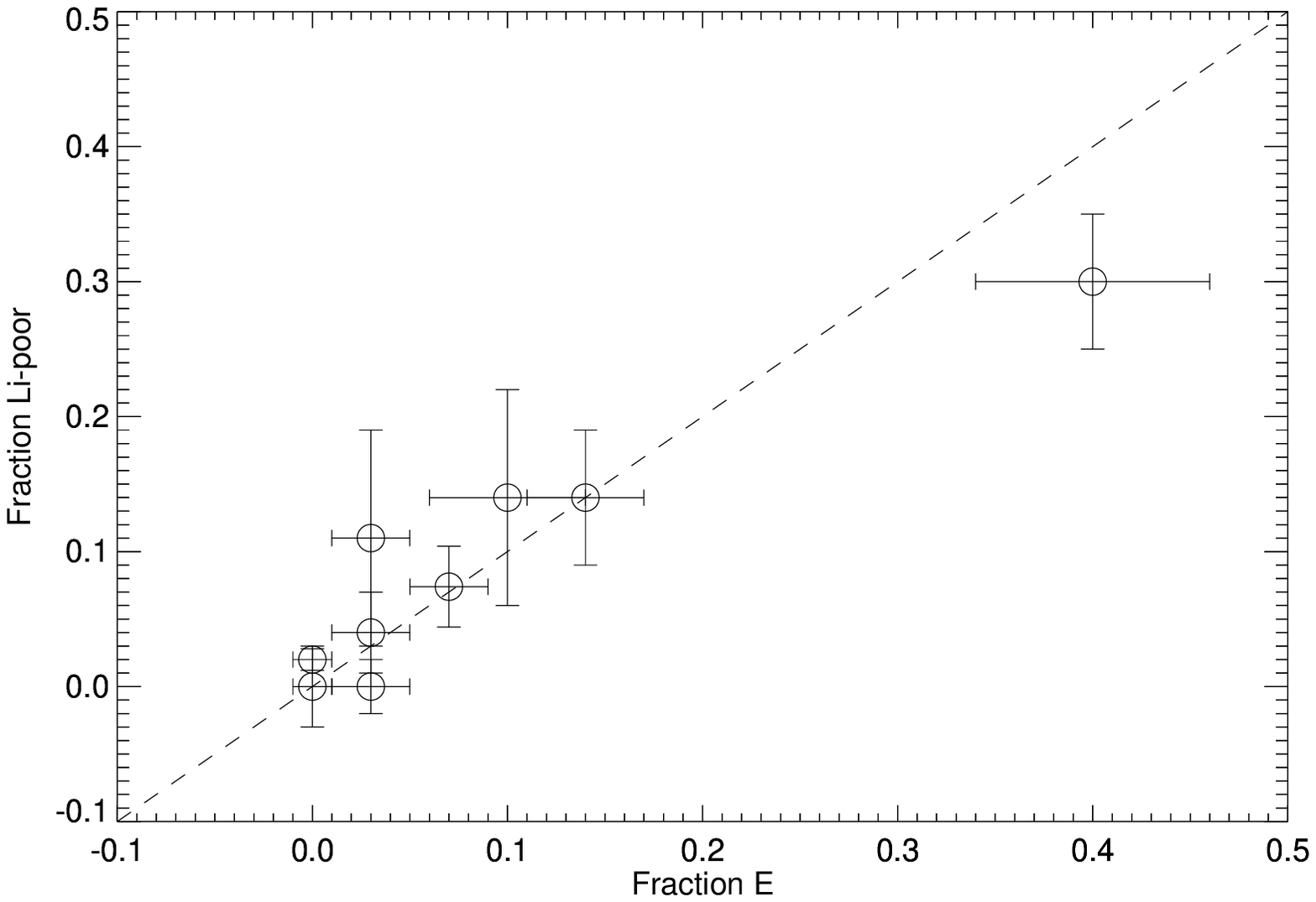}
\caption{Run of the fraction of Li-poor stars with the fraction of E-stars. Dashed line represents equality}
\label{fig:li_fraction}       
\end{figure}

Considering all the GCs for which Li and p-capture element abundances have been determined (Table~\ref{tab:lithium})\footnote{Given the primordial Li scatter in NGC~104, which is unrelated to the multiple population scenarios, this GC was omitted from the present discussion.}, there is a significant correlation between the fraction of stars with extreme composition along the [Na/O] anticorrelation (E-stars, following the definition by \citealt{Carretta2009a}) and the fraction of Li-poor stars (see Fig.~\ref{fig:li_fraction}). The Pearson correlation coefficient is 0.94 over nine GCs, with an extremely low probability that this is a random result. The general picture emerging from the analysis of all the GCs for which Li has been determined in conjunction with p-capture elements suggests that {\it the more massive the cluster, the larger the Li variation}. In other words, the Li production across the different stellar generations is larger (more efficient) in relatively low-mass GCs. In this framework, the cluster metallicity does also play a role: NGC~362 is much less massive than NGC~2808 and has a metal content more than a factor of two higher than NGC~1904, which indicates that Li production is less efficient in more massive and more metal-poor systems. Moreover, the fraction of Li-poor stars in GCs is anticorrelated with the dilution of the stars with intermediate composition along the [Na/O] anti-correlation (I-stars, following the definition by \citealt{Carretta2009a}), with a Pearson correlation coefficient of $r=-0.74$. This means that the larger is the fraction of extreme Li-poor stars, the lower is the dilution for the I population. This suggests that extreme Li-poor and I populations are not independent, at least in some clusters. There is also a strong negative correlation ($r=-0.83$) between the fraction of Li-poor stars and the difference in Li abundance between SG and FG stars: the larger the Li content detected in SG  stars, the lower the fraction of Li-poor stars in GC. Both these aspects might be simply reflecting the dependence on the GC (current) mass: however, there is no evidence for a direct correlation between the fraction of Li-poor star and the cluster mass.

A further complication to this composite mosaic is added by the recent analysis of Li, Na, and Al in 199 RGB stars in the peculiar GC $\omega$~Cen=NGC~5139 by \cite{Mucciarelli2018}. Here, only the most metal-poor component displays a clear Li-Na (and Al) anti-correlation, and from the point of view of Li, the cluster hosts at least 4 populations: FG stars coexist with Li-rich SG stars, with Li-poor SG stars, and with the anomalous (metal-rich) population that is only characterized by Li depletion. 

\subsection{Lithium and multiple population: some considerations}

\begin{figure}[htb]
\centering
\includegraphics[width=0.9\textwidth]{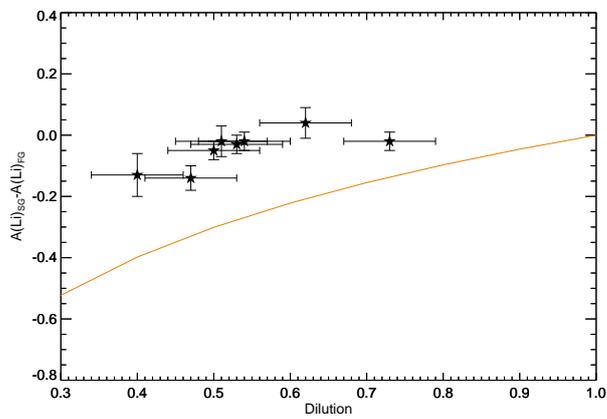}
\caption{Differences in Li abundances between first and intermediate second-generation stars as a funcion of dilution factors. Data are for clusters: NGC~362, NGC~1904, NGC~2808, NGC~5904, NGC~6121, NGC~6218, NGC~6397, NGC~6752 (see Table~\ref{tab:lithium} for the corresponding references). The orange continuous line is the dilution process calculated under the assumption that there is no Li production within the polluters.}
\label{fig:li_dilution}       
\end{figure}

\begin{figure}[htb]
\centering
\includegraphics[width=0.9\textwidth]{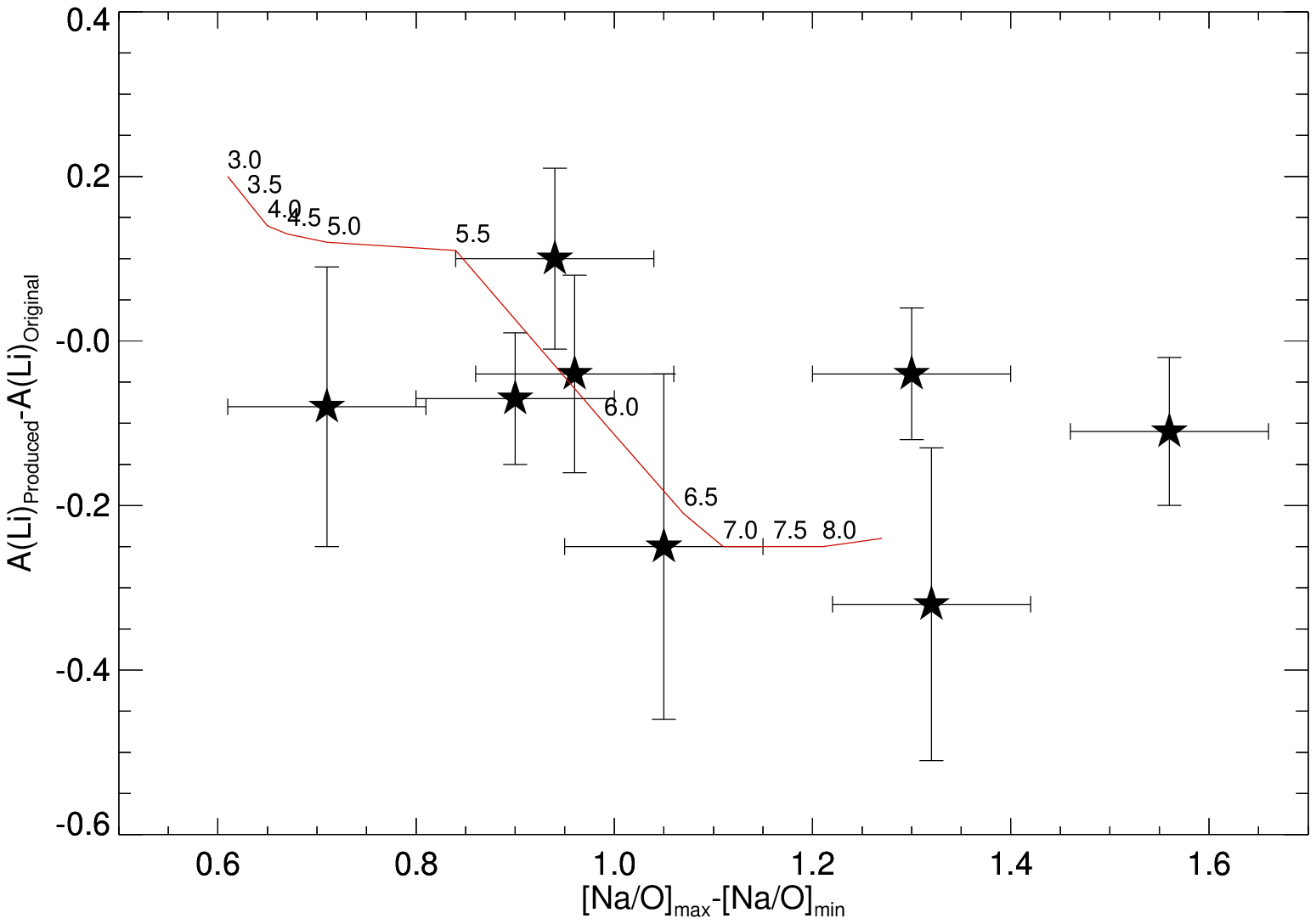}
\caption{Difference between Li abundances in the ejecta of the polluters and original Li abundances as a function of difference between the [Na/O] ratio in the polluter and in the original material from which GC formed in a number of GCs. Red lines are predictions from models by \citet{D'Antona2012}. These authors give Li production for stars of different masses. Since stars over a range of mass are likely to be involved in any pollution mechanism, we produced average values from these predictions weighting results for different masses using a \citet{Kroupa2002} mass function; the average was done over a mass range of [-2, 2]~M$_\odot$\ around each value of the mass. The labels written over the line indicates the mean values considered.}
\label{fig:li_production}       
\end{figure}

The argument related to the Li production within the polluters have been often overlooked in the literature, despite being a crucial diagnostic that should be considered to disentangle the stellar source of the polluters. At first glance, it might appear that there is a sort of ``conspiracy'' in that the Li production within the polluters has to be exactly at the same level of the original (primordial) Li content, i.e., A(Li) $\approx$ 2--2.2 dex. How much is this exactly the case depends on the abundance of Lithium in the diluting material (see Section 8.1). Usually it is assumed that the diluting material has the original Li abundance; we should remark that this is not obvious, because this depends on its origin; for instance, in Sect. 8.1.1 we will consider the case of diluting material from interacting intermediate mass binaries \citep{Vanbeveren2012} that should be Li-poor. Anyhow, in Fig.~\ref{fig:li_dilution}, we plot the difference in Li abundances for FG and SG (only intermediate) stars as a function of the dilution factors (where dilution 0 means pure ejecta and dilution 1 is for pristine material), as obtained from Na and O abundances and anti-correlations published by Carretta and collaborators. The orange curve represents the dilution under the assumption that there is no Li production within the polluters: as it can be seen from the plot, this curve does not reproduce the observed points, corroborating the indication that we must call for Li production in the polluters. In this context it is noteworthy that in case the diluting (pristine) material is Li-free, as it could be in the case of interacting binaries as source of diluters, the Li production would have been even more efficient.

If we now consider that the relationship between the observed Li abundance, the Li production and the dilution factor can be expressed as
$$ {Li}_{prod} = \log{[(10^{Li_{SG}} - dil\times 10^{Li_{FG}})/(1-dil)]}$$ 
and we can derive the amount of Li that has to be produced as a function of the dilution - in the hypothesis that the diluting material has the original Li abundance. This is the quantity that we have considered in Fig.~\ref{fig:li_production}, where we plot the difference in the Li production with respect to the original value as a function of the difference in ratios [Na/O]$_{\max}$ and [Na/O]$_{\min}$ from the Carretta's papers (\citealt{Carretta2009a}, \citeyear{Carretta2013a}). There is a hint for an anti-correlation between the two quantities, although the scatter is quite large. The argument is not so much different - but the relation should be tighter - if we assume that the diluting material is Li-poor.

The only polluter proposed so far able to produce Lithium are AGB stars: on the whole there is a reasonable agreement of the required Li production and at least some of the AGB model predictions (red solid line in Fig.~\ref{fig:li_production}), that is, those by \citet{D'Antona2012}. However, this agreement depends on details of the AGB models that suffer dramatic uncertainties, which include, but are not limited to: (i) the treatment of convection, (ii) the mass loss law, and (iii) the nuclear reaction rates. There are numerous works in the literature where these topics have been extensively discussed (see e.g. \citealt{Ventura2009}; \citealt{D'Orazi2013}; \citealt{doherty2014}, and references therein). The relevant aspect in this context is that the Li production, as revealed in SG intermediate stars of GCs, seems to point towards AGB model details that are in agreement with the lesson learnt from Na, O, Mg and Al nucleosynthesis. Thus, we require an enhancement in the $\alpha_{MLT}$ parameter (defined as the ratio between the characteristic size of the convective elements and the pressure scale-height). Note that the standard values between 1.7 and 2.05 are calibrated on the Sun, whereas for instance observations suggest that values up to $\sim$ 2.6 are needed to reproduce massive AGB stars in the MCs; \citep{mcsaveney2007}. Moreover, in order to have significant Li production in intermediate-mass AGB stars the mass loss has to be fast (otherwise the Li is burned again within the stellar interiors), following the approach by \cite{Bloecker1995}. This mass loss is higher than the formula by \cite{Vassiliadis1993}, reducing the lifetime of the star, and consequently the number of the thermal pulses. This last choice is now further corroborated by recent investigations such as e.g., the counts of massive (above 3 M$_{\odot}$) AGB stars in the SMC (\citealt{Pastorelli2019}, and references therein). Other critical points are however still kept alive, such as for example the efficiency of the third dredge-up and the number of thermal pulses, which would cause significant CNO and s-process element variation even in stars with masses around $\sim$ 5 M$_\odot$ (Marigo et al., in preparation).


\section{Evidence from dynamics}
\label{Sec:6}

Another important piece of evidence is provided by dynamical considerations coming from both the gas and the stars. Many complex processes are involved in the evolution of GCs, in particular in their tubulent early stages where the characteristics of multiple populations are set. In this Section we will analyse the main issues related to the dynamical properties of these two components with the aim of discussing the involved problematics and to identify possible signatures of the formation process of multiple populations imprinted in the kinematics of the stars observed today. 

\subsection{Gas dynamics}

GCs are currently deprived of gas, so that all the available information on the properties of the gaseous component from which all the stellar populations originated must be deduced from the chemical composition of stars survived till the present day. Although a commonly accepted picture of the GC formation process is still missing, it is possible to define some general characteristics of the gas-rich cloud in which proto-GCs formed.

A working hypothesis is that GCs formed in conditions resembling those occurring today in starburst galaxies where several massive clusters are observed to form. In particular, observations in the most massive star forming regions in the Local Group (like e.g. NGC~346 in the SMC, and NGC~604 in M~33) show a stellar complex forming in a cavity surrounded by a large amount of gas. Hydrodynamical simulations of turbulent molecular clouds suggest that the final star cluster form from the hierarchical merger of several sub-clumps, composed by both stars and residual gas \citep{Zamoraaviles2014}. In this scenario, a first question is whether multiple populations formed in independent clumps before their merging or this process formed a single homogeneous FG \citep{Elmegreen17}. Unfortunately, the hierarchical or monolithical origin of star clusters is still matter of debate \citep{Bonnell2011,Banerjee2015} and hydrodynamical simulations performed till now are far from providing detailed predictions for the chemical enrichment of the individual clumps.

An important constraint is set by the formation and evolution of the most massive FG stars. Indeed, massive O stars emit a large fraction of high-energy photons able to ionize the intra-cluster medium thus preventing star formation during their entire life \citep{Bodenheimer1979}. However, such massive stars often form in the dense centers of stars-forming region, possibly as a result of competitive accretion or primordial mass segregation \citep{Bonnell2003,Mcmillan2007}. Under this condition, most of the ionizing power of UV photons is absorbed by the gas flowing onto the star limiting the erosion of the neutral/molecular gas to a small region surrounding the source \citep{Dale2011}. Magnetic fields could also act as a further feedback agent, reheating the gas at epochs of the order of a few Myr \citep{Balsara2008}. In this case, little is known about their impact on the evolution of the intra-cluster medium.

At the end of their evolution (after $3\div 30$ Myr depending on their mass) massive ($8<M/M_{\odot}<25$) stars explode as SNe II whose feedback have enough energy ($\sim~10^{51}$~erg) to clean the intra-cluster medium from residual gas. An effect of the gas expulsion is that it acts as a net loss of potential energy occurring in a timescale much shorter than the dynamical time (the characteristic timescale over which stars react to potential changes). In this situation the cluster is off-virial equilibrium and reacts with a sudden expansion. The potential change exerts a mechanical work on stars pushing up their orbital energies. Those stars reaching the level of the inner Lagrangian point can evaporate from the system. A similar process has been advocated by \citet{Dercole2008} to explain the puzzling predominance of SG stars (the so-called "mass-budget problem"). In this case, the loss of the ejecta of FG SNe II induce an early episode of loss of FG stars (characterized by higher energies and therefore more prone to evaporation) over a timescale long enough to allow the formation of the SG. The material expelled by massive stars in the SN II explosion is enriched in Fe and $\alpha$-elements and may contaminate the composition of the surrounding gas. This poses a strong problem for many of the pollution/dilution models developed so far. Indeed, in all the models predicting a SG forming from the ejecta of massive ($8<M/M_{\odot}<20$) stars, the polluted gas is released in the same time interval of SNe II explosion. It is therefore necessary to recycle these ejecta into SG stars before the explosion of the first SN expels or contaminate the gas. 

The same fate is expected for the pristine gas which has been hypothesized to be a source of the dilution necessary to explain the extent of the Na-O anticorrelation. Note that the SN feedbacks could be unable to unbind the primordial gas if the proto-cluster has a mass $>10^{7}~M_{\odot}$ \citep{Leigh2013}. 
So, a way to retain the pristine gas while expelling the SNe II ejecta before these two chemically different gas mix is needed. \citet{Dercole2016} proposed a scenario in which GCs form in the disc of dwarf galaxies at high-redshift ($z\sim 2$) which later merge with the MW. The wind powered by SNe II create a bubble which breaks out the disc, releasing most of the Fe-, $\alpha-$rich material out of the galaxy. Only a negligible fraction of this gas mix with the gas in the disc which therefore maintains its primordial composition. When the bubble reaches a critical radius ($\sim$700 pc) the wind ram pressure is not able to balance the gravitational attraction of the cluster and the bubble contracts falling onto the potential well where the SG is forming. The entire process should last $\sim$40-50 Myr after the end of the SNe II explosion, but this timescale could be shortened assuming the cluster in motion with respect to the surrounding medium. In this case the boundaries of the gaseous bubble feel a different brake from the surrounding gas, creating a relative motion with the stellar component (which is instead not subject to any friction). Of course, most of these considerations rely on simplified simulations where spherical geometry is assumed to reduce the problem to a 1D treatment. The situation could be different in 3D simulations where the gas could find some escape route through the inhomogeneities of the surrounding medium altering both the duration of the gas removal and the mixing efficiency \citep{Calura2015}. Note that alternative sources of dilution have been proposed e.g. from outflows of non-conservative mass-transfer in binaries, which can occur after SNe II explosion and are therefore not affected by such an effect (see Section 8.1). 
 
Another issue related to the gas dynamics is linked to the duration of the star formation burst forming the SG. A relatively extended period is needed to explain the spread observed in the anticorrelation plots (see Fig. \ref{fig:abu2808a}), which cannot be explained only in terms of observational uncertainties. However, if SG forms in an extended interval of time, its massive stars would have enough time to explode leading to the same problem discussed above.

For this reason, most of the scenarios proposed so far require either a fast and inhomogeneous mixing of material expelled by massive stars (\citealt{Bastian2013, Elmegreen17, Gieles2018}) or an initial mass function for the SG characterized by a cutoff at high-mass preventing the explosion of SG SNe II \citep{Dercole2008, Charbonnel2014}. An alternative hypothesis is that SG massive stars are quickly ejected from the cluster before explosion as a result of their interactions in three- and four-body interactions. This effect could be accelerated by {\it i)} primordial mass segregation and {\it ii)} the fast segregation and decoupling of these massive stars from the rest of the cluster \citep[the so-called "Spitzer instability";][]{Spitzer1969}.

In the above discussion the feedback provided by SNe Ia has been overlooked. These explosions are indeed twice more energetic than core-collapse SNe being more efficient in expelling the gas from the cluster polluting the intra-cluster medium of Fe and Fe-peak elements. SNe Ia can explode at any time provided that a sizeable population of binaries involving white dwarfs are available. Given the uncertainties on the progenitor of these explosions (both single-degenerate or double-degenerate models have pro and cons), a precise constraint on the SNe Ia timing is not available, but theoretical models predict a rise of the explosion rate between 70 and 400 Myr, depending on the adopted model, from the formation of the FG \citep{Wang2012}. This timescale constitutes a limit for those models involving a pollution from intermediate-mass stars.

\subsection{Stellar dynamics}

\begin{figure}[htb]
\includegraphics[width=1.0\textwidth]{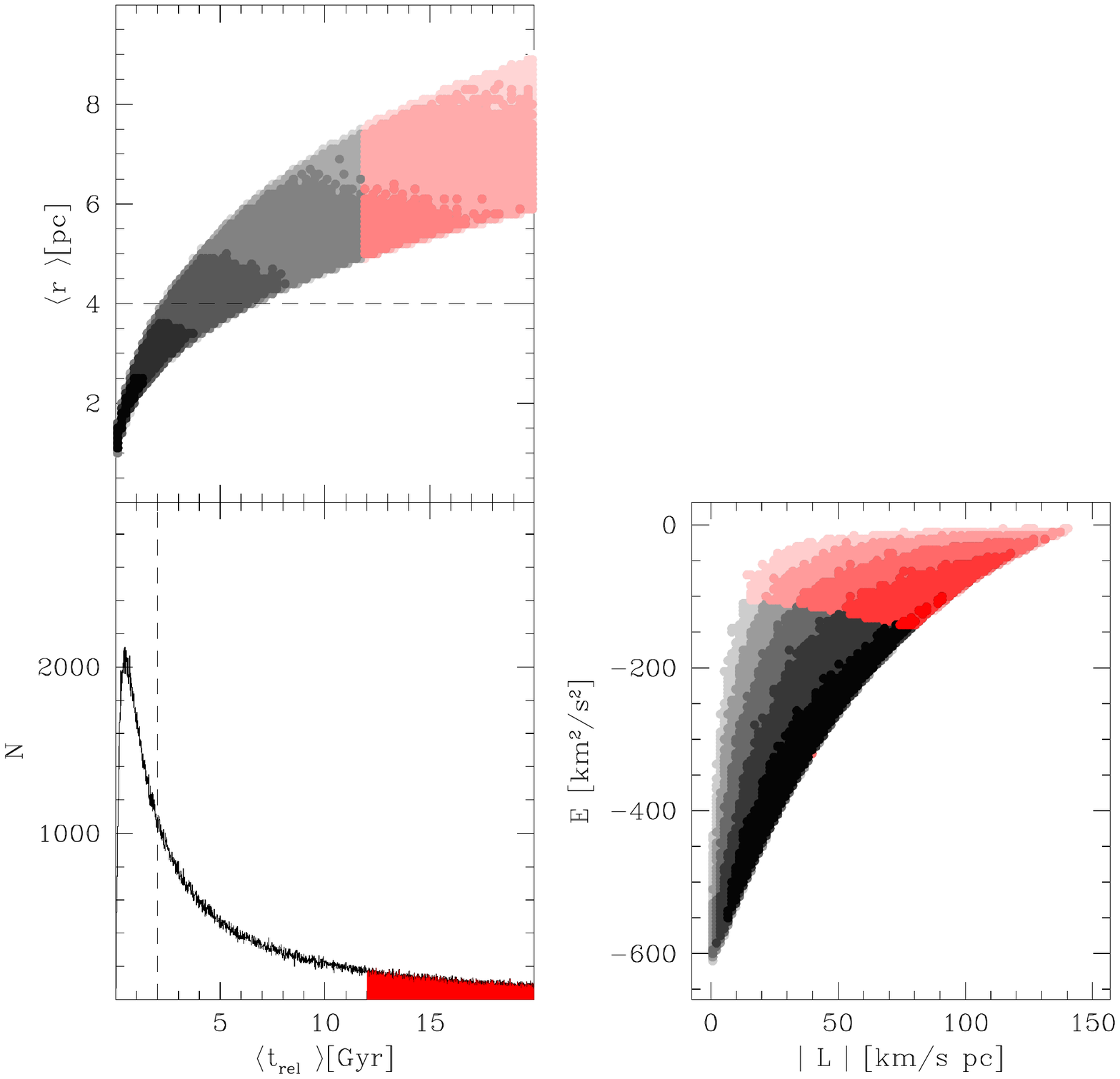}
\caption{Bottom left: Distribution of orbit-averaged relaxation times in a simulated cluster with $M=5\times 10^{5}~M_{\odot},~m=0.5~M_{\odot},~r_{h}=4~pc$ and a \citet{King1966} model profile with $W_{0}=5$. The half-mass relaxation time is indicated by the dashed line. Top left: distribution of orbit-averaged relaxation times as a function of the mean distance from the cluster center. Right: Distribution of particles in the E-L plane. The red area mark the region occupied by particles with $t>12~Gyr$. Darker contours indicate regions with increasing density in logarithmic steps.}
\label{fig:trel}       
\end{figure}

After the initial turbulent phase of their formation, FG and SG stars are expected to have different distribution in the phase-space determined by the corresponding properties of the gas from which they originate. Different scenarios predict different properties of multiple populations. In particular, in those scenarios predicting a SG formed by the gas collected in a cooling flow of low-velocity winds, SG stars are expected to be more centrally concentrated and with a lower mean kinetic energy (i.e. velocity dispersion at the same distance from the center) with respect to FG \citep{Dercole2008}. Moreover, any pre-existing angular momentum possessed by the cooling gas is conserved during the collapse thus leading to an increase of the systemic rotation speed \citep{Bekki2010,Bekki2011}. A larger radial anisotropy is also expected for SG stars since, forming from a cloud off-virial equilibrium, should retain information on the radial motion in their final velocity \citep{Lynden-Bell1967}. Because of the many uncertainties in the theory of star formation, little is known on the expected initial mass functions of the two populations, although the basic principle of competitive accretion predicts that concentrated populations (like the SG) are expected to have on average a bottom-heavy initial mass function \citep{Bonnell2011}. Scenarios predicting a SG formed from material ejected in-situ in the central cluster region \citep{Decressin2007, Gieles2018} predict similar characteristics with the remarkable exception that the two generations should share the same rotation pattern. Finally, if the chemical peculiarities of SG stars comes from an accretion of gas onto their proto-stellar disks \citep{Bastian2013}, they are expected to be constituted by those low-mass stars spending most of their lives in the cluster core. So, while SG stars in this scenario should share some of the structural and dynamical properties predicted by other models (concentration and low velocity dispersion), they should be characterized by a high-mass cutoff in their mass function, radial anisotropy and low rotational velocity \citep[at odds with what predicted by other scenarios;][]{Henaultbrunet2015}.

Unfortunately, in the subsequent evolution (constituting more than 99\% of the entire cluster life) stars of both populations interact between them and with the Galactic tidal field, moving across the phase-space. Differences between the structural and kinematic properties of FG and SG can be created or erased by the above effects.

Rotation is expected to be present since the formation of GCs because of the large-scale torques present in the original cloud \citep{Mapelli2017}. Additionally, a small degree of rotation can emerge in the outermost regions because of the interaction with the tidal field. Indeed, the Coriolis force produced by the joint motion of the star and the cluster is directed inward/outward according to the retrograde/prograde motion of the star. So, stars on prograde orbits are more easily expelled leaving a retrograde rotation close to the tidal radius \citep{Henon1970,Keenan1975,Read2006,Tiongco2016b}. FG stars, mainly located at large radii should be more prone to this effect than SG ones. In the same way, a certain degree of radial anisotropy can be primordial in all populations as a consequence of the violent relaxation occurring in the first stage of cluster formation \citep{Lynden-Bell1967}. During the subsequent evolution, radial anisotropy can develop outside the core where a significant number of stars are ejected by close encounters occurring in the central region \citep{Lyndenbell1968}. This effect is however reversed in the outermost regions where stars on tangential orbits are protected from evaporation by the angular momentum barrier (at the tidal radius, only stars with positive radial velocity $v_{r}=\sqrt{E-L^{2}/2 r_{t}}$ can escape; \citealt{Oh1992,Tiongco2016a}). So, SG stars spending most of their lives in the central region should develop radial anisotropy more efficiently than FG ones, while the opposite trend is expected in the outermost portion of the cluster. 
Another effect produced by dynamical evolution is on the mass function: stars with low masses indeed tend to acquire energy in collisions as a result of the tendency toward kinetic energy equipartition. They therefore are more prone to evaporate, leading to a flattening of the mass function as the mass-loss process proceeds. Numerical simulations show that 
in a tidally limited single population cluster the rate at which stars of different masses evaporate is a unique function of the fraction of lost mass, with larger mass-loss rate leading to flatter mass functions \citep{Vesperini1997,Baumgardt2003}.
On the other hand, clusters starting from a compact structure develop strong mass segregation before expanding up to the tidal boundary, and are characterized by a more efficient depletion of their mass function \citep{Trenti2010}.
The SG should be less exposed to tidal stress during its entire evolution and started its evolution in an underfilling configuration. N-body simulations by \citet{Vesperini2018} have shown that the balance between the two above effects leads to only marginal differences in the present-day mass functions of FG and SG.

Among the dynamical processes that erase primordial kinematical differences the dominant one is two-body relaxation. In particular, long-range interactions lead stars to exchange kinetic energy so that after a timescale compared to the relaxation time stellar orbital energies are randomized. The effect of two-body relaxation is therefore to homogenize the kinematic properties of FG and SG stars erasing the signatures left by their different formation mechanism. As a rule of thumb, the timescale needed for a star
to lose memory of its original motion is:
$$
t_{rel} = \frac{v^{2}}{D[(\Delta v_{\parallel})^{2}]}\sim 0.063 \frac{\sigma^{3}}{G^{2} m \rho \ln{\Lambda}}
$$
\noindent where $v$\ is the initial velocity, $D[(v_{\parallel})^{2}]$\ is the diffusion coefficient responsible for the spread in the velocity component parallel to the motion, $G$\ is the Newton gravitational constant, $m$\ is the mean mass of cluster stars, $\rho$\ is the density, $\sigma = \langle v^{2} \rangle ^{1/2}$\ is the 3D velocity dispersion and $\ln{\Lambda}$\ is the Coulomb logarithm \citep{Spitzer1969}. From the above formula it is immediately apparent that this quantity varies within the cluster (since both $\rho$\ and $\sigma$\ are functions of the distance from the center). In any realistic model \citep[e.g.][]{King1966} the relaxation time is shorter in the center and increases at large distances. This is a consequence of the largest number of interactions occurring in the dense central region accelerating the exchange of kinetic energy among stars. In a real cluster the situation is however more complex since stars vary their distance from the cluster center along their orbits passing through regions characterized by different efficiency of interaction. Usually, to have a gross estimate of the collisional status of an entire system as a function of its general parameters, the above timescale is integrated over the half-mass radius to obtain the so-called "half-mass relaxation time":
$$t_{rh}=\frac{0.138}{m~\ln{\Lambda}} \sqrt{\frac{M~r_{h}^{3}}{G}}$$
\noindent On average, in clusters with an age larger than their corresponding half-mass relaxation time, the process of two-body relaxation had enough time to randomize stellar orbits. At a first look, all the Galactic GCs lie in this regime with the only exceptions of $\omega$~Cen=NGC~5139, NGC~2419 and Pal 14. Note however that the above calculation consider the cluster as a whole while stars follow different kind of orbits. Thus, the distribution of individual orbit-averaged relaxation times, while peaked at short timescales have a long tail extending in the range exceeding the cluster age. This is shown in Fig. \ref{fig:trel} where the distribution of orbit-averaged relaxation time for a sample of $10^{6}$ synthetic particles in a \citet{King1966} potential with a typical GC mass and size ($M=5\times 10^{5}~M_{\odot}~r_{h}=4~pc$) is shown. While the half-mass relaxation time of this simulation is $t_{rh}=2~Gyr$, $\sim 29\%$ of the particles have orbits characterized by an average relaxation time longer than 12 Gyr being therefore only marginally affected by two-body relaxation. As expected, these stars are those with orbits confined in the outermost region of the cluster and occupy a peculiar region of the energy-angular momentum plane characterized by large energies and modulus of the angular momentum. So, the differences between the primordial FG and SG distribution in this portion of the phase-space (determining the rotation and anisotropy at large radii) are expected to be preserved even after an Hubble time and visible today.

Beside two-body relaxation, the interaction with the tidal field, while possibly creating small differences between the two populations close to the tidal radius (see above), also contributes to erase primordial differences. Indeed, the presence of the tidal field accelerates the process of mass-loss both imposing an energy cut and through the energy perturbations produced by disk/bulge shocks \citep{Henon1971,Ostriker1972,Aguilar1988}. This process carries away angular momentum, so that the larger is the fraction of lost stars, the smaller is the residual rotation \citep{Tiongco2017}. In a similar way, clusters losing a significant fraction of stars shrink, thus decreasing their half-mass relaxation time and boosting the effect of two-body relaxation.

The effect of dynamical evolution on the kinematic and structural properties of multiple populations has been studied using detailed N-body simulations by \citet{Henaultbrunet2015} \citep[see also][]{Tiongco2019} who found that, by assuming reasonable initial conditions for two formation scenarios, the final rotation pattern of FG and SG should show opposite trends similar to those set at the beginning of the simulation. A conservation of the segregation of SG in the innermost region has been also noticed by \citet{Vesperini2013} using a set of N-body simulations. They found that a complete dynamical mixing between the two populations occurs only in the most evolved GCs, while in many of them the SG should remain more concentrated than FG still today. A similar consideration holds for the velocity dispersion of the two populations. Indeed, in any system at equilibrium the density and the velocity dispersion of each population ($n_{pop}$ and $\sigma_{pop}$, respectively) are univocally connected by the Jeans equation:
$$\frac{d n_{pop}\sigma_{pop}^{2}}{d r}=-n_{pop} g$$
\noindent where $g=G M(<r)/r^{2}$\ is the gravitational acceleration at the radius $r$, $M(<r)$\ is the mass enclosed within $r$\ and an isotropic distribution is considered. So, the difference in the present-day radial distribution of the two populations naturally reflects into a difference in the corresponding velocity dispersions. Similar results have been obtained also by \citet{Mastrobuonobattisti2013} who analysed the evolution of an N-body simulation tailored to $\omega$~Cen=NGC~5139 and found that the initial concentration, flattening, small dispersion and rotation of the SG are preserved after 12 Gyr of evolution.

Summarizing, regardless of the adopted scenario for the formation of multiple populations, SG stars should appear more concentrated, with a smaller velocity dispersion (measured at the same distance from the cluster center), a smaller fraction of binaries and a significant radial anisotropy at intermediate radii. A larger rotation amplitude is also expected for this population if the scenarios involving an original cooling flow are correct.
 
From the observational side, the concentration of SG stars has been proved in almost all GCs (e.g. \citealt{Sollima2007a,Lardo2011}, with only a few possible exceptions, NGC~6362, NGC~6093=M~80, NGC~7078=M~15,  \citealt{Dalessandro2014,Dalessandro2018b,Larsen2015}, respectively. For M~15 see however also \citealt{Nardiello2018a}). This difference does not seem to be associated with any velocity dispersion difference, although some tentative evidence has been proposed in some cluster \citep{Bellazzini2012,Dalessandro2018c}. In this regard, note that velocity dispersion profiles suffer from uncertainties which are several times larger than those of projected density because of the limited sample of radial velocities available. The first evidence of differences in the anisotropy profile of FG/SG stars have been put forward thanks to the accurate proper motions obtained through HST in 47 Tuc \citep{Richer2013,Bellini2015,Milone2018} and NGC~362 \citep{Libralato2018}, with the SG displaying a larger degree of radial anisotropy with respect to the FG. The only available evidence to date of differences in the rotation pattern of different generations of stars is provided by \citet{Pancino2007} (in $\omega$~Cen=NGC~5139) and \citet{Cordero2017} (M~13=NGC~6205) who found opposite results: while SG stars in $\omega$~Cen=NGC~5139 share the same rotation pattern of FG ones, in M 13 they have an average rotation amplitude which is larger than the rest of the cluster stars. All this evidence agree with the expectations of the theoretical models exposed above. Another evidence related to structural differences between FG/SG is provided by \citet{Milone2012e} who found that SG stars in NGC~2808 have a flatter mass function with respect to FG ones. 
Consider however, that the mass function measured in a limited radial range is not representative of the global mass function, so that it is hard to interpret this evidence without a complete modelling of the dynamical evolution of this cluster accounting for this observational bias. Moreover, a consensus on the star formation theory determining the shape of the initial mass function is missing \citep[see e.g. ][]{Adams1996, Chabrier2014} so that it is not clear if the turbulent environment where multiple populations formed could have lead to primordial differences in their initial mass functions which left traces on their present-day mass functions.

Another aspect poorly investigated till now regards the fraction of massive remnants retained by GCs. In the commonly accepted scenario, black holes and neutron stars formed after SN II explosions should receive natal kicks resulting from the off-center onset of the deflagration process. Models of asymmetric SN II explosions predict kick velocity distributions characterized by dispersions of $\sigma_{k}=80-100$ km~s$^{-1}$, i.e. larger than the cluster escape speed, so they are expected to be ejected outside the cluster after their formation \citep{Drukier1996,Moody2009}. Assuming a \citet{Plummer1911} model and a Maxwellian distribution of velocities truncated at the cluster escape speed, the fraction of neutron stars/black holes which can be retained by a cluster with mass $M$ and Plummer radius $r_{0}$ is: 
$$ f_{ret}=\left[Errf(x)-\frac{2}{\sqrt{\pi}}~x~\exp{(-x^{2})}\right]$$
where: 
$$x=\left(\frac{3\pi}{32}+\sqrt{2^{2/3}-1}\frac{r_{h}~\sigma_{k}^{2}}{G~M}\right)^{-1/2}$$
The above formula indicates that the retention fraction is a rapidly increasing function of the ratio $M/r_{h}$. Assuming $r_{h}=4~pc$ and $M=5\times 10^{5}$~M$_\odot$, the above relation predicts a retention of less than 1\% of massive remnants. On the other hand, this fraction increases to 18\% if GCs were an order of magnitude more massive at their birth, as required by some of the formation scenarios of multiple populations. While the most massive remnants (e.g. black holes with a mass contrast $>$10 with respect to the mean cluster mass) are expected to quickly evaporate as a result of the Spitzer instability, the less massive neutron stars will be retained more efficiently than the other less massive stars till the present day. Unfortunately, assuming a standard initial mass function (IMF), the mass contained in neutron stars will never exceed a few percent of the total mass, so that a proof of this scenario cannot be obtained from dynamical considerations. However, an increased retention of neutron star could help to explain the high fraction of millisecond pulsars observed in GCs \citep[exceeding by a factor 100-1000 over the field population,][]{Verbunt1989} which would be otherwise difficult to be explained if all neutron stars were expelled by natal kicks.


\section{Binaries}
\label{Sec:7}

\begin{figure}[htb]
\includegraphics[width=\textwidth]{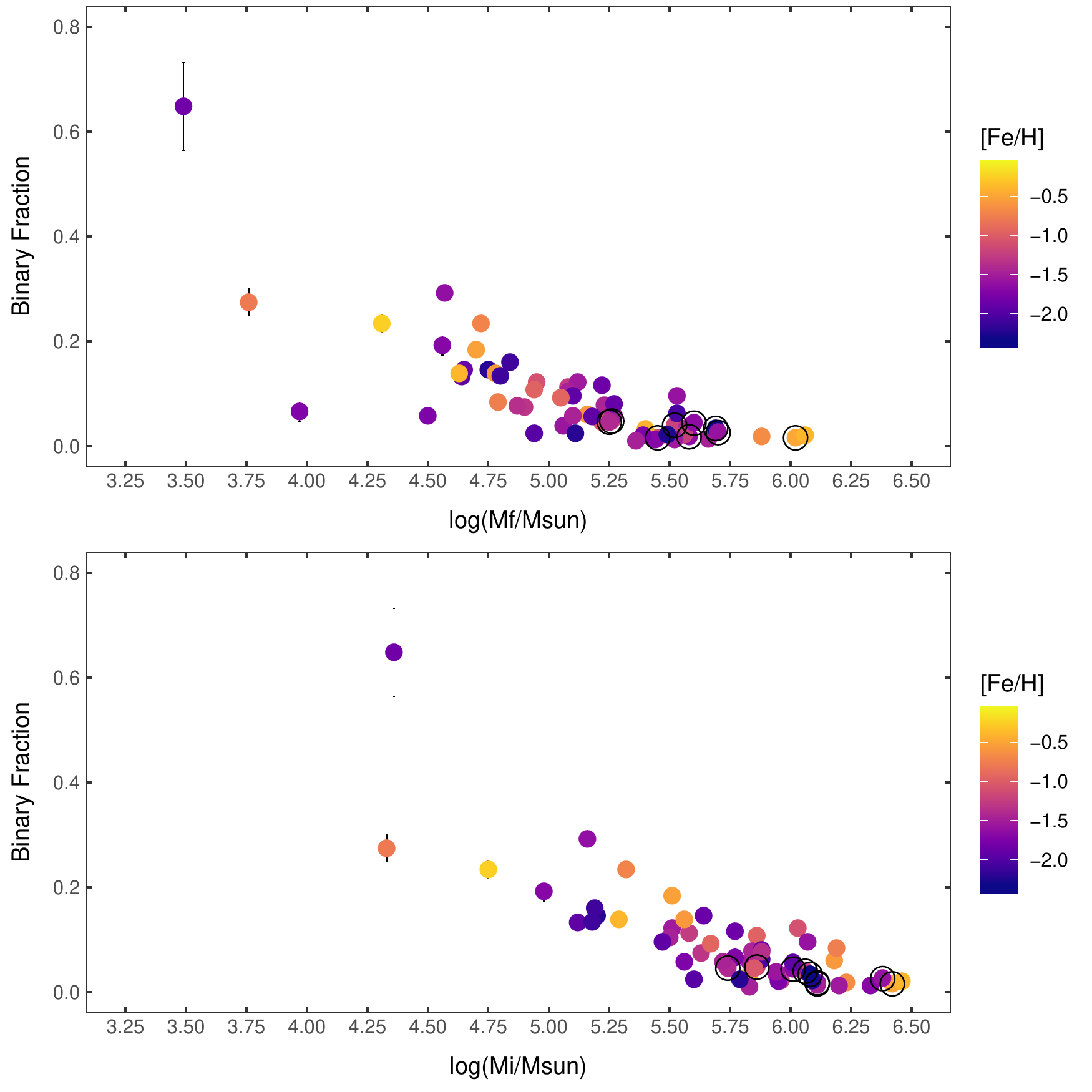}
\caption{Upper panel: run of binaries within a cluster \citep{Milone2012c, Milone2016} and the cluster mass from \citet{Baumgardt2019}. Lower panel: the same, but using the initial mass from \citet{Baumgardt2019}. Circled symbols are for Type~II clusters defined as in \citet{Milone2017}. Colours code metallicity (see scale on the right of the plot) }
\label{fig:mass_bin}       
\end{figure}

It is well established that a very large proportion, if not the majority, of stars in the field are in multiple systems, with the binary fraction increasing as a function of stellar mass (see e.g. \citealt{Moe2017}). Metallicity and environmental density seem to play a role in the incidence and orbital parameters of double systems. The binary fraction seems to increase with decreasing metallicity \citep{Moe2018}. On the other hand dense systems seem to disrupt these objects and thus decrease their incidence (see e.g. \citealt{Duchene2018}).

Binaries play an important role in our understanding of GCs. They are a source of heating, and thus they are relevant to the study of GC dynamics. Many of the exotic objects (e.g. Blue Straggler Stars  - BSS, CH-stars, cataclysmic variables, milli-second pulsars, X-ray binaries, etc.) found in GCs are the result of the evolution of a binary system. Accurate accounting for binaries has bearings on the derivation of the cluster mass and luminosity function.

Photometric searches for binaries in GCs have been undertaken since the early '90s, looking either for eclipsing binaries (e.g. \citealt{Yan1996}) or for stars on the so-called binary sequence (located on the red side of the MS -- e.g. \citealt{Bolte1992,Rubenstein1997,Bellazzini2002,Sollima2007b,Milone2012c,Milone2016} just to name a few). The earlier method is limited to systems with large orbital inclination and tends to favour short orbital periods (which however are more common in clusters than in the field, see below for a discussion), but it provides information about the periods and can be applied to any evolutionary stage. The latter is more complete in terms of the binary census, but is limited to the MS stage and to binaries with high mass-ratios, and provides no information on the binary orbital parameters.

Radial velocity monitoring is another avenue to characterize the binary population. While the availability of spectrographs with high-multiplexing capabilities like e.g. FLAMES@VLT has made this kind of search reasonably efficient for GC giants, the statistics is based still on samples several orders of magnitude smaller than that of photometry. The use of MUSE has also shown promise in this field (see e.g. \citealt{Giesers2019}).
The method is biased towards shorter periods and large orbital inclinations, but it can provide information on the orbital parameters and on the composition of the binary stars. 

\subsection{Overall frequency of binaries GCs}

\begin{figure}[htb]
\includegraphics[width=\textwidth]{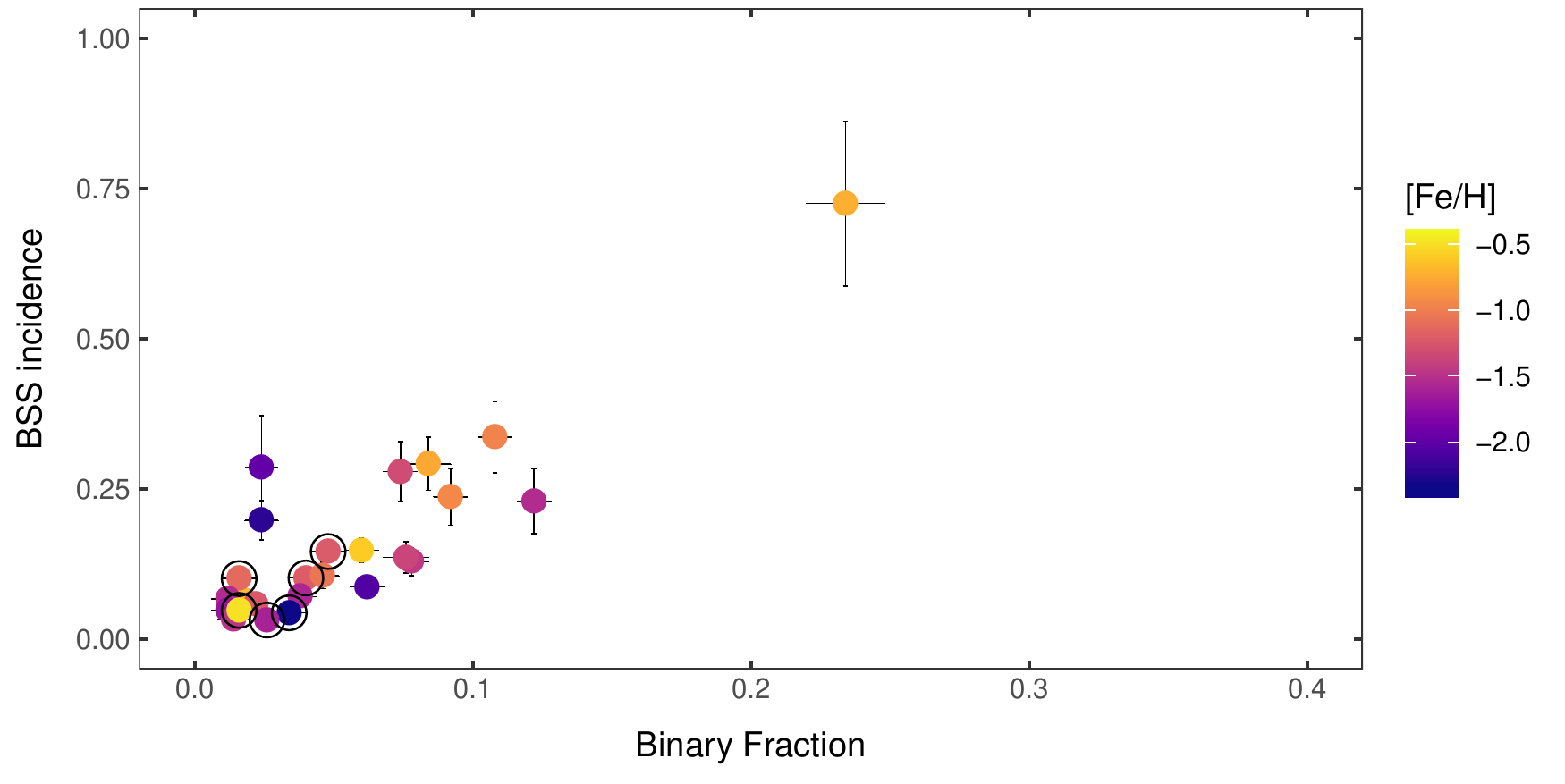}
\caption{Correlation between the fraction of binaries \citep{Milone2012c, Milone2016} and the incidence of BSS (number of blue stragglers per 100~$L_\odot$ in the same area: \citealt{Moretti2008}). Circled symbols are Type~II clusters defined as in \citet{Milone2017}.  Colours code metallicity (see scale on the right of the plot)}
\label{fig:bin_bss}       
\end{figure}

\subsubsection{Observational evidence}

Observations agree in finding binary fractions among GC stars generally lower than found in the field for stars of similar kind. This is consistent with the general expectation that a concentrated environment tends to disrupt binaries \citep{Heggie75}. The measured overall fractions show however considerable variations. \citet{Milone2016} investigated the monovariate relations of the MS binary fraction and various cluster parameters for 59 GCs, and found an anti-correlation with cluster luminosity and a correlation with BSS incidence, confirming earlier findings reported by \citet{Sollima2010} and \citet{Sollima2008} on the basis of smaller samples, who suggested that cluster mass might be one of the driving parameters to the binary fraction.

Figure \ref{fig:mass_bin} shows the run of the binary fractions  (we will be using the total binary fraction from the HST-WFC field listed in \citealt{Milone2012c,Milone2016}) and the initial and present time  cluster masses \citep{Baumgardt2019}. The quantities are clearly anti-correlated, with an effect that is more pronounced when initial rather than final masses are used (Spearman correlation coefficients -0.81 and -0.77 respectively). On the other hand, Milone et al. found a moderate anti-correlation of the binary fraction with core relaxation time. 

\subsubsection{Evolution of binary systems in GCs}

The estimate of the fraction of expected binaries in a GC and of their characteristics requires rather extensive and detailed N-body simulations. In particular, such a complexity is due to the fact that binary orbital periods (hours to few tens of years) are much smaller than the characteristic dynamical time of stars ($10^{5}$ to $10^{10}$~yr). This imposes an upper limit to the time-step of the simulation when a close encounter involving a binary star is going to occur. For these reasons, predictions on the dependence of the binary fraction on the various GC parameters were performed using Monte Carlo \citep{Ivanova2005,Fregeau2007,Fregeau2009} and simplified analytical \citep{Sollima2008} calculations. Predictions from N-body simulations have been provided by \citet{Hurley2007} and \citet{Trenti2007}. In all these last studies however, to reduce computation time, only those binaries with binding energy larger than the average kinetic energy of single stars are considered.

The basic idea is that the lower incidence of binaries found in clusters with respect to the field is due to their being high density environments, with high velocity dispersion and thus large typical relative velocities. Therefore, close encounters of the double system with a third stellar object are much more likely than in the field. In these events, the outcome depends on the relative energies of the involved parties. Encounters where the binding energy of the binary exceeds the kinetic energy of the third star will tend to make the system more bound, while in the opposite case the binary will become looser (or possibly be disrupted). This is the so-called Heggie's Law \citep{Heggie75}, which states that in an environment such as a GC the effect of three body encounters over time will make soft binaries become softer and hard binaries become harder \citep[see also][for a review]{Heggie2003}.

Binary ionization can happen under two conditions: {\it i}) the relative velocity of the binary and the incoming star is rather large, exceeding the so-called critical velocity \citep{Hut1983}, which depends on the binary binding energy and on the masses of the three stars involved. For stars of similar mass, this is very similar to the binary orbital velocity (save for a shape factor which accounts for details of the encounter, including inclination of the encounter and eccentricity of the orbit); {\it ii}) the ionizing star must get close enough for the collision to take place, a distance comparable to the separation of the double system.

\begin{figure}[htb]
\includegraphics[width=\textwidth]{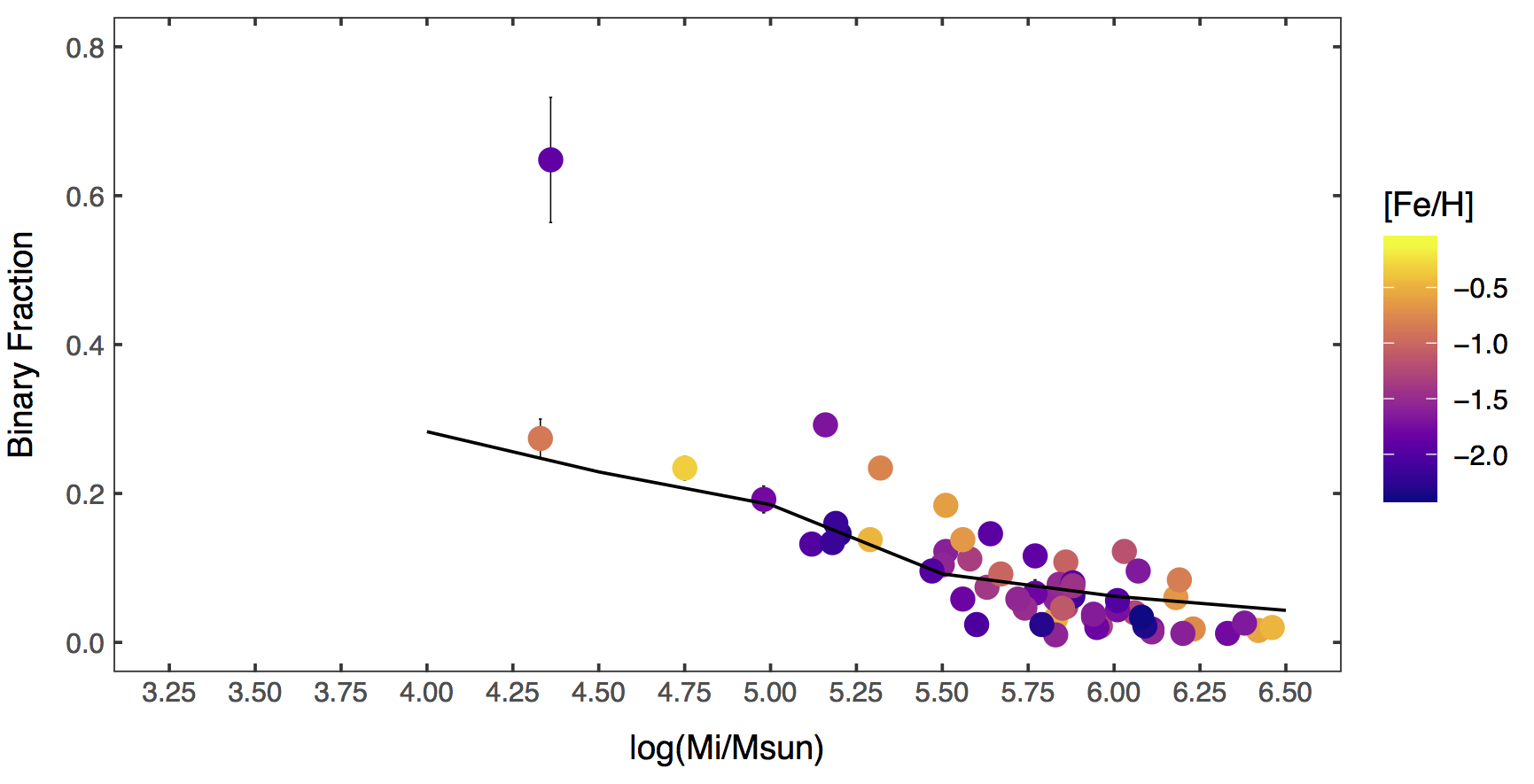}
\caption{Run of the fraction of binaries within a cluster \citep{Milone2012c, Milone2016} with the cluster initial mass from \citet{Baumgardt2019}. Circled symbols are for Type~II clusters defined as in \citet{Milone2017}.  Colours code metallicity (see scale on the right of the plot). The black line is the prediction with obtained with a simple model for binary destruction - see Section 7.3.}
\label{fig:mass_bintoy}       
\end{figure}

\subsection{Frequency of binaries in different stellar populations}




A surprisingly small number of observational studies have attempted to study binaries in different GC stellar populations.

\citet{D'Orazi2010a} took an indirect approach, deriving the incidence of Ba stars (which are known to belong to binary systems of rather short period) among the FG and SG in 15 Galactic GC and finding that their fraction in FG stars is similar to the field, but much smaller in the SG. They also reported on the binary fraction based on long term radial velocity monitoring for the cluster NGC~6121, finding a binary fraction in FG stars over one order of magnitude larger than in SG stars.


\citet{Lucatello2015} used radial velocity monitoring to derive the binary fraction in FG and SG stars in 10 GCs. The reported binary incidences in the two population are 4.9$\pm $1.3\% among FG stars and 1.2$\pm $ 0.4\% among SG stars. They then report that the binary fraction in FG is four times larger than in SG, under the assumption that the period distributions in the two populations is the same. They conclude that such finding suggests that SG stars were born in a denser environment than FG stars. It is worth noticing, that, as discussed before, the denser an environment the more binaries are ionized but also the period distribution of the surviving ones is skewed toward shorter periods. Therefore, the binary fraction detection efficiency from searches of spectroscopic binaries is expected to be lower for FG than SG binaries, given that the latter likely have a period distribution more skewed toward shorter periods, and thus the ratio between the binary fraction in the two populations is expected to be even larger than the one detected.


\citet{Dalessandro2018a} monitored the radial velocity of over 500 members of the low mass cluster NGC~6362. They also found that the incidence of binaries was over an order of magnitude higher among FG stars than in SG stars.

These observational findings are a good match for theoretical predictions. \citet{Vesperini2011} used an hybrid analytical-numerical approach to follow the evolution of the binary population in the context of multiple populations. They found that one of the consequences of the SG forming in a more centrally concentrated environment than the FG was indeed a lower binary fraction in the former with respect to the latter. The reason behind this difference is the increased disruption rate that SG binaries experience as a consequence of a larger number of stellar encounters in the high density environment of their birth. The above finding has been confirmed by N-body simulations by \citet{Hong2015,Hong2016} who also found that such a difference is expected to be observable only for those binaries above a critical separation, while tight binaries (e.g. those producing X-ray binaries) should be less affected by such an effect.

As a caveat, we remind that \citet{Hong2019} find that the present time spatial distribution is affected by the differences in the binary fractions, and that the relative incidence of FG and SG binaries might very well show considerable radial dependence even after the spatial distribution of single stars from the two populations become identical.

Given the different incidence of binaries in FG and SG discussed above, one can wonder if the trend observed in Fig.~\ref{fig:mass_bin} could be interpreted as a simple consequence of the decreasing fraction of FG stars with increasing mass (see Fig.~\ref{fig:mass_f1}.) High mass clusters are dominated by SG, where binaries are very rare, while low mass clusters are mostly FG, which has a much larger binary fraction. Fig.~\ref{fig:f1_bin} shows the run of the binary fraction, of the frequency of BSS with respect to subgiant branch stars, and an averaged one as a function of the FG fraction. The upper panel, as discussed, shows that the fractions of binaries and of FG stars is correlated, however a simple calculation shows that such trend is not reproducible by just changing the FG fraction while keeping the fractions of binaries in FG and SG stars constant, but requires that each of these quantities themselves also varies with the FG fraction (and hence with the mass).


\subsection{Toy model and the different populations}

\begin{figure}[htb]
\includegraphics[width=\textwidth]{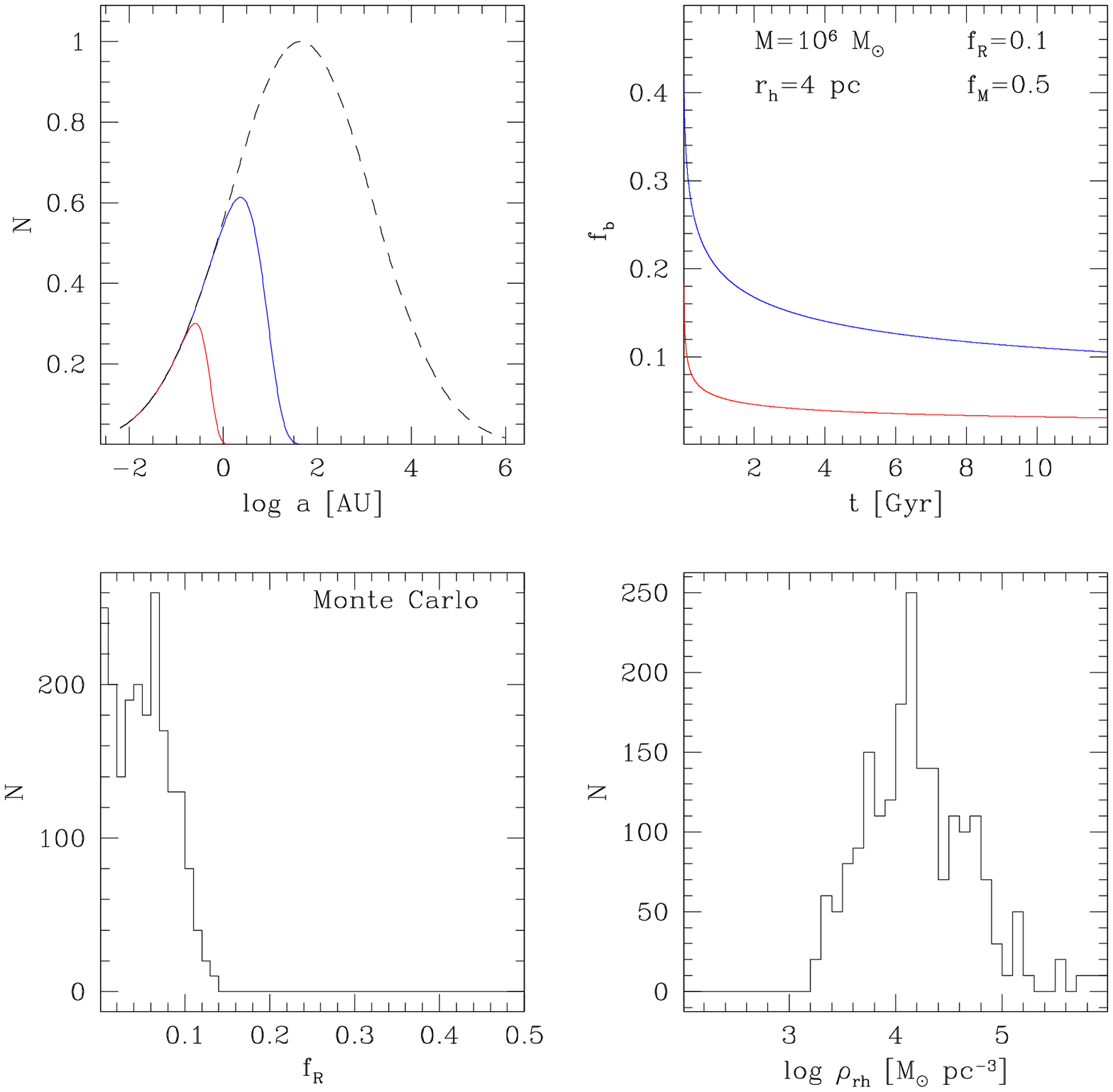}
\caption{Top panels: semi-major axes distribution (top-left) and binary fraction as a function of time (top-right) for the toy model simulation with $M=10^{6}M_{\odot}~r_{h}=4~pc,~f_{M}=0.5$ and $f_{R}=0.1$. Blue and red lines indicate predictions for the FG and SG, respectively, while the black dashed line indicate the original distribution at birth. Bottom panels: distribution of $f_{R}$ (bottom-left) and $log~\rho_{rh}$ for the Monte Carlo toy model simulations corresponding to final FG binary fractions 5\%$<f_{b,FG}<$10\% and ratio $f_{b,SG}/f_{b,FG}<0.2$.}
\label{fig:bintoy}       
\end{figure}

In order to interpret the implications of the binary fraction within GCs in the context of multiple populations, we may use a simple toy model. For the sake of simplicity we will assume that, at its formation, the distribution of FG and SG ($\rho_{FG}$ and $\rho_{SG}$, respectively) can be represented by the superposition of two \citet{Plummer1911} models as a function of radius (r) within the cluster, with different masses and characteristic radii. An initial fraction of 50\% of binaries are distributed with the same radial distribution of their parent populations and with the period/semi-major axes distribution of field stars \citep[$g(a)$ taken from][]{Raghavan2010}, where $a$\ is the semi-major axis. We consider equal-mass stars with mass of $m=0.5~M_{\odot}$, no mass segregation between binaries and single stars and no dynamical evolution. These are no doubt incredibly simplistic approximations, however the underlying assumption is that the main driver of the evolution of the binary fractions in both populations is the process of ionization mainly occurring at early stages, while the subsequent evolution and its details act as second order effects.

The local ionization rate can be calculated from the relation \citep{Hut1983}:
$$\Delta(a,r)=\frac{1}{N_{b}}\frac{d N_{b}}{d t}=\frac{3~\rho(r)~\pi~a^{2}~ \sigma(r)~R(a,r)}{\sqrt{2}~m} $$
where:
$$R(a,r)=1.64/[(1+0.2/x)(1+\exp{(x)})]$$
and $x=\frac{G~m}{2~a~\sigma^{2}}$ is the hardness parameter, $G$ being the gravitational constant. Note that in the above formula $N_{b}(t)$ is the number of binaries at the time $t$, $M$ is the cluster mass,  $\rho=\rho_{FG}+\rho_{SG}$ is the overall mass density profile and $\sigma$ is the velocity dispersion in any one direction which includes the contribution of both populations (calculated by solving the isotropic Jeans equation). For a given semi-major axis $a$, the average ionization rate $k$ of FG/SG binaries can be calculated by integrating over the corresponding density profiles: 
$$k_{FG}(a)=\frac{1}{(1-f_{M})~M}\int_{0}^{+\infty} 4 \pi r^{2} \rho_{FG} \Delta(a,r) dr$$
$$k_{SG}(a)=\frac{1}{f_{M}~M}\int_{0}^{+\infty} 4 \pi r^{2} \rho_{SG} \Delta(a,r) dr$$
while the fraction of surviving binaries at the time $t$ is given by: 
$$\eta~(t)=\frac{N_{b}(t)}{N_{b}(0)}=\int g(a)~\exp{[-k~t]}~da$$
The fraction of binaries of each generation is then calculated from the above quantity as:
$$ f_{b}(t)=\frac{\eta~ (t) f_{b}(0)}{(1-\eta~ (t)~f_{b}(0)+1}$$
In the above calculation, there are four free parameters: the global mass and half-mass radius of the system, and the ratio between the masses ($f_{M}$) and characteristic radii ($f_{R}$) of FG/SG. The evolution of the binary fraction of FG and SG assuming $f_{M}=0.5$, $f_{R}=0.1$, $r_{h}=4~pc$ and $M=10^{6}~M_{\odot}$ is shown in the top-right panel of Fig. \ref{fig:bintoy}. It is apparent that the fractions of binaries of both populations are mainly set in the first few 100 Myr with only a negligible evolution at later time. This is a consequence of the adopted period distribution which peaks at periods of $\sim~300$~yrs, so that most of the binaries in both populations are soft and are quickly destroyed, while the remaining hard binaries survive for a long time.

The above toy model does not account for the dynamical evolution of the system which certainly affects both populations. However, N-body simulations by \citet{Dercole2008} show that in a realistic cluster the FG roughly maintains a constant size, while SG expands further reducing the ionization efficiency. So, the structural evolution occurring over a long timescale should play only a second-order role in decreasing the number of binaries. The consequence of the above consideration is that the binary fractions of FG/SG contain crucial information on the early stage of cluster evolution when the maximum efficiency of binary ionization determined these fractions \citep[see also][]{Fregeau2009}.

We randomly extracted the four involved parameters over a wide range ($0<f_{M}<1$; $0<f_{R}<1$; $6<\log{M/M_{\odot}}<7.5$; $0<\log{r_{h}/pc} <1.5$) and calculated the corresponding fraction of FG/SG binaries after 12 Gyr of evolution. We found that the final ratio $f_{b,SG}/f_{b,FG}$ is a unique function of the $f_{R}$ parameter, and it is almost insensitive to the other parameters. In particular, the values of $f_{b,SG}/f_{b,FG}<0.2$ measured in real GCs can be obtained only assuming $f_{R}<0.15$ i.e. a SG forming in a volume $\sim~300$ times smaller than that of FG (see the bottom panels of Fig.\ref{fig:bintoy}). Moreover, the general value $5\%<f_{b,FG}<10\%$ can be obtained only assuming initial half-mass densities ($\rho_{hm} \equiv \frac{3M}{8\pi r_{h}^{3}}$) in the range $3.2<\log{\rho_{hm}}<5.2$.

In the top-left panel of Fig. \ref{fig:bintoy} the distribution of semi-major axes of the surviving binaries in our toy model is also shown. Note that the cutoff occurs for both populations at rather small values. This could explain the correlation between the overall binary fraction and the BSS incidence, whose precursors must be systems with relatively small separations ($<$200~R$_\odot$, that is $<$1~au).

Of course, the above calculation is approximated and relies on the strong assumption that binaries in GCs form in the same fashion as in the field. Moreover, other complex processes (like binary-binary interactions, segregation of binaries, exchanges between binary components, hardening/softening, coalescence, tidal capture, stellar evolution and tidal field effects, etc.) could have non-negligible effects in shaping the long-term evolution of the binary fraction in both populations. However, the above exercise provides an example of the unique information on the original properties of GCs retrievable from the properties of FG/SG binaries.

\subsection{Blue stragglers, CH/Ba stars, and the contribution of binaries to the fraction of stars with chemical anomalies}

Internal mixing (related e.g. to rotation) or heavy mass-loss in binary systems may cause significant variations of the surface abundances, in particular for Carbon and Nitrogen. This is indeed expected \citep{Sarna1996} and observed in the case of a fraction of the BSS (\citealt{Sandage1953}: \citealt{Ferraro2006})\footnote{BSS may also be produced by collision in the dense core of GCs. In that case, there should not be large chemical anomalies \citep{Lombardi1995}. However, the majority of BSS in both globular and open clusters are likely the aftermath of the evolution of primordial binaries (see e.g. \citealt{Piotto2004}).} and Ba-CH stars \citep{D'Orazi2010c}. While we expect that only a minority of stars in a cluster are BSS or Ba-CH stars - or the result of their evolution - caution should be exerted when considering cases where the fraction of N-enriched stars is very low as evidence for multiple populations. In the following we will try to have a first rough estimate of the incidence of such objects on number counts of stars along the RGB.

Those binaries with separation of the order of or smaller than 1~au are expected to interact during the evolution along the red giant branch producing mass-transfer BSS \citep{McCrea1964} or along the asymptotic giant branch producing Ba-stars or CH stars \citep{McClure1980}. This separation should roughly corresponds to initial periods of the order of 1 year. If we then consider the period distribution for field binaries by \citet{Raghavan2010}, we end up with a fraction of interacting binaries that is about 14\% of the total. Since binaries make up some 46\% of the F-G spectral type stars in the solar neighbourhood, we expect that about 6\% of the stars may show some abundance anomalies related to being binaries. This is similar to the fraction of BSS per main sequence star in open clusters, that is between 7 and 10\% for clusters older than 1~Gyr (\citealt{DeMarchi2006, Ahumada2007}, with the caveat that an improved demographics of BSS in open clusters should be derived using the Gaia data on membership in clusters) while the fraction of Ba-stars is of the order of 2\% \citep{Luck1991}.

\begin{figure}[htb]
\includegraphics[width=1.0\textwidth]{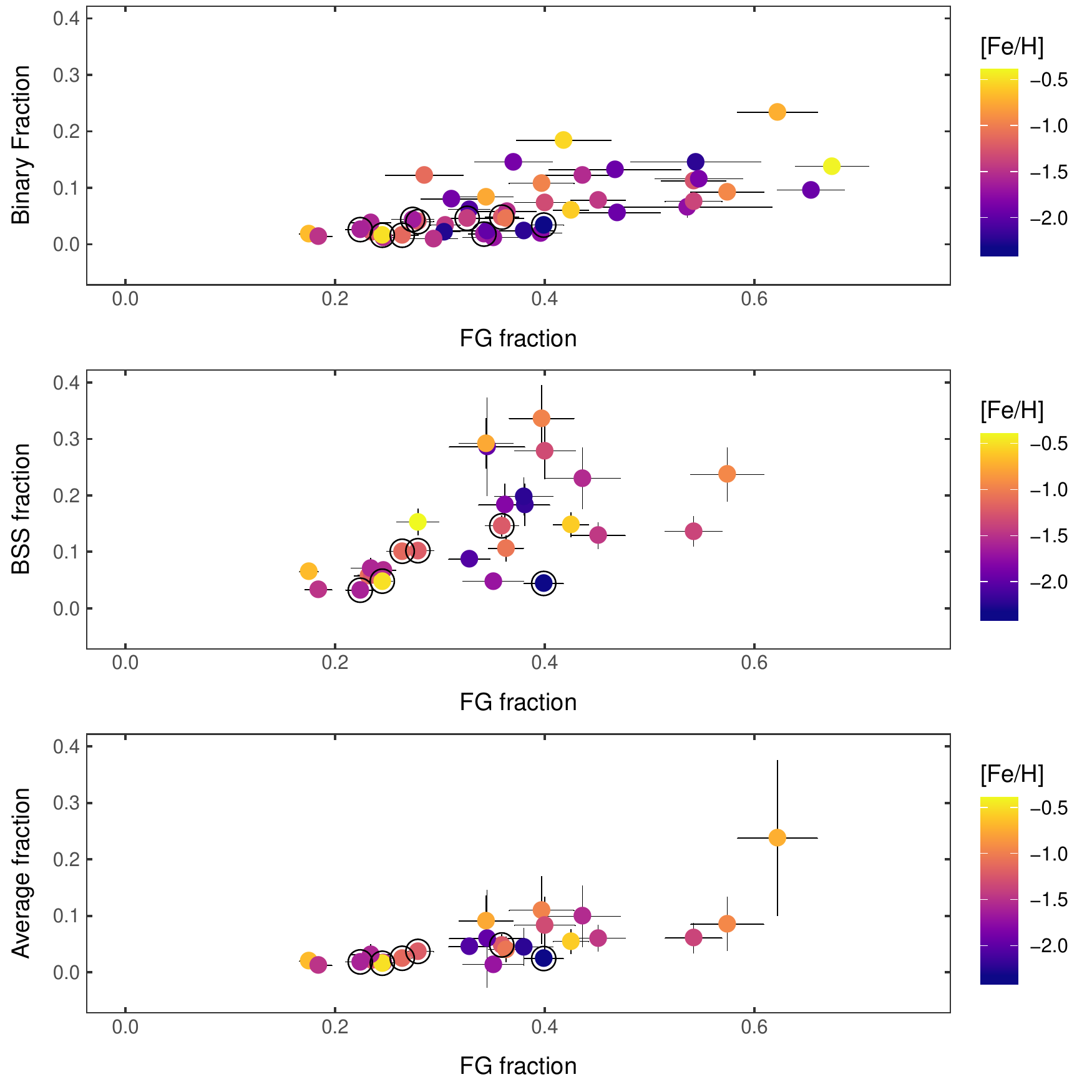}
\caption{Upper: run of the fraction of binaries within a cluster \citep{Milone2012c, Milone2016} and the fraction of First Generation stars from \citet{Milone2017}. Circled symbols are for Type~II clusters according to the classification by \citet{Milone2017}. Colours code metallicity (see scale on the right of the plot)}. Middle panel: the same, but with the incidence of BSS \citep{Moretti2008}. Lower panel: the same, but with the average of the fraction of binaries and of the incidence of BSS {\bf for individual clusters. This last was divided by 3 before making the average to scale it similarly to the binary fraction.}
\label{fig:f1_bin}       
\end{figure}

On the other hand, the fraction of BSS declines with cluster mass/luminosity down to a fraction more than an order of magnitude smaller in massive GCs \citep{Piotto2004, DeMarchi2006, Moretti2008}. As shown in Fig.~\ref{fig:bin_bss} there is indeed a good correlation between the incidence of BSS and the fraction of binaries in a cluster, that may be interpreted in terms of the evolution of primordial binaries, which is affected by the stellar encounters (see e.g. \citealt{Davies2004}). This decline parallels that in the binary fraction of binaries in FG/SG stars (see previous subsection). This good agreement can be used to reduce the scatter in the relation between the fraction of first generation stars and of binaries, as shown in Fig.~\ref{fig:f1_bin}, where we compare results obtained using only the binary fraction determined from main sequence stars, the incidence of BSS, and averaging the two.

Once H at center is exhausted, BSS in old clusters are expected to evolve along the RGB, where they may then mix with single stars in the colour-magnitude diagram. Since they are more massive, they should be slightly bluer and evolve somewhat faster, and then they should be under-represented along the RGB. To evaluate this last term, we may use evolutionary tracks (e.g. the BASTI ones, \citealt{Pietrinferni2006}). While a complete analysis is beyond the purposes of this review, we found that e.g. for $\alpha-$enhanced tracks with Z=0.004, a star with 1.3~M$_\odot$\ employs 60\% of the time of a star with 0.9~M$_\odot$\ to evolve from an absolute $M_V=2$\ to the tip of the RGB. We then expect that the post-BSS may be of the order of 6\% of the RGB stars in an old open cluster, and likely an order of magnitude less in a typical GC; Ba/CH-stars are expected to be about a factor of five less frequent. This fraction should then be multiplied by the fraction of those BSS that are C-poor and N-rich, that is about 15\%  \citep{Ferraro2006}. Evolved BSS may then generate a small ($\sim 1$\%) population of chemical anomalous stars in old open clusters that should be taken into account when searching for evidence of multiple populations within them. On the other hand, we expect that the impact of these objects within most GCs is negligible.



\section{GCs and the halo}
\label{Sec:8}

\subsection{Mass budget}

The peculiar nucleosynthesis observed in GCs, so much different from that typically observed in the field of galaxies, suggests that the material from which SG stars form is processed in the interior of only a fraction of the original stars present in a cluster. Since the SG stars typically makes up the majority of cluster stars (at least in massive clusters), this suggests that the original mass of the FG stars should be much larger than currently observed. This became known as the mass budget issue \citep[e.g.][]{Prantzos2006, Decressin2007, Carretta2010c, Renzini2015, Larsen2014, Bastian2018}. Data presented earlier in this review are actually a bit different from those used in previous discussions, so it might be worth to revisit this item. In the following, we will examine the mass budget issue separately for Type~I and Type~II  GCs, following the scheme adopted in previous sections.


\subsubsection{The case of Type~I GCs}

In a simple schematic view, SG stars are made of a mix of the ejecta from a fraction of the FG stars ($M_{ej}$) and of pristine material. We call dilution ($d$) the fraction of pristine material in the material used to form SG stars. The total mass available $M_A$ to form SG stars is $M_{ej}/(1-d)$. We call mass budget factor $b=M_{SG}/M_A$ the ratio between the observed mass in the SG stars and the available mass. If $b>1$, then the fraction of FG stars lost from the cluster since its origin should be larger than the average fraction of stars lost from the cluster, that is, FG stars should have been lost from the cluster more efficiently than SG stars. This likely occurred very early in the history of the clusters.

In order to quantify $b$\ we need to know the fraction of FG and SG (as represented e.g. by the fraction of FG stars ($f_{FG}$) in a cluster, the average dilution factor, make some assumption about the initial mass function, and estimate how much mass is locked into remnants of the FG stars, and is then not available to form the SG stars.

Hereinafter, we will consider that the IMF is represented by a power law between 0.25 and 60~M$_\odot$, with exponent $\alpha$\ between -1.7 and -2.3 (see \citealt{Beuther2007}; and \citealt{Hosek2019} and references therein); here $\alpha=-2.3$\ represents the Salpeter mass function and note that with this assumption, the lower extreme of integration provides a result very similar to the \citet{Kroupa2002} mass function. Once subtracted the mass locked in remnants, the mass given back to the interstellar medium by a FG star may be represented by:
\begin{equation}
M_{back}=0.894~M_*-0.434~{\rm M}_\odot
\end{equation}
for $0.9<M_*<9$~M$_\odot$, and:
\begin{equation}
M_{back}=0.9~M_*-0.5~{\rm M}_\odot
\end{equation}
for $M_*>9$~M$_\odot$ (see e.g. \citealt{Cummings2018}).

With these assumptions, the fraction of mass given back to the interstellar medium by intermediate mass stars ($3<M_*<9$~M$_\odot$) in units of the initial mass of first generation stars is 0.155 for $\alpha=-1.7$, 0.153 for $\alpha=-2.0$, and 0.120 if $\alpha=-2.3$.  The same values for massive stars ($9<M_*<60$~M$_\odot$) are 0.463 for $\alpha=-1.7$, 0.290 for $\alpha=-2.0$, and 0.142 if $\alpha=-2.3$. In this schematic view, the FG stars that pollute the ISM from which SG stars form cover a relatively large range of masses, that are likely characterized by different yields. We do not discuss this point in detail here because we are only interested in the mass budget issue. Here, we simply assume that once properly weighted, the material given back to the interstellar material has the appropriate composition to generate SG stars, after an appropriate dilution with pristine material; see however Section 5.3 for a case where the variation of the yield as a function of mass was considered in more detail.

Using the work by \citet{Milone2017} and \citet{Baumgardt2019}, we find that the fraction of FG stars depend on the initial mass of the cluster, being about $f_{FG}=0.6$ for clusters with an initial mass of about $2\times 10^5$~M$_\odot$, $f_{FG}=0.36$ for clusters with an initial mass of about $10^6$~M$_\odot$, and $f_{FG}=0.2$ for clusters with an initial mass of about $3\times 10^6$~M$_\odot$. 

We will then consider separately two different groups of SG stars, one characterized by a value of the dilution factor of about 0.5, and a second one characterized by a much smaller dilution factor, say about 0.05. The first group corresponds to the I population, and the second one to the E population (see Section 5, where we considered the E population defined by \citealt{Carretta2009a} in the context of the Li abundances). As we have seen in Sect. 5, the I population has a Li abundance not too different from that of the FG stars. This requires production of Li in the polluter. Since the only polluter known able to produce Li on a relatively short timescale is the intermediate mass AGB stars, in our estimates of the mass budget we will assume that the I population is produced by diluting the ejecta of these stars. On the other hand, there is no similar constraint for the E population or at least, for the fraction of the E stars that do not have Li. Besides, the high He abundances related to this population are more easily produced by supermassive stars or fast rotating massive stars (even if the latter cannot produce material depleted in Mg, one of the signatures of the E population). We will then assume that the E population is produced by these stars. We notice that the fraction of E stars $f_E$ over the total is null for small GCs ($2\times 10^5$~M$_\odot$), about 0.05 in clusters with $10^6$~M$_\odot$, and about 0.2 for clusters with an initial mass of about $3\times 10^6$~M$_\odot$. We also notice that the fraction of I stars in a cluster is then $f_I=1-f_{FG}-f_E$. Finally, in this approach we should have two separate mass budget factors, one for the I stars ($b_I$) and the other for the E stars ($b_E$).

\begin{table}[htb]
\caption{Mass budget factors for clusters of different mass}
\begin{tabular}{lcccc}
\hline
$M_{in}$ (M$_\odot$)& $\alpha$ & $2.0\times 10^5$ & $1.0\times 10^6$ & $3.0\times 10^6$ \\
\hline
$f_{FG}$ &               &       0.60       &       0.36       &       0.20       \\
$f_E$    &               &       0.00       &       0.05       &       0.20       \\
$M_{FG}$ (M$_\odot$) &               & $1.2\times 10^5$ & $3.6\times 10^5$ & $0.6\times 10^6$ \\   
$M_I$  (M$_\odot$)   &               & $0.8\times 10^5$ & $5.9\times 10^5$ & $1.8\times 10^6$ \\
$M_E$  (M$_\odot$)   &               &        0         & $0.5\times 10^5$ & $0.6\times 10^6$ \\
\\
$b_I$    & -1.7 &     2.0        &       5.0        &       9.1        \\
$b_I$    & -2.0 &     2.1        &       5.0        &       9.2        \\
$b_I$    & -2.3 &     2.6        &       6.4        &      11.8        \\
\\
$b_E$    & -1.7 &     0.0        &       0.3        &       2.1        \\
$b_E$    & -2.0 &     0.0        &       0.5        &       3.3        \\
$b_E$    & -2.3 &     0.0        &       0.9        &       6.5        \\
\\
$M_{start}$ (M$_\odot$) & -1.7 & $2.42\times 10^5$ & $1.79\times 10^6$ & $5.45\times 10^6$ \\
$M_{start}$ (M$_\odot$) & -2.0 & $2.46\times 10^5$ & $1.81\times 10^6$ & $5.54\times 10^6$ \\
$M_{start}$ (M$_\odot$) & -2.3 & $3.14\times 10^5$ & $2.31\times 10^6$ & $7.06\times 10^6$ \\
\\
$M_{gas,I}$ (M$_\odot$) &               & $4.24\times 10^4$ & $3.13\times 10^5$ & $9.54\times 10^5$ \\   
$M_{gas,E}$ (M$_\odot$) &               &        0.0        & $2.50\times 10^3$ & $3.00\times 10^4$ \\
\hline
\end{tabular}
\label{tab:massbudget}
\end{table}

We may then combine these different assumptions to derive the values for the mass budget factor. We will consider clusters in three bins of mass, because they have different fractions of FG, I and E stars. Results are given in Table~\ref{tab:massbudget}. Inspection of this table shows that the mass budget factor actually has quite small values (between 2 and 3) for low mass clusters and raises to large values (around 10) for massive clusters. Also, the mass budget factor is larger for the I population than for the E one. This implies that in the scenario considered here, where I stars are polluted by the ejecta of massive AGB stars and E stars by those of fast rotating massive stars (though a contribution by supermassive stars could not be excluded), we found that no more than half of the ejecta of this second group of stars are enough to provide the mass locked into E stars.  Hence, the constraint for the initial mass in the FG required to produce the SG stars ($M_{start}=b_I\times M_{FG}$) is determined by the I stars. These values are listed in the bottom part of Table~\ref{tab:massbudget}. It might appear a bit surprising at first look, but these values are within a factor of 1.5 to 3.5 the values of $M_{in}$, mainly depending on the cluster mass. This is due to the combination of the rather large value of dilution appropriate to I stars and of the small fraction of FG stars in massive clusters. Anyhow, these values indicate that in the scenario here considered, very early Type~I GCs should not need to be enormously more massive than at the end of the formation of the SG. We should emphasize that this result is obtained because in this scenario we separate the production of I stars (from the ejecta of AGB stars) from the production of E stars (from more massive stars). 

Another interesting point concerns the total mass of diluting material required. Expressed in units of the original mass $M_{start}$, this quantity is about 15\% for the formation of I stars and it ranges from 0 up to 0.5\% for the formation of the E stars (this second value is actually so low that does not provide any strong requirement). The first one is more challenging. Where does this diluting mass come from? As first possibility, it may be a simple consequence of stellar evolution; dilution would then be a natural process, likely governed by simple statistical laws, with no need of any specific hydrodynamical model.

We consider here the possibility that dilution is provided by mass loss from single stars and/or interacting binaries. The first case has been examined by \citet{Gratton2010c}, who concluded that the wind from young main sequence stars may provide at most some 1-1.5\% (that is, a tenth of what is needed) of the original mass as diluting material on a timescale of a few $10^7-10^8$~yr. More promising is the case of close binaries that have a Roche lobe overflow or develop a common envelope, as proposed by \citet{Vanbeveren2012}: they might have non-conservative evolution and lose a substantial fraction of their mass before material in the envelope is nuclearly processed. The material lost in this phase is possibly available as diluting material. As noticed by \citet{Vanbeveren2012} and considered previously by \citet{deMink2009}, actually part of this material is enriched in helium and nitrogen and possibly depleted in carbon and oxygen\footnote{There is evidence that Algol systems - that are interacting intermediate mass binaries - are depleted in C \citep{Tomkin1993, Sarna1996}.} and may be considered as polluting rather than diluting material. Population synthesis models based on detailed computations of binary evolution over a range of parameters (mass, separation, mass ratios) are required; \citet{Vanbeveren2012} provided a first exploration. One of the assumptions made by them is that 50\% of the stars in the mass range 3 to 9~M$_\odot$ are in binaries with period less than 10 years and with a mass fraction $q>0.1$. This is very similar to what found by \citet{Moe2017}, who reviewed the incidence of binaries among field stars and concluded that 50\% of those in the mass range 3--9~M$_\odot$ and virtually all the O-stars have a companion with a mass ratio larger than 0.1 and period less that 5000 days, so that they should evolve through a phase of mass transfer through Roche lobe overflow. The timescale of mass loss from O-type binaries (that is the original proposal of \citealt{deMink2009} for the polluting material) is very similar to that of core-collapse SNe: since there is very little trace of contamination by the ejecta of these SNe, it is very difficult that mass loss from massive binaries contribute here. But \citet{Vanbeveren2012} computations suggest that in the case of intermediate mass stars, binaries might indeed provide the required dilution, at least so far as the O-Na and Mg-Al anti-correlations are considered\footnote{Note that the mass budget values discussed above should be revised in this scenario because only a fraction of the massive AGB stars should contribute to nucleosynthesis. On the other hand, in this scenario the diluting material was already present in the GC since its birth.}. Part of this diluting material is slightly enriched in He, but this might perhaps not be a serious concern because the final effect on the He abundances is limited. The computations by \citet{Vanbeveren2012} should be repeated with updated stellar evolutionary code. Moreover, binary distributions more appropriate for the case of GCs must be considered, taking into account both ``ionization'' as well as hardening processes \citep{Heggie2003}. Since binaries properties are found to depend on the cluster mass (see Sect.~\ref{Sec:7}) this parameter might influence the dilution if this is the way it is generated. An important aspect not considered by \citet{Vanbeveren2012} is how much Lithium is preserved in the matter lost by these binaries. Actually Li is preserved only in the outer 0.03~M$_\odot$ of a 5~M$_\odot$ star (D'Antona and Cassisi, private communications), so that we may consider the ejecta of intermediate mass binaries to be almost Li-free. This possibly calls for a dilution parameter for Li different from that needed for O, though not in the way required to explain observations of e.g. NGC~6752 \citep{Pasquini2005}.  


We conclude that at present the origin of the diluting material for Type~I GCs is not yet well understood, and it is still possible that we need a substantial reservoir of pristine gas that is later accreted on the cluster \citep{Dercole2011, Dercole2016, Calura2019}, though these last scenarios might have difficulty to produce the right amount of gas at the right moment (see e.g. \citealt{Renzini2015}). However, this concern is not applicable for individual cases, such as e.g. NGC~2808, the archetypical GC considered in \citet{Dercole2016}, that may well have its own peculiar history.

\begin{figure}[htb]
\includegraphics[width=\textwidth]{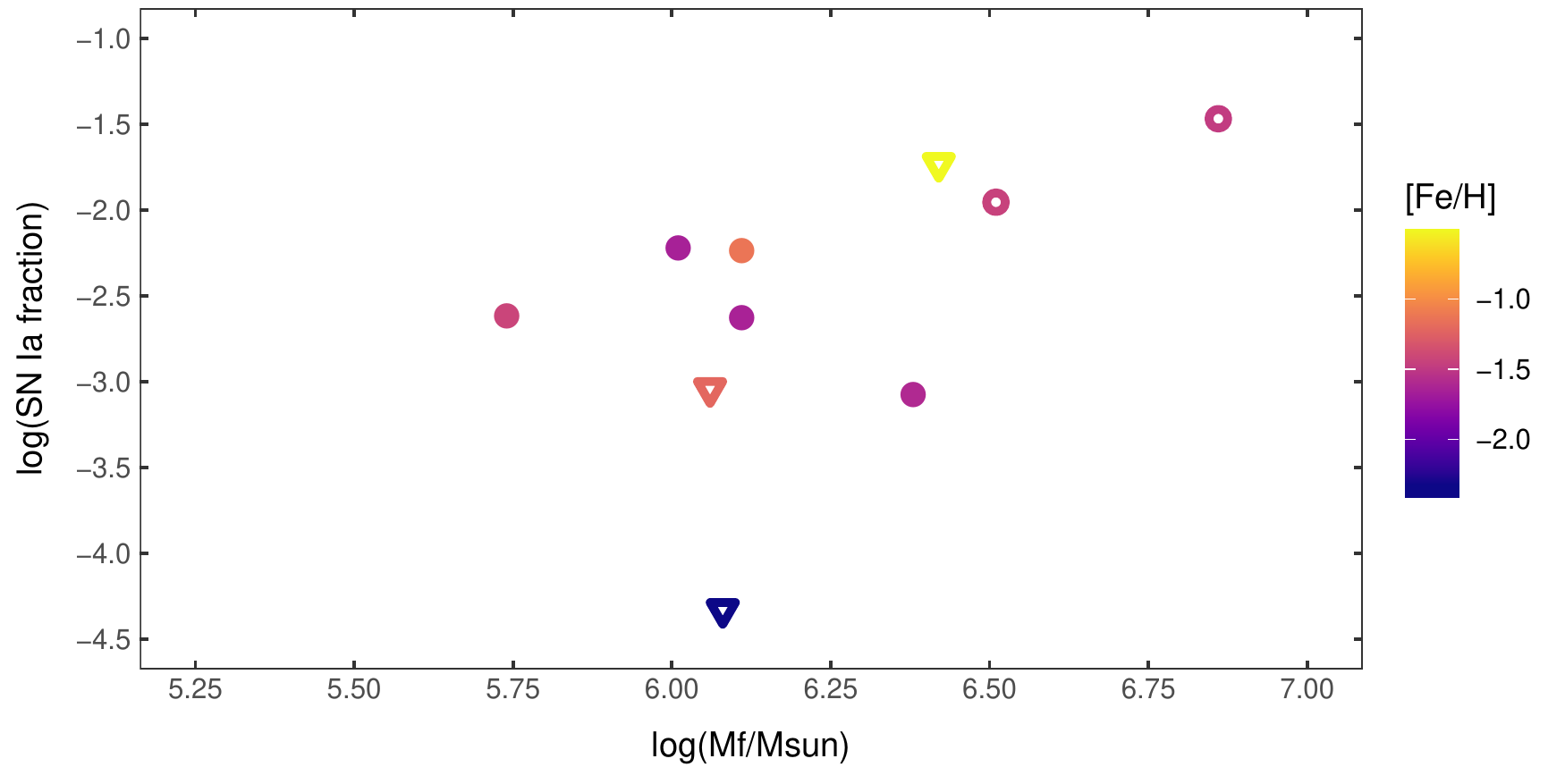}
\caption{Fraction of the ejecta of core collapse SNe retained by Type~II GCs. Open circles (for NGC~5139 and NGC~6715) indicate uncertain quantities, inverted triangles indicate upper limits. Colours code metallicity (see scale on the right of the plot)}
\label{fig:sne}       
\end{figure}

\subsubsection{The case of Type~II GCs}

\begin{table}[htb]
\begin{centering}
\caption{Number of Type~II SNe compatible with Fe abundance spread in Type~II clusters}
\begin{tabular}{cccccrrrrr}
\hline
NGC~&[Fe/H]&$\log{M_{\rm in}}$&f(typeII)&M(typeII)&d[Fe/H]&ref&dM(Fe)&nSN&f(SN)\\
\hline									
362  & -1.26 & 6.06 & 0.075 & 8.6E+04 &$<$0.050 & 1 & $<$0.7 & $<$10 &$<$4.2E-4\\
1261 & -1.27 & 5.86 & 0.038 & 2.8E+04 &         &   &	      &    &          \\
1851 & -1.18 & 6.11 & 0.300 & 3.9E+05 &   0.065 & 2 &    5.2 &    74 &   2.7E-3\\
5139 & -1.53 & 6.86 & 0.640 & 4.6E+06 &   0.300 & 3 &  172.3 &  2461 &   1.6E-2\\
5286 & -1.69 & 6.11 & 0.167 & 2.2E+05 &   0.140 & 4 &    2.1 &    30 &   1.1E-3\\
6388 & -0.55 & 6.42 & 0.299 & 7.9E+05 &$<$0.050 & 5 &$<$34.2 &$<$489 &$<$8.7E-3\\
6656 & -1.70 & 6.01 & 0.403 & 4.1E+05 &   0.150 & 6 &    4.3 &    61 &   2.8E-3\\
6715 & -1.49 & 6.51 & 0.460 & 1.5E+06 &   0.150 & 7 &   25.1 &   359 &   5.2E-3\\
6934 & -1.47 & 5.74 & 0.067 & 3.7E+04 &   0.200 & 8 &    0.9 &    13 &   1.1E-3\\
7078 & -2.37 & 6.08 & 0.050 & 6.0E+04 &$<$0.050 & 9 & $<$0.1 &  $<$1 &$<$2.2E-5\\
7089 & -1.65 & 6.38 & 0.043 & 1.0E+05 &   0.170 &10 &    1.4 &    20 &   3.9E-4\\
\hline    
\end{tabular}
\end{centering}
References: 1. \citet{Carretta2013a}; 2. \citet{Gratton2012b}; 3. \citet{Johnson2015}; 4. \citet{Marino2015}; 5. \citet{Carretta2018a}; 6. \citet{Marino2011}; 7. \citet{Carretta2010b}; 8. \citet{Marino2018}; 9. \citet{Carretta2009c}; 10. \citet{Yong2014a}. \\
Note: f(typeII) is the fraction of stars belonging to the population of stars occupying a region of high N but low He abundances in the chromosome diagram \citep{Milone2017}. The values of f(typeII) for NGC~1851 and NGC~6715 have been corrected for misprints in the original paper
\label{tab:sn}
\end{table}

In order to better understand the origin of the abundance anomalies observed in Type~II clusters, we may try to set some quantities. We will first focus on the variation of the abundance of Fe that likely implies the capability of these clusters to retain a (small) fraction of the ejecta of supernovae (SNe); the same argument can also be used to quantify the inability of Type~I clusters to do the same. We first notice that in those clusters where there is variation of Fe abundances, very similar results are also obtained for the $\alpha-$elements Si and Ca: this includes NGC~5286 \citep{Marino2015}, NGC~6273 \citep{Johnson2015}, NGC~6656=M~22 \citep{Marino2011}, NGC~7089 \citep{Yong2014a}, NGC~6715=M~54 \citep{Carretta2010b}, NGC~5139=$\omega$~Cen \citep{Johnson2010, Johnson2015, Marino2011b}. This is not what is expected if the observed Fe is produced in thermonuclear SNe, and rather argues for core collapse SNe.

Second, the fraction of the SNe ejecta that is retained by a Type~II cluster is observed to be a function of the cluster mass. To show this, we collected relevant data in Table~\ref{tab:sn}. They include the metallicity \citep{Harris1996}, the initial mass of the clusters \citep{Baumgardt2019}, the fraction of Type~II stars \citep{Milone2017}, and the offset in [Fe/H] between the normal (metal-poor) stars and those that are metal-enriched from a number of literature references. Combined with the solar Fe content \citep{Asplund2009}, these quantities allow to estimate how much additional Fe is needed to reproduce the observed abundance spread. If we now assume that each core collapse SN produces a given amount of Fe (we assumed 0.07~M$_\odot$: \citealt{Umeda2017}), we may estimate the number of SNe whose ejecta may reproduce the observed abundance spread. Finally, this number can be compared to the total number of SNe that are expected to explode in a young GC. This last value actually depends on the adopted mass function. If we consider a \citet{Kroupa2002} IMF, the rule of thumb is a SN every 100~M$_\odot$. We may then estimate the fraction of the SN ejecta that is incorporated in the Type~II stars. These values are indeed very small, the largest one being less than 2\% for $\omega$~Cen=NGC~5139 (very similar values are actually cited by \citealt{Renzini2015} and are given by \citealt{Marino2019}). There is a roughly linear correlation between this quantity and the mass of the cluster (see Fig.~\ref{fig:sne}). This agrees with the naive idea that the deepest is the potential well of a GC, the highest should be its ability to retain SN ejecta. The very small fraction of SN ejecta that can be kept within a GC may obviously be related to their very large kinetic energy: the ejecta of a single SN have in fact a kinetic energy comparable to the whole potential energy of the residual gas within a GC. Since the production of Fe in SNe is primary, stringent constraints are obtained for the most metal-poor clusters, such as NGC~7078=M~15, for which a single core collapse SN should produce a detectable star-to-star variation in the Fe abundances.

We may repeat similar arguments for the production of CNO and $s-$process elements. Data for CNO are scarce, because derivation of the total abundance from spectroscopy is difficult. This is exemplified by the case of NGC~1851, for which \citet{Yong2015} obtained a large difference of 0.6~dex between the two populations using spectroscopy, while smaller offsets of $\sim 0.15$~dex have been considered to justify the distributions of colour and magnitudes of subgiant and horizontal branch stars \citep{Gratton2012b}. There is better agreement at a value of $\sim 0.3$~dex between the various determinations for NGC~6656=M~22 \citep{Marino2011b, Alves-Brito2012, Gratton2014}. On the whole, this data might perhaps be compatible with the production by core collapse SNe, being not too far from the spread seen for Fe. On the other hand, the variation in the abundances of the second peak $s-$process elements (Ba, La, etc.) between FG and Type~II one is well established at a value in the range 0.4--0.7 dex in most Type~II GCs \citep{Marino2011b, D'Orazi2011, Carretta2013b, Johnson2015, Marino2015, Yong2016, Marino2018}. An even larger spread is observed in the most metal-rich population of $\omega$~Cen=NGC~5139 \citep{Smith2000, D'Orazi2011}, where a quite large enhancement of the third-peak element Pb is also observed \citep{D'Orazi2011}. The production of the heavy $s-$process elements calls for a significant contribution by the thermal pulse phase in moderate mass ($\gtrsim$ 2-4~M$_\odot$) AGB stars (see Section 2.4). However, in order to reproduce the observed pattern, we should consider in addition to these stars, also the contribution to O by the core collapse SNe and moreover a substantial dilution by unprocessed material, that reduces the abundance offset between FG and Type~II stars by an order of magnitude. This large dilution implies a conspicuous reservoir of gas with pristine composition available. The mass of this reservoir is of the order of the initial mass for those clusters where the fraction of Type~II stars is $<0.1$, that is roughly half of the Type~II clusters, but it is as much as 5 times larger in clusters such as NGC~1851 and NGC~6656=M~22, that have a large fraction of Type~II stars. We notice that this large dilution implies that the first Type~II stars should have a composition similar to that of the FG stars in the cluster. On the other hand, while such a large reservoir of gas may well be present in GC forming in dwarf galaxies, their intervention just at the right moment is a clear difficulty for the scenarios of formation of Type~II GCs.

If is interesting to note that none of the clusters studied so far shows Fe spread without $s$-process spread. This might be surprising given that the $s$-process and Fe variation come from independent nucleosynthetic sites which have different timescales ($\sim 200-500$~Myr vs a few Myr respectively) and hence if re-accretion would occur on a timescale shorter than a few hundreds Myr, it would produce Fe abundance variations with not a significant variation in the $s-$process elements. The number of clusters with known Fe spread and well studied $s$-process abundances is rather small, and this could then be an artifact of small number statistics; however, it is intriguing to think that this could be hinting at some mechanism that led to the formation of the Fe enriched populations only with a considerable time lag with respect to the original population. This looks easier to explain within the context of a dwarf satellite (see e.g. \citealt{D'Antona2016,Dercole2016,Bekki2016})

\subsubsection{Conclusions about mass budget}

In order to explain the chemical composition of GCs, we should assume that the original mass involved in the star formation episodes that finally led to the formation of present-day GCs was significantly larger than their final mass and that the ejecta from the polluters were diluted by a large amount of gas with primordial composition, especially in Type~II clusters. Since the presence of diluting material seems a general feature, scenarios for the formation of GCs should explain its origin in a simple way. As considered in Section 8.1, the mass loss from chemically unevolved binary stars might possibly explain the less demanding case of Type~I clusters. On the other hand, Type~II GCs might have formed far from the center of the MW (see Fig.~\ref{fig:rapo_massin} and discussion in Sect. 3.4) within satellites that had a chemical evolution independent of the main stream of the Galaxy for quite a long time, possibly helping in providing the required reservoir of diluting gas to be used for further generations. In a more limited way, something similar might have occurred also for Type~I GCs too, or at least for a fraction of them (see e.g. \citealt{Bekki2007}). In this framework, we might think of a ``normal'' mass-dependent evolutionary sequence for isolated structures (that observed in Type~II GCs) that was interrupted quite early by interactions with the MW in Type~I GCs. However, the reproduction of the right time scales at which the dilution and star formation episodes occurs needs much more elaboration, so that we are still far from a satisfactory model. 

We finally notice here that Type~II clusters must possibly be thought as an extension to low masses of the generic existence of a central massive object that contains a mean fraction $\sim 0.2$\% of the total mass of a galaxy \citep{Ferrarese2006}. This underlines that GCs might possibly be a heterogeneous class of transition objects between normal stellar clusters and the very compact objects at the center of galaxies.

\subsection{GC stars in the field}

Globular clusters were disrupted in the Galactic halo and even more in the bulge: \citet{Baumgardt2018} estimated that at least 80\% of the original population of GCs is now dissolved, and that the remaining GCs have lost a significant fraction of their stars. There is observable evidence of the loss of these stars. \citet{Grillmair2006} discovered a long stream in the halo, named GD-1 stream, which is found to be narrow, cold and metal-poor \citep{Huang2018}. These features strongly point towards an origin from a stripped or disrupted GC \citep{Koposov2010}, even if the progenitor is no longer detectable. The same tale is told by other very narrow stellar streams \citep{Grillmair2009}. Very small dispersions in the estimated metallicity (e.g. $\sigma_{\rm [Fe/H]} < 0.1$ dex) are usually taken as evidence that the stream resulted from the disruption of a globular cluster, rather than a dwarf galaxy. Although among the iron complex GCs currently known (e.g., $\omega$~Cen=NGC~5139, M~54=NGC~6715, M~22=NGC~6656) the spread in [Fe/H] can reach 0.3 dex, the narrow nature of several identified streams strongly indicates that probably they originated from lower mass, mono-metallic GCs (see also \citealt{Veljanoski2018}).

Thus, GCs were disrupted. But GCs also are currently in disruption, as clearly shown by the famous tidal tails associated to the globular cluster Pal~5 \citep{Odenkirchen2001, Odenkirchen2003}. GCs are mostly found in the halo, so they obviously contribute to the Galactic halo, but the actual contribution is not limited to the current $\sim 2\%$ of the total mass enclosed in GCs. Depending on the formation environment, the dynamical evolution of GCs is subject to a number of external processes, apart from the internal mechanism of evolution. Critical to cluster disruption are shocks due to the interaction with any irregularity in the gravitational potential (see the introduction in \citealt{Webb2018} and references therein). As a consequence, how actually GCs contributed to the formation of the halo by releasing stars that become unbound over almost a full Hubble time is a difficult question, because we have very limited knowledge of the environment where GCs started their evolution about 0.5-1 Gyr after the Big Bang. The consequence is that we are limited by a number of assumptions and we have to rely on indirect probes to evaluate the contribution of GCs to the halo formation. 

It is likely that at the early phases of Galaxy formation the impact of collision with giant molecolar clouds was more relevant than today, and this is a chief mechanism to generate GC shredding and disruption \citep{Webb2018}. However, an estimate of the GMC distribution in the proto-MW is an educated guess, at best. The same gravitational potential of the early Galaxy is not known and its temporal evolution could have had important impact in the tidal stress exerted on proto-GCs \citep{Li2019}. Many related questions remains unanswered, such as: was the thickening of an early disk simultaneous to the formation of halo and bulge? Was the influence of the mass distribution in central Galactic regions enough to affect orbits of the just formed GCs? Were the major merger(s) occurring about 10 Gyr ago accompanied by the formation and/or disruption of GCs or did the falling satellites of the Galaxy simply release their population of associated GCs into the halo of the main Galaxy? 

An attempt to include the GC formation in a cosmological context has been made in the past using sub-grid resolution post processing of cosmological simulations \citep[e.g. ][]{Beasley2002,Prieto2008,Griffen2010}. In particular, \citet{ReinaCampos2018}, adopting a simplified recipe for the formation of GCs from molecular cores, suggests that a relatively small mass-loss occurs over the subsequent evolution. On the other hand, their simplified treatment of dynamical evolution as well as the unknown initial structure of proto-GCs make this conclusion weak.

Despite these issues being largely without a firm answer, we expect a contribution from GCs simply because they lose mass. Mass loss is expected in particular at early phases \citep{Lynden-Bell1967, Baumgardt2008, Vesperini2010}, but GCs are in general dynamically evolved systems, so that mass loss is predicted to occur over their whole lifetime (e.g. \citealt{McLaughlin2008, Webb2018, Baumgardt2019}), so it is possible that many clusters dissolved, as corroborated by the lack of GCs with large ratios of halo-mass to Jacobi radius in the near dissolution region (see \citealt{Baumgardt2010}).

Early and even recent studies on the contribution of GCs to the halo focussed on dwarf galaxies or the entire systems of GCs captured from dwarfs  (e.g. \citealt{Lin1992, FusiPecci1995, Forbes2010}), rather than on the contribution to halo stars from  GCs. The main reason is that it is relatively simple to distinguish between the chemical pattern of GCs and dwarf galaxies when the tagging is made using abundances of $\alpha-$elements, as usually occurred in these works. However, on the high$-\alpha$ plateau, GC stars are often superimposed to other Galactic components, such as stars formed in-situ, accreted or even kicked out (see \citealt{Sheffield2012}). $\alpha-$elements may be able to resolve GC stars from dSph stars, to some extent, but not so from field halo MW stars. Fortunately, one can use another diagnostic provided by the chemical pattern of SG stars in GCs, since this signature is unambigously unique among old stellar systems.

A first attempt was made in \citet{Carretta2010c} by comparing Na abundances in field stars with SG stars in GCs. They found a small fraction of stars with SG signature, 6 Na-rich stars out of 144 examined. After excluding 4 objects (likely binary stars) a fraction of 1.4\% resulted. This number was doubled to 2.8\% by considering the typical ratio of FG to SG stars in present-day GCs.

More systematic studies later found similar fractions of SG stars by looking for large N excesses in metal-poor halo stars with low resolution spectra acquired in the Sloan survey. \citet{Martell2010} selected from the SEGUE survey 49 relatively CN-strong, CH-weak stars out of an initial sample of 1958 giant (likely halo) stars, corresponding to a fraction 2.5\%. Similar results were obtained in \citet{Martell2011} by selecting from SEGUE-2 spectra of more distant RGB stars and retrieving a fraction of 3\% of stars with SG composition, in good agreement with more serendipitous discoveries based on the O abundances, like in the study by \citet{Ramirez2012}. Out of a sample of 67 halo stars, the latter authors found 2 O-poor stars for which \citet{Nissen2010} obtained very high Na abundances. This being exactly the pattern along the Na-O anti-correlation in GCs, \citet{Ramirez2012} estimated a fraction of $3 \pm 2\%$ unless one of the two O-poor stars is revealed to be polluted by a mass transfer events, as hinted by its large abundance of Barium and Yttrium.

Also higher temperature ranges of the burning regions, sampled by the Mg-Al anti-correlation, have been used to trace the possible origin of some field stars back to GCs. Using abundances from the Gaia-ESO survey, \citet{Lind2015} found one halo star with high Aluminum and large Mg depletion out of few hundreds of examined field stars. They considered this finding not inconsistent with the estimates by \citet{Martell2010, Martell2011}, especially because the Mg-Al anti-correlation is not found in all GCs, in contrast to the C-N and Na-O anti-correlations, but only in the most massive and/or metal-poor ones \citep{Carretta2009b, Meszaros2015}. This occurrence led \citet{Fernandez-Trincado2017} to conclude that the finding from APOGEE survey of Mg-poor, Al-rich stars mostly in the metal-rich regime was inconsistent with an origin from GCs. Note that by itself the deficiency in Mg can be also viewed as the distinctive signature of stars shredded from accreted dwarf galaxy. However, no large enhancement in Al is expected in dSph's stars (e.g. \citealt{Shetrone2001}).

In view of these uncertainties, currently one of the most used approaches for tracking SG stars in the halo remains the selection according to N enhancement. \citet{Martell2016} used APOGEE spectra to find 5 stars with N and Al enhancements out of 253 halo giants, after discarding stars with high C abundances and evidence of binarity. The resulting fraction of 2\% of stars with SG pattern agrees with previous results. Two important caveats have to be considered. First, the normal evolution of low-mass stars naturally distribute them along a C-N anti-correlation as surface abundances are changed by the dredge-up and the extra-mixing episode on the RGB \citep{Gratton2000,Martell2008}. Care must be then exercised to pick up stars that are real outliers at any evolutionary phase. Second, as pointed out by  \citet{Smith2015}, the comparison of the C, N, O, and Na pattern observed in GCs reveal an apparent decoupling between the C-N and Na-O anti-correlation, which in turn is related to processing at different temperatures and may occur in different stars altogether. 

To these, we add another caveat based on the kinematic signatures of these stars that can now be provided using Gaia astrometric information. We note that almost all the candidate SG stars in the field found in previously mentioned and other recent studies \citep{Fernandez-Trincado2016, Tang2019} show orbits characterised by high eccentricities. Peculiar chemical abundances, especially concerning $\alpha-$elements such as Mg, and a dominance of high eccentricity orbits have been recently found as the distinctive chemo-dynamical signature of the massive accretion event that about 8-10 Gyr ago brought the so called Gaia-Enceladus dwarf \citep{Helmi2018} into the MW \citep{Haywood2018, Simion2019, Iorio2019, Mackereth2019}. While in a couple of cases preliminary orbit comparison allows to claim that candidate stars lost by GCs were compatible with an origin from the massive GC $\omega$~Cen=NGC~5139 \citep{Lind2015, Fernandez-Trincado2016} we caution that a full chemical characterization with all the species involved in multiple population in GCs would provide a more clear cut clue to the origin of these stars, since GCs and remnants of past accretion events cleanly separate in the abundance space.

Taking these caveats into account, various indicators concur to assess the estimates of the fraction of SG in the field in a range from 2 to 4--5\%. Considering the typical ratios of FG to SG stars observed in present-day GCs, the total contribution of GCs to the mass budget of the Galactic halo critically depends on the assumptions made for the mass loss at early phases. Strong mass loss, namely if GCs were initially $>10$\ times more massive (the value proposed for massive GCs in Sect.~\ref{Sec:8}) and were able to lose about 90\% of their mass at early times, would inflate the observed estimates to fractions from 13-17\% up to 40-50\% of stars in the halo originally formed in GCs \citep{Vesperini2010, Schaerer2011, Martell2011, Martell2016, Gratton2012}. Without this strong mass loss, a limited fraction of about 4-5\% of the halo in mass would be originated in GCs \citep{Martell2011}. A simple formalism was recently presented by \citet{Koch2019} to estimate the fraction of the Galactic halo in stars originally born in GCs, as inferred by the observed fraction of 2.6\% of field stars with CN-strong (SG) signature from SDSS-IV DR14. Their derived value of 11\% stems from a number of different assumptions clearly discussed in their work, with the early mass loss rate from GCs and the number of completely dissolved GCs being the major still unknown relevant factors, poorly constrained by observations.






\section{Conclusions and open issues}
\label{Sec:9}

\begin{figure}[htb]
    \centering
    \includegraphics[width=1.0\textwidth]{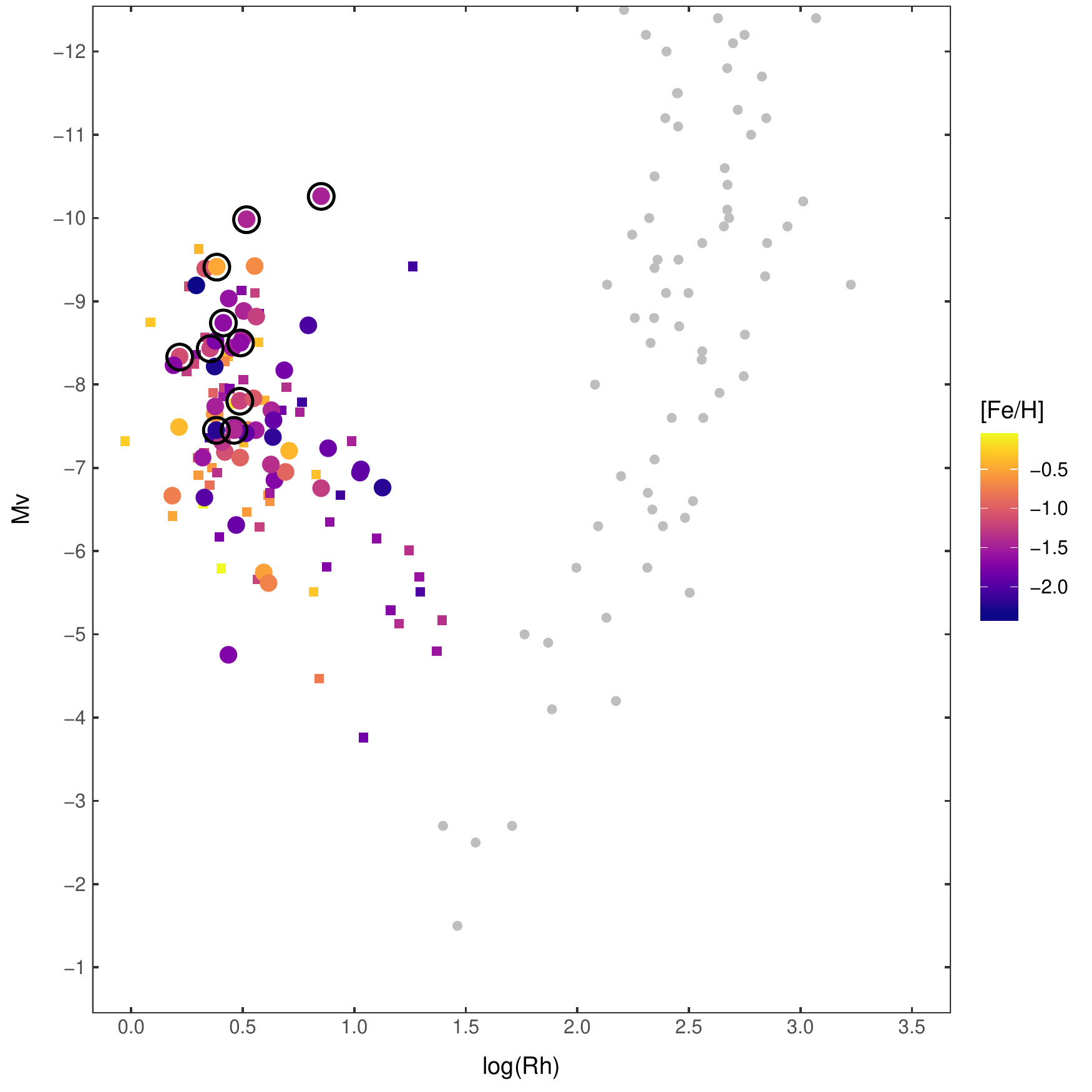}
    \caption{Relation between M$_V$ and half-mass radius for dwarf galaxies (grey circles) and GCs. Filled circles indicate Type~I and Type~II (circled symbols) clusters, while squares indicate clusters for which the type has not been determined. Colours code metallicity (see scale on the right of the plot). Data are taken by \citet{McConnachie2012} for dwarfs, \citet{Harris1996} and \citet{Baumgardt2018} for GCs.}
    \label{fig:simil_leamans}
\end{figure}

We reviewed current knowledge about the chemical composition of GCs. They show a quite high level of complexity that may be understood by considering that they are transition objects between single stellar population clusters (the open clusters) and fully developed cases of chemical evolution (galaxies). From a structural point of view, GCs clearly differentiate from most dwarf galaxies (see Fig.~\ref{fig:simil_leamans}), being much more compact and lacking dark matter. However, as shown e.g. by \citet{Dabringhausen2008}, they merge into the sequence of compact galaxies.

The complexity of GC chemistry has so far defied attempts to explain all observed features in a single scenario \citep[e.g.][]{Renzini2015, Bastian2018}. This might be perhaps attributed to the coexistence of different polluters and different diluters. While there is a vast literature concerning possible polluters (see e.g. the review by \citealt{Bastian2018}), less attention has been paid to the diluters. We suggest that there might actually be two different diluting mechanisms active in GCs. A first one may be considered as ``intrinsic'' to GCs, that is a mechanism that is present in all GCs with more or less similar trend, likely modulated by the mass or initial density. This mechanism might be related to the usual anti-correlations found in GCs. While mass-loss by single stars might play some role \citep{Gratton2010c}, the most likely candidate for this dilution are interacting binaries \citep{Vanbeveren2012}. This mechanism acts on the same timescale of the polluter and the mass is lost through winds at similar speed, and it does not then need any special history: it should repeat self-similarly from cluster-to-cluster and be essentially universal to GCs. The second mechanism is ``extrinsic'' and it is only relevant for massive GCs, being in particular related to Type~II GCs. This extrinsic mechanism shows a wide range of variation from cluster-to-cluster both on the amount of involved mass and in the timescale where it occurs, that is however typically longer than that of the first mechanism. This second dilution mechanism is responsible for the variety of the characteristics of massive clusters, as can be obtained from the chromosome map \citep{Milone2017, Marino2018}. On the whole, the best candidate for this second mechanism is some variety of the mass re-accretion considered by \citet{Dercole2011, Dercole2016}, and most likely related to the formation within a satellite \citep{Bekki2003, Bekki2007, Bekki2016, Dercole2016}. We suggest that massive clusters may represent a continuum from cases where there was no re-accretion at all (e.g. 47~Tuc=NGC~104), others where the re-accretion (if any) occurred on a quite short time-scale (e.g. NGC~2808 or NGC~2419), to others where the time-scale was longer but with very different amounts of mass re-accreted (see e.g. the comparison between NGC~362 and ~NGC~1851 or NGC~6656=M~22), to finally complex cases with several re-accretion episodes (such as $\omega$~Cen=NGC~5139). This last clearly recall the case of nuclear star clusters \citep{Bekki2003}. Note that when we compare NGC~2808 and NGC~2419 with the other type I GCs, a number of peculiarities makes them unique: e.g. variations of the abundances of K and Sc and a large spread in He abundance. These facts suggest the contribution by a class of polluters not relevant for other GCs of this class. A full analysis of the chemical composition of the K- and Sc-rich stars might reveal if they coincide with the super He-rich stars and what is their Li-content. This last point is crucial to establish the nature of the polluter. Though there is no evidence of variation of Fe and total CNO content, the chromosome diagrams of NGC~2808 and NGC~2419 are complex and different from those of typical type~I clusters, indicating the presence of several different populations, perhaps each one associated to different polluters and diluters. Their mass and position in the Galaxy is more similar to type II rather than type I clusters. These peculiarities suggest that they may be considered as a special class of GCs and that they may be discussed in the context of Type~II GCs. 

Summarizing, the interplay between the two different diluting mechanisms, coupled with the possibility of different polluters (supermassive stars: \citealt{Gieles2018}; fast rotating massive stars: \citealt{Decressin2007}; massive AGB stars: \citealt{Ventura2001}) might help to understand the variety of chemical evolution observed in GCs.

Besides complexity, there are a number of problematic points still open and we name here the most relevant in our opinion. First, while nucleosynthesis in massive AGB stars has some of the properties useful to describe many but likely not all) of the chemical peculiarities observed in GCs, theoretical models are still not robust enough to unequivocally predict the relevant yields. These crucially depend on details that are poorly understood, including convection, mass loss, and the same nuclear cross sections. A suitable combination of these parameters and of their dependence on stellar mass is indeed able to produce the right nucleosynthesis (see e.g. \citealt{Ventura2001, D'Antona2016}); however, it is not at all obvious that this combination is really the correct one to be considered (see e.g. \citealt{karakas2014}). While some recent progress has been made in the positive direction (see e.g. the case of the mass loss rate: \citealt{Pastorelli2019}), it is still too early for a definite conclusion. Second, while there is clear indication that Lithium should be produced in the polluters at least for the case of the widespread intermediate population (and this argues for the important role played by massive AGB stars), this production should mimic the original abundance, strongly constraining models. Again, this production requires that particular recipes are adopted for the evolution of these stars. Third, the timescale involved in the re-accretion events needed to explain Type~II clusters is quite constrained by the fact that variation in the abundances of $s-$process elements seems very common, while this is not the case e.g. for the variation of Fe, and there is no known case where there is variation of Fe abundances but not of the $s-$process elements. While this might be simply the consequence of a limited statistics, there may be something more basic behind this fact.

Finally, we wish to stress the importance of the implications related to the binary frequencies in the different populations of globular clusters. GCs are usually thought as very high density environments where a number of exotic objects may form. The difference in binary frequency between first and second generation stars indicate that the very dense environment is actually the one at the origin of the second generation; there the density is two order of magnitudes larger than for the first generation. In addition, in both cases there is a strong (roughly linear) dependence on the initial mass. This suggests that the extremely high densities that may be at the origin of the exotic objects are likely related to the second generations in very massive clusters or even in their big brothers, the nuclear star clusters and the compact galaxies. This is the most favourable ambient for the runaway growth of massive or even super-massive compact objects \citep{Ferrarese2006, Gieles2018}. This underlines the general importance of a better understanding of the formation of GCs.

\begin{acknowledgements}
This work has made use of BaSTI web tools and of TOPCAT \citep{Taylor2017}. We thank Alessio Mucciarelli for having provided us with unpublished results, and Leo Girardi and Emanuele Dalessandro. We also wish to thank Nate Bastian, Simon Campbell, Santi Cassisi,  Franca D'Antona, Enrico Vesperini, and an anonymous referee for having read a draft version of the review and having provided very useful comments. Finally, we wish to thank Frank Schulz that made the many editing steps required to have this review publishable.
\end{acknowledgements}

\bibliographystyle{spbasic}      
\bibliography{Gratton_review.bib}   

%
%



\section{Appendix 1: Summary of data for Milky Way GCs}

This Appendix collects data for galactic GCs used in this review. We give information on the references used for the columns in Table~\ref{tab:A1}, \ref{tab:A1b}, and \ref{tab:A2} in the following.

\noindent For Table~\ref{tab:A1}: 
\begin{itemize}
\item Col 1: Designation
\item Col 2-3: $R_{\rm per}$ and $R_{\rm apo}$\ in kpc from \citet{Baumgardt2019}
\item Col 4-5-6-7-8-9: dMY, Ymed, Ymax, Ymax-Ymed, delta(B-V), delta(V-I) from \citet{Gratton2010b}
\end{itemize}

\noindent For Table~\ref{tab:A1b}: 
\begin{itemize}
\item Col 1: Designation
\item Col 2-3-4-5: dY2g1G, err, dYmax, err from \citet{Milone2018} and \citet{Zennaro2019}
\item Col 6: [Fe/H] from \citet{Carretta2019}
\item Col 7-8: d[Fe/H], err from \citet{Bragaglia2010a} (NGC~6402 from \citealt{Johnson2019})
\item Col 9-10: $\log{M_{\rm fin}}$, $\log{M_{\rm in}}$ from \citet{Baumgardt2019}
\end{itemize}

\noindent For Table~\ref{tab:A2}:
\begin{itemize}
\item Col 1: Designation
\item Col 2-3: IQR(Na/O),Source: (1) \citet{Gratton2006, Gratton2007, Carretta2007, Carretta2009a, Carretta2009b, Carretta2010, Carretta2011, Carretta2013a, Carretta2014, Carretta2015, Carretta2017, Bragaglia2015, Bragaglia2017, Carretta2018} (2) \citet{Villanova2016}; (3) \citet{SanRoman2015}; (4) \citet{Boberg2015, Boberg2016}; (5) \citet{Marino2011}; (6) \citet{Kraft1992}; (7) \citet{Marino2015}; (8) \citet{Smith2002}; (9) \citet{Mucciarelli2013}; (10) \citet{Pancino2017}; (11) \citet{Koch2014}; (12) \citet{Johnson2016};  (13) \citet{Johnson2017a}; (14) \citet{Johnson2017b}; (15) \citet{Johnson2015}; (16) \citet{Yong2014a}; (17) \citet{Feltzing2009}; (18) \citet{Mucciarelli2016, Massari2017}; (19) \citet{Munoz2017}; (20) \citet{Villanova2017}; (21) \citet{Marino2009}; (22) \citet{O'Malley2017}; (23) \citet{Kacharov2013}; (24) \citet{Kraft1998}; (25) \citet{Yong2014b}; (26) \citet{Caliskan2012}; (27) \citet{Cohen2004}; (28) \citet{Villanova2013}; (29) \citet{Munoz2018}; (30) \citet{Mucciarelli2018};  (31) \citet{Johnson2018}; (32) \citet{Sbordone2007}; (33) \citet{Johnson2019}
\item Col 4: IQR(Al/Mg) from \citet{Carretta2010} and others determination from this group; only 1 digit: \citet{Meszaros2015}
\item Col 5-6-7-8-9: dRGB, err, f(FG), err, GC type from \citet{Milone2017} and \citet{Zennaro2019}
\item Col 10: spectroscopic d[Al/Mg] from \citet{Milone2018} 
\end{itemize}



\begin{table}
\caption{Main parameters for selected GCs}
\label{tab:A1}
\begin{tabular}{lrrrrrrrr}
\hline 
  \multicolumn{1}{c}{Name} &
  \multicolumn{1}{c}{R$_{per}$} &
  \multicolumn{1}{c}{R$_{apo}$} &
  \multicolumn{1}{c}{dMY} &
  \multicolumn{1}{c}{Y$^{med}$} &
  \multicolumn{1}{c}{Y$^{\max}$} &
  \multicolumn{1}{r}{Y$^{\max}-$}&   
  \multicolumn{1}{c}{$\Delta$} &
  \multicolumn{1}{c}{$\Delta$} \\
  \multicolumn{1}{c}{} &
  \multicolumn{1}{c}{(kpc)} &
  \multicolumn{1}{c}{(kpc)}&
  \multicolumn{1}{c}{} &
  \multicolumn{1}{c}{} &
  \multicolumn{1}{c}{} &
  \multicolumn{1}{r}{Y$^{med}$} &
  \multicolumn{1}{c}{$(B-V)$} &
  \multicolumn{1}{c}{$(V-I)$} \\
\hline
  NGC~104  &  5.46 &  7.44 & 0.0   & 0.234 & 0.234 & 0.0   & 0.0   & 0.0   \\
  NGC~288  &  3.33 & 13.01 & 0.016 & 0.280 & 0.292 & 0.012 & 0.005 & 0.007 \\
  NGC~362  &  1.05 & 12.48 & 0.059 & 0.243 & 0.289 & 0.046 & 0.018 & 0.025 \\
  NGC~1261 &  1.41 & 19.93 & 0.068 & 0.244 & 0.297 & 0.053 & 0.021 & 0.029 \\
  Eridanus  & 33.56 &134.93 &       &       &       &       &       &       \\
  Pal  2    &  2.49 & 39.41 &       &       &       &       &       &       \\
  NGC~1851 &  0.83 & 19.13 & 0.063 & 0.247 & 0.295 & 0.048 & 0.019 & 0.027 \\
  NGC~1904 &  0.82 & 19.49 & 0.055 & 0.274 & 0.317 & 0.043 & 0.018 & 0.022 \\
  NGC~2298 &  1.86 & 17.74 & 0.0   & 0.249 & 0.249 & 0.000 & 0.0   & 0.0   \\
  NGC~2419 & 16.52 & 90.96 &       &       &       &       &       &       \\
  NGC~2808 &  0.97 & 14.72 & 0.08  & 0.273 & 0.334 & 0.061 & 0.024 & 0.034 \\
  Pal  3    & 65.31 &124.42 &       &       &       &       &       &       \\
  NGC~3201 &  8.15 & 23.54 & 0.032 & 0.253 & 0.278 & 0.025 & 0.010 & 0.013 \\
  Pal  4    & 23.66 &111.35 &       &       &       &       &       &       \\
  NGC~4147 &  1.92 & 24.57 & 0.014 & 0.247 & 0.258 & 0.011 & 0.005 & 0.006 \\
  NGC~4372 &  2.94 &  7.20 & 0.026 & 0.254 & 0.275 & 0.021 & 0.009 & 0.010 \\
  Rup  106  &  4.71 & 35.18 &       &       &       &       &       &       \\
  NGC~4590 &  8.86 & 29.20 & 0.036 & 0.228 & 0.257 & 0.029 & 0.012 & 0.013 \\
  NGC~4833 &  0.79 &  7.39 & 0.078 & 0.246 & 0.308 & 0.062 & 0.026 & 0.030 \\
  NGC~5024 &  9.09 & 21.96 & 0.0   & 0.241 & 0.241 & 0.0   & 0.0   & 0.0   \\
  NGC~5053 & 10.28 & 17.69 & 0.017 & 0.234 & 0.248 & 0.014 & 0.006 & 0.006 \\
  NGC~5139 &  1.35 &  7.00 &       &       &       &       &       &       \\
  NGC~5272 &  5.44 & 15.14 & 0.030 & 0.249 & 0.272 & 0.023 & 0.009 & 0.012 \\
  NGC~5286 &  1.16 & 13.27 &       &       &       &       &       &       \\
  NGC~5466 &  7.95 & 65.53 & 0.0   & 0.223 & 0.223 & 0.000 & 0.0   & 0.0  \\
  NGC~5634 &  4.27 & 23.91 &       &       &       &       &       &       \\
  NGC~5694 &  3.98 & 66.04 & 0.028 & 0.246 & 0.268 & 0.022 & 0.009 & 0.010 \\
  IC  4499  &  6.38 & 27.67 &       &       &       &       &       &       \\
  NGC~5824 & 15.17 & 38.26 & 0.066 & 0.246 & 0.299 & 0.053 & 0.022 & 0.025 \\
  Pal  5    & 17.40 & 24.30 &       &       &       &       &       &       \\
  NGC~5897 &  2.86 &  9.31 & 0.0   & 0.253 & 0.253 & 0.000 & 0.0   & 0.0   \\
  NGC~5904 &  2.90 & 24.20 & 0.040 & 0.262 & 0.293 & 0.031 & 0.012 & 0.017 \\
  NGC~5927 &  3.99 &  5.42 & 0.011 & 0.250 & 0.256 & 0.006 & 0.002 & 0.004 \\
  NGC~5946 &  0.83 &  5.82 & 0.054 & 0.258 & 0.300 & 0.042 & 0.017 & 0.023 \\
  NGC~5986 &  0.67 &  5.05 & 0.085 & 0.261 & 0.328 & 0.067 & 0.027 & 0.034 \\
  Lynga  7  &  1.91 &  4.56 &       &       &       &       &       &       \\
  Pal  14   &  3.90 & 94.81 &       &       &       &       &       &       \\
  NGC~6093 &  0.35 &  3.52 & 0.087 & 0.252 & 0.321 & 0.069 & 0.028 & 0.034 \\
  NGC~6101 & 11.37 & 46.89 & 0.0   & 0.250 & 0.250 & 0.000 & 0.0   & 0.0   \\
  NGC~6121 &  0.55 &  6.16 & 0.030 & 0.238 & 0.261 & 0.023 & 0.009 & 0.013 \\
  NGC~6139 &  1.34 &  3.52 &       &       &       &       &       &       \\
  NGC~6144 &  2.27 &  3.36 &       &       &       &       &       &       \\
\hline
\end{tabular}
\addtocounter{table}{-1}
\end{table}

\begin{table}
\caption{Cont...}
\begin{tabular}{lrrrrrrrr}
\hline 
  \multicolumn{1}{c}{Name} &
  \multicolumn{1}{c}{R$_{per}$} &
  \multicolumn{1}{c}{R$_{apo}$} &
  \multicolumn{1}{c}{dMY} &
  \multicolumn{1}{c}{Y$^{med}$} &
  \multicolumn{1}{c}{Y$^{\max}$} &
  \multicolumn{1}{r}{Y$^{\max}-$}&   
  \multicolumn{1}{c}{$\Delta$} &
  \multicolumn{1}{c}{$\Delta$} \\
  \multicolumn{1}{c}{} &
  \multicolumn{1}{c}{(kpc)} &
  \multicolumn{1}{c}{(kpc)}&
  \multicolumn{1}{c}{} &
  \multicolumn{1}{c}{} &
  \multicolumn{1}{c}{} &
  \multicolumn{1}{r}{Y$^{med}$} &
  \multicolumn{1}{c}{$(B-V)$} &
  \multicolumn{1}{c}{$(V-I)$} \\
\hline
  Ter  3    &  2.26 &  3.25 &       &       &       &       &       &      \\
  NGC~6171 &  1.02 &  3.65 & 0.005 & 0.232 & 0.235 & 0.003 & 0.001 & 0.002\\
  1636-283  &  0.48 &  2.75 &       &       &       &       &       &      \\
  NGC~6205 &  1.55 &  8.32 & 0.082 & 0.260 & 0.325 & 0.065 & 0.026 & 0.033\\
  NGC~6218 &  2.35 &  4.79 & 0.026 & 0.270 & 0.290 & 0.020 & 0.008 & 0.011\\
  NGC~6229 &  1.94 & 30.94 &       &       &       &       &       &       \\
  NGC~6235 &  1.53 & 19.42 &       &       &       &       &       &       \\
  NGC~6254 &  1.97 &  4.58 & 0.051 & 0.287 & 0.327 & 0.040 & 0.016 & 0.021\\
  Pal  15   &  1.68 & 51.54 &       &       &       &       &       &      \\
  NGC~6266 &  0.83 &  2.36 & 0.068 & 0.262 & 0.314 & 0.052 & 0.020 & 0.029\\
  NGC~6273 &  1.22 &  3.33 & 0.104 & 0.261 & 0.344 & 0.083 & 0.034 & 0.041\\
  NGC~6284 &  1.28 &  7.35 & 0.028 & 0.278 & 0.300 & 0.022 & 0.009 & 0.012\\
  NGC~6287 &  1.25 &  5.86 & 0.027 & 0.230 & 0.252 & 0.022 & 0.009 & 0.010\\
  NGC~6304 &  1.77 &  3.01 &       &       &       &       &       &       \\
  NGC~6316 &  1.45 &  4.79 &       &       &       &       &       &       \\
  NGC~6333 &  1.16 &  6.65 &       &       &       &       &       &       \\
  NGC~6341 &  1.00 & 10.53 & 0.027 & 0.239 & 0.260 & 0.021 & 0.009 & 0.009\\
  NGC~6342 &  1.12 &  1.88 & 0.027 & 0.250 & 0.267 & 0.017 & 0.006 & 0.011\\
  NGC~6352 &  2.98 &  3.59 & 0.0   & 0.258 & 0.258 & 0.000 & 0.0   & 0.0  \\
  NGC~6356 &  3.17 &  8.35 &       &       &       &       &       &       \\
  IC  1257  &  2.01 & 18.05 &       &       &       &       &       &      \\
  NGC~6362 &  2.52 &  5.14 & 0.030 & 0.237 & 0.260 & 0.023 & 0.009 & 0.013\\
  NGC~6366 &  2.04 &  5.43 & 0.007 & 0.248 & 0.252 & 0.004 & 0.002 & 0.003\\
  Ter  4    &  0.41 &  1.45 &       &       &       &       &       &      \\
  Liller  1 &  0.14 &  1.07 &       &       &       &       &       &      \\
  NGC~6380 &  0.33 &  2.38 &       &       &       &       &       &       \\
  Ter  2    &  0.18 &  1.12 &       &       &       &       &       &      \\
  NGC~6388 &  1.11 &  3.79 & 0.060 & 0.250 & 0.287 & 0.037 & 0.014 & 0.023\\
  NGC~6397 &  2.63 &  6.23 & 0.0   & 0.262 & 0.262 & 0.000 & 0.0   & 0.0  \\
  NGC~6401 &  0.60 &  2.04 &       &       &       &       &       &       \\
  NGC~6402 &  0.65 &  4.35 &       &       &       &       &       &       \\
  Pal  6    &  0.40 &  3.71 &       &       &       &       &       &      \\
  NGC~6426 & 26.84 &215.31 &       &       &       &       &       &       \\
  Ter  5    &  0.82 &  2.83 &       &       &       &       &       &      \\
  NGC~6440 &  0.30 &  1.53 &       &       &       &       &       &       \\
  NGC~6441 &  1.00 &  3.91 & 0.053 & 0.250 & 0.283 & 0.033 & 0.012 & 0.021\\
  UKS  1    &  0.25 &  1.06 &       &       &       &       &       &      \\
  NGC~6496 &  4.02 & 11.54 & 0.003 & 0.250 & 0.252 & 0.002 & 0.001 & 0.001\\
  Djorg  2  &  0.82 &  2.85 &       &       &       &       &       &      \\
  NGC~6517 &  0.50 &  4.24 &       &       &       &       &       &       \\
  Ter  10   &  0.94 &  2.41 &       &       &       &       &       &      \\
  NGC~6528 &  0.41 &  1.61 &       &       &       &       &       &       \\
\hline
\end{tabular}
\addtocounter{table}{-1}
\end{table}

\begin{table}
\caption{Cont...}
\begin{tabular}{lrrrrrrrr}
\hline 
  \multicolumn{1}{c}{Name} &
  \multicolumn{1}{c}{R$_{per}$} &
  \multicolumn{1}{c}{R$_{apo}$} &
  \multicolumn{1}{c}{dMY} &
  \multicolumn{1}{c}{Y$^{med}$} &
  \multicolumn{1}{c}{Y$^{\max}$} &
  \multicolumn{1}{r}{Y$^{\max}-$}&   
  \multicolumn{1}{c}{$\Delta$} &
  \multicolumn{1}{c}{$\Delta$} \\
  \multicolumn{1}{c}{} &
  \multicolumn{1}{c}{(kpc)} &
  \multicolumn{1}{c}{(kpc)}&
  \multicolumn{1}{c}{} &
  \multicolumn{1}{c}{} &
  \multicolumn{1}{c}{} &
  \multicolumn{1}{r}{Y$^{med}$} &
  \multicolumn{1}{c}{$(B-V)$} &
  \multicolumn{1}{c}{$(V-I)$} \\
\hline
  NGC~6535 &  1.01 &  4.47 & 0.018 & 0.272 & 0.286 & 0.014 & 0.006 & 0.007\\
  NGC~6541 &  1.76 &  3.64 & 0.071 & 0.252 & 0.308 & 0.056 & 0.023 & 0.028\\
  NGC~6544 &  0.62 &  5.48 & 0.017 & 0.273 & 0.286 & 0.013 & 0.005 & 0.007\\
  NGC~6553 &  1.29 &  2.35 &       &       &       &       &       &       \\
  IC   1276 &  3.47 &  5.76 &       &       &       &       &       &      \\
  Ter  12   &  2.99 &  5.82 &       &       &       &       &       &      \\
  NGC~6569 &  1.84 &  2.94 &       &       &       &       &       &       \\
  NGC~6584 &  2.10 & 19.25 & 0.046 & 0.235 & 0.271 & 0.036 & 0.014 & 0.019\\
  NGC~6624 &  0.46 &  1.56 & 0.017 & 0.250 & 0.260 & 0.010 & 0.004 & 0.006\\
  NGC~6626 &  0.57 &  2.90 &       &       &       &       &       &       \\
  NGC~6637 &  0.73 &  2.07 & 0.010 & 0.247 & 0.254 & 0.007 & 0.003 & 0.004\\
  NGC~6638 &  0.40 &  2.94 &       &       &       &       &       &       \\
  NGC~6642 &  0.37 &  2.11 &       &       &       &       &       &       \\
  NGC~6652 &  0.65 &  3.66 & 0.0   & 0.242 & 0.242 & 0.000 & 0.0   & 0.0  \\
  NGC~6656 &  2.96 &  9.45 & 0.021 & 0.252 & 0.269 & 0.017 & 0.007 & 0.008\\
  Pal  8    &  2.29 &  5.58 &       &       &       &       &       &      \\
  NGC~6681 &  0.84 &  4.97 & 0.048 & 0.262 & 0.299 & 0.037 & 0.015 & 0.019\\
  NGC~6712 &  0.45 &  4.77 &       &       &       &       &       &       \\
  NGC~6715 & 12.58 & 36.93 &       &       &       &       &       &       \\
  NGC~6717 &  0.89 &  2.72 & 0.0   & 0.261 & 0.261 & 0.000 & 0.0   & 0.0  \\
  NGC~6723 &  2.08 &  2.84 & 0.062 & 0.241 & 0.287 & 0.046 & 0.018 & 0.026\\
  NGC~6749 &  1.60 &  5.07 &       &       &       &       &       &       \\
  NGC~6752 &  3.23 &  5.37 & 0.076 & 0.268 & 0.327 & 0.059 & 0.024 & 0.031\\
  NGC~6760 &  1.90 &  5.67 &       &       &       &       &       &       \\
  NGC~6779 &  0.97 & 12.30 & 0.031 & 0.247 & 0.271 & 0.024 & 0.010 & 0.012\\
  Ter  7    & 13.14 & 44.72 &       &       &       &       &       &      \\
  Pal  10   &  4.01 &  7.02 &       &       &       &       &       &      \\
  Arp  2    & 18.46 & 60.87 &       &       &       &       &       &      \\
  NGC~6809 &  1.59 &  5.54 & 0.032 & 0.250 & 0.275 & 0.025 & 0.011 & 0.012\\
  Ter  8    & 16.23 & 53.86 &       &       &       &       &       &      \\
  Pal  11   &  5.43 &  9.16 &       &       &       &       &       &      \\
  NGC~6838 &  4.77 &  7.08 & 0.0   & 0.227 & 0.227 & 0.000 & 0.0   & 0.0  \\
  NGC~6864 &  2.06 & 17.98 &       &       &       &       &       &       \\
  NGC~6934 &  2.60 & 39.52 & 0.058 & 0.238 & 0.283 & 0.045 & 0.018 & 0.023\\
  NGC~6981 &  1.29 & 24.01 & 0.047 & 0.238 & 0.274 & 0.036 & 0.015 & 0.019\\
  NGC~7006 &  2.07 & 55.50 &       &       &       &       &       &       \\
  NGC~7078 &  3.57 & 10.39 & 0.091 & 0.232 & 0.305 & 0.073 & 0.031 & 0.032\\
  NGC~7089 &  0.56 & 16.80 & 0.097 & 0.253 & 0.330 & 0.077 & 0.031 & 0.039\\
  NGC~7099 &  1.49 &  8.15 & 0.005 & 0.245 & 0.249 & 0.004 & 0.002 & 0.002\\
  Pal  12   & 15.75 & 71.17 &       &       &       &       &       &      \\
  Pal  13   &  9.04 & 67.47 &       &       &       &       &       &      \\
  NGC~7492 &  4.27 & 28.23 &       &       &       &       &       &       \\
\hline
\end{tabular}
\end{table}

\begin{table}
\caption{Additional parameters for selected GCs}
\begin{tabular}{lrrrrrrrrr}
\hline 
  \multicolumn{1}{c}{Name} &
  \multicolumn{1}{c}{dY2$_{g1G}$} &
  \multicolumn{1}{c}{err} &
  \multicolumn{1}{c}{dYmax} &
  \multicolumn{1}{c}{err} &
  \multicolumn{1}{c}{[Fe/H]} &
  \multicolumn{1}{c}{d[Fe/H]} &
  \multicolumn{1}{c}{err} &
  \multicolumn{1}{c}{Log} &
  \multicolumn{1}{c}{Log} \\
  \multicolumn{1}{c}{} &
  \multicolumn{1}{c}{} &
  \multicolumn{1}{c}{} &
  \multicolumn{1}{c}{} &
  \multicolumn{1}{c}{} &
  \multicolumn{1}{c}{} &
  \multicolumn{1}{c}{} &
  \multicolumn{1}{c}{} &
  \multicolumn{1}{c}{(M$_{fin}$)} &
  \multicolumn{1}{c}{(M$_{in}$)}  \\
\hline
  NGC~104  &0.011 & 0.005 & 0.049 & 0.005 & -0.72 & 0.012 & 0.010 & 5.88 & 6.23 \\
  NGC~288  &0.015 & 0.010 & 0.016 & 0.012 & -1.32 & 0.007 & 0.013 & 5.08 & 5.58 \\
  NGC~362  &0.008 & 0.006 & 0.026 & 0.008 & -1.26 &       &       & 5.52 & 6.06 \\
  NGC~1261 &0.004 & 0.004 & 0.019 & 0.007 & -1.27 &       &       & 5.26 & 5.86 \\
  Eridanus  &     &       &       &       & -1.43 &       &       & 4.04 & 4.41 \\
  Pal  2    &     &       &       &       & -1.42 &       &       & 5.36 & 6.05 \\
  NGC~1851 & 0.007 & 0.005 & 0.025 & 0.006 & -1.18 &       &       & 5.45 & 6.11 \\
  NGC~1904 &        &       &       &       & -1.60 & 0.013 & 0.015 & 5.23 & 6.08 \\
  NGC~2298 & -0.003 & 0.009 & 0.011 & 0.012 & -1.92 &       &       & 4.65 & 5.64 \\
  NGC~2419 &        &       & 0.19  &       & -2.15 &       &       & 6.09 & 6.40 \\
  NGC~2808 & 0.048 & 0.005 & 0.124 & 0.007 & -1.14 & 0.070 & 0.018 & 5.91 & 6.36 \\
  Pal  3    &     &       &       &       & -1.63 &       &       & 4.36 & 4.70 \\
  NGC~3201 & -0.001 & 0.013 & 0.028 & 0.032 & -1.59 &-0.026 & 0.013 & 5.12 & 5.51 \\
  Pal  4    &     &       &       &       & -1.41 &       &       & 4.43 & 4.80 \\
  NGC~4147 &        &       &       &       & -1.80 &       &       & 4.50 & 5.56 \\
  NGC~4372 &        &       &       &       & -2.17 &       &       & 5.34 & 5.81 \\
  Rup  106  &     &       &       &       & -1.68 &       &       & 4.57 & 5.16 \\
  NGC~4590 & 0.007 & 0.009 & 0.012 & 0.009 & -2.23 &-0.021 & 0.018 & 5.09 & 5.48 \\
  NGC~4833 & 0.016 & 0.008 & 0.051 & 0.009 & -1.85 &       &       & 5.24 & 6.05 \\
  NGC~5024 & 0.013 & 0.007 & 0.044 & 0.008 & -2.10 &       &       & 5.53 & 5.88 \\
  NGC~5053 & -0.002 & 0.013 & 0.004 & 0.025 & -2.27 &       &       & 4.75 & 5.20 \\
  NGC~5139 & 0.033 & 0.006 & 0.090 & 0.010 & -1.53 &       &       & 6.53 & 6.86 \\
  NGC~5272 & 0.016 & 0.005 & 0.041 & 0.009 & -1.50 &       &       & 5.57 & 5.94 \\
  NGC~5286 & 0.007 & 0.006 & 0.044 & 0.004 & -1.69 &       &       & 5.58 & 6.11 \\
  NGC~5466 & 0.002 & 0.017 & 0.007 & 0.024 & -1.98 &       &       & 4.64 & 5.12 \\
  NGC~5634 &        &       &       &       & -1.88 &       &       & 5.33 & 5.77 \\
  NGC~5694 &        &       &       &       & -1.98 &       &       & 5.56 & 5.92 \\
  IC  4499  &0.004 & 0.006 & 0.017 & 0.008 & -1.53 &       &       & 5.08 & 5.50 \\
  NGC~5824 &        &       &       &       & -1.91 &       &       & 5.87 & 6.19 \\
  Pal  5    &     &       &       &       & -1.41 &       &       & 4.22 & 4.86 \\
  NGC~5897 &        &       &       &       & -1.90 &       &       & 5.22 & 5.77 \\
  NGC~5904 & 0.012 & 0.004 & 0.037 & 0.007 & -1.29 &-0.001 & 0.010 & 5.56 & 5.96 \\
  NGC~5927 & 0.011 & 0.004 & 0.055 & 0.015 & -0.49 &       &       & 5.40 & 5.83 \\
  NGC~5946 &        &       &       &       & -1.29 &       &       & 5.06 & 6.02 \\
  NGC~5986 & 0.005 & 0.006 & 0.031 & 0.012 & -1.59 &       &       & 5.52 & 6.20 \\
  Lynga  7  &     &       &       &       & -1.01 &       &       & 5.00 & 5.72 \\
  Pal  14   &     &       &       &       & -1.41 &       &       & 4.19 & 5.07 \\
  NGC~6093 & 0.011 & 0.008 & 0.027 & 0.012 & -1.75 &       &       & 5.44 & 6.33 \\
  NGC~6101 & 0.005 & 0.010 & 0.017 & 0.011 & -1.98 &       &       & 5.10 & 5.47 \\
  NGC~6121 & 0.009 & 0.006 & 0.014 & 0.006 & -1.16 & 0.008 & 0.012 & 4.95 & 6.03 \\
  NGC~6139 &        &       &       &       & -1.65 &       &       & 5.53 & 6.07 \\
  NGC~6144 & 0.009 & 0.011 & 0.017 & 0.013 & -1.76 &       &       & 4.72 & 5.61 \\
  Ter  3    &     &       &       &       & -0.74 &       &       & 4.70 & 5.88 \\
\hline
\end{tabular}
\label{tab:A1b}
\addtocounter{table}{-1}
\end{table}

\begin{table}
\caption{Cont...}
\begin{tabular}{lrrrrrrrrr}
\hline 
  \multicolumn{1}{c}{Name} &
  \multicolumn{1}{c}{dY2$_{g1G}$} &
  \multicolumn{1}{c}{err} &
  \multicolumn{1}{c}{dYmax} &
  \multicolumn{1}{c}{err} &
  \multicolumn{1}{c}{[Fe/H]} &
  \multicolumn{1}{c}{d[Fe/H]} &
  \multicolumn{1}{c}{err} &
  \multicolumn{1}{c}{Log} &
  \multicolumn{1}{c}{Log} \\
  \multicolumn{1}{c}{} &
  \multicolumn{1}{c}{} &
  \multicolumn{1}{c}{} &
  \multicolumn{1}{c}{} &
  \multicolumn{1}{c}{} &
  \multicolumn{1}{c}{} &
  \multicolumn{1}{c}{} &
  \multicolumn{1}{c}{} &
  \multicolumn{1}{c}{(M$_{fin}$)} &
  \multicolumn{1}{c}{(M$_{in}$)}  \\
\hline
NGC~6171 & 0.019 & 0.011 & 0.024 & 0.014 & -1.02 & 0.018 & 0.025 & 4.94 & 5.86 \\
  1636-283  &     &       &       &       & -1.50 &       &       &      &      \\
  NGC~6205 & 0.020 & 0.004 & 0.052 & 0.004 & -1.53 &       &       & 5.66 & 6.11 \\
  NGC~6218 & 0.009 & 0.007 & 0.011 & 0.011 & -1.37 & 0.000 & 0.009 & 4.90 & 5.63 \\
  NGC~6229 &        &       &       &       & -1.47 &       &       & 5.46 & 6.02 \\
  NGC~6235 &        &       &       &       & -1.28 &       &       & 5.04 & 5.49 \\
  NGC~6254 & 0.006 & 0.008 & 0.029 & 0.011 & -1.56 &-0.002 & 0.015 & 5.27 & 5.83 \\
  Pal  15   &      &       &       &       & -2.07 &       &       & 4.62 & 5.62 \\
  NGC~6266 &        &       &       &       & -1.18 &       &       & 5.82 & 6.29 \\
  NGC~6273 &        &       &       &       & -1.74 &       &       & 5.81 & 6.27 \\
  NGC~6284 &        &       &       &       & -1.26 &       &       & 5.39 & 5.96 \\
  NGC~6287 &        &       &       &       & -2.10 &       &       & 5.12 & 5.85 \\
  NGC~6304 & 0.008 & 0.005 & 0.025 & 0.006 & -0.45 &       &       & 5.16 & 5.80 \\
  NGC~6316 &        &       &       &       & -0.45 &       &       & 5.57 & 6.12 \\
  NGC~6333 &        &       &       &       & -1.77 &       &       & 5.48 & 6.05 \\
  NGC~6341 & 0.022 & 0.004 & 0.039 & 0.006 & -2.31 &       &       & 5.49 & 6.09 \\
  NGC~6342 &        &       &       &       & -0.55 &       &       & 4.78 & 5.89 \\
  NGC~6352 & 0.019 & 0.014 & 0.027 & 0.006 & -0.64 &       &       & 4.78 & 5.56 \\
  NGC~6356 &        &       &       &       & -0.40 &       &       & 5.57 & 6.04 \\
  IC  1257  &     &       &       &       & -1.70 &       &       & 4.80 & 5.64 \\
  NGC~6362 & 0.003 & 0.011 & 0.004 & 0.011 & -0.99 &       &       & 5.05 & 5.67 \\
  NGC~6366 & 0.022 & 0.010 & 0.011 & 0.015 & -0.59 &       &       & 4.70 & 5.51 \\
  Ter  4    &      &       &       &       & -1.41 &       &       & 4.88 & 6.19 \\
  Liller  1 &      &       &       &       & -0.33 &       &       & 5.81 & 6.49 \\
  NGC~6380 &        &       &       &       & -0.75 &       &       & 5.48 & 6.32 \\
  Ter  2    &     &       &       &       & -0.70 &       &       & 4.52 & 6.35 \\
  NGC~6388 & 0.019 & 0.007 & 0.067 & 0.009 & -0.55 & 0.038 & 0.042 & 6.02 & 6.42 \\
  NGC~6397 & 0.006 & 0.009 & 0.008 & 0.011 & -2.02 &       &       & 4.94 & 5.60 \\
  NGC~6401 &        &       &       &       & -1.02 &       &       & 5.45 & 6.20 \\
  NGC~6402 &        &       &       &       & -1.28 & 0.029 & 0.021 & 5.87 & 6.38 \\
  Pal  6    &     &       &       &       & -0.91 &       &       & 5.13 & 6.24 \\
  NGC~6426 &       &       &       &       & -2.15 &       &       & 4.84 & 5.19 \\
  Ter  5    &     &       &       &       & -0.23 &       &       & 5.59 & 6.13 \\
  NGC~6440 &        &       &       &       & -0.36 &       &       & 5.58 & 6.37 \\
  NGC~6441 & 0.029 & 0.006 & 0.081 & 0.022 & -0.46 &       &       & 6.06 & 6.46 \\
  UKS  1    &     &       &       &       & -0.64 &       &       & 4.88 & 6.29 \\
  NGC~6496 & 0.009 & 0.011 & 0.021 & 0.006 & -0.46 &       &       & 4.63 & 5.29 \\
  Djorg  2  &       &       &       &       & -0.65 &       &       & 4.79 & 5.96 \\
  NGC~6517 &        &       &       &       & -1.23 &       &       & 5.56 & 6.26 \\
  Ter  10   &      &       &       &       & -1.79 &       &       & 4.72 & 5.88 \\
  NGC~6528 &        &       &       &       & -0.11 &       &       & 4.97 & 6.19 \\
  NGC~6535 & 0.003 & 0.021 & 0.003 & 0.022 & -1.79 &       &       & 3.97 & 5.77 \\
\hline
\end{tabular}
\addtocounter{table}{-1}
\end{table}

\begin{table}
\caption{Cont...}
\begin{tabular}{lrrrrrrrrr}
\hline 
  \multicolumn{1}{c}{Name} &
  \multicolumn{1}{c}{dY2$_{g1G}$} &
  \multicolumn{1}{c}{err} &
  \multicolumn{1}{c}{dYmax} &
  \multicolumn{1}{c}{err} &
  \multicolumn{1}{c}{[Fe/H]} &
  \multicolumn{1}{c}{d[Fe/H]} &
  \multicolumn{1}{c}{err} &
  \multicolumn{1}{c}{Log} &
  \multicolumn{1}{c}{Log} \\
  \multicolumn{1}{c}{} &
  \multicolumn{1}{c}{} &
  \multicolumn{1}{c}{} &
  \multicolumn{1}{c}{} &
  \multicolumn{1}{c}{} &
  \multicolumn{1}{c}{} &
  \multicolumn{1}{c}{} &
  \multicolumn{1}{c}{} &
  \multicolumn{1}{c}{(M$_{fin}$)} &
  \multicolumn{1}{c}{(M$_{in}$)}  \\
\hline
  NGC~6541 & 0.024 & 0.005 & 0.045 & 0.006 & -1.81 &       &       & 5.39 & 5.95 \\
  NGC~6544 &        &       &       &       & -1.40 &       &       & 5.06 & 6.09 \\
  NGC~6553 &        &       &       &       & -0.18 &       &       & 5.52 & 6.07 \\
  IC   1276 &     &       &       &       & -0.75 &       &       & 4.96 & 5.55 \\
  Ter  12   &     &       &       &       & -0.50 &       &       & 3.13 & 5.18 \\
  NGC~6569 &        &       &       &       & -0.76 &       &       & 5.36 & 5.96 \\
  NGC~6584 & 0.0   & 0.007 & 0.015 & 0.011 & -1.50 &       &       & 5.23 & 5.84 \\
  NGC~6624 & 0.010 & 0.004 & 0.022 & 0.003 & -0.44 &       &       & 4.88 & 6.18 \\
  NGC~6626 &        &       &       &       & -1.32 &       &       & 5.47 & 6.21 \\
  NGC~6637 & 0.004 & 0.006 & 0.011 & 0.005 & -0.64 &       &       & 5.16 & 6.18 \\
  NGC~6638 &        &       &       &       & -0.95 &       &       & 5.26 & 6.27 \\
  NGC~6642 &        &       &       &       & -1.26 &       &       & 4.58 & 6.27 \\
  NGC~6652 & 0.008 & 0.007 & 0.017 & 0.011 & -0.81 &       &       & 4.79 & 6.19 \\
  NGC~6656 & 0.005 & 0.008 & 0.041 & 0.012 & -1.70 &       &       & 5.60 & 6.01 \\
  Pal  8    &     &       &       &       & -0.37 &       &       & 4.75 & 5.61 \\
  NGC~6681 & 0.009 & 0.008 & 0.029 & 0.015 & -1.62 &       &       & 5.06 & 5.94 \\
  NGC~6712 &        &       &       &       & -1.02 &       &       & 5.07 & 6.21 \\
  NGC~6715 & 0.012 & 0.003 & 0.052 & 0.012 & -1.49 &       &       & 6.19 & 6.51 \\
  NGC~6717 & 0.003 & 0.006 & 0.003 & 0.009 & -1.26 &       &       & 4.24 & 5.87 \\
  NGC~6723 & 0.005 & 0.006 & 0.024 & 0.007 & -1.10 &       &       & 5.22 & 5.85 \\
  NGC~6749 &        &       &       &       & -1.60 &       &       & 4.90 & 5.68 \\
  NGC~6752 & 0.015 & 0.005 & 0.042 & 0.004 & -1.54 &-0.014 & 0.014 & 5.36 & 5.83 \\
  NGC~6760 &        &       &       &       & -0.40 &       &       & 5.43 & 5.92 \\
  NGC~6779 & 0.011 & 0.007 & 0.031 & 0.008 & -1.98 &       &       & 5.18 & 6.01 \\
  Ter  7    &     &       &       &       & -0.32 &       &       & 4.31 & 4.75 \\
  Pal  10   &     &       &       &       & -0.10 &       &       & 4.74 & 5.38 \\
  Arp  2    &     &       &       &       & -1.75 &       &       & 4.56 & 4.98 \\
  NGC~6809 & 0.014 & 0.008 & 0.026 & 0.015 & -1.94 &-0.009 & 0.013 & 5.27 & 5.88 \\
  Ter  8    &     &       &       &       & -2.16 &       &       & 4.80 & 5.18 \\
  Pal  11   &     &       &       &       & -0.40 &       &       & 4.78 & 5.34 \\
  NGC~6838 & 0.005 & 0.009 & 0.024 & 0.010 & -0.78 &-0.005 & 0.016 & 4.72 & 5.32 \\
  NGC~6864 &        &       &       &       & -1.29 &       &       & 5.60 & 6.07 \\
  NGC~6934 & 0.006 & 0.003 & 0.018 & 0.004 & -1.47 &       &       & 5.25 & 5.74 \\
  NGC~6981 & 0.011 & 0.006 & 0.017 & 0.006 & -1.42 &       &       & 4.87 & 5.88 \\
  NGC~7006 &        &       &       &       & -1.52 &       &       & 5.10 & 5.72 \\
  NGC~7078 & 0.021 & 0.009 & 0.069 & 0.006 & -2.37 &-0.008 & 0.022 & 5.69 & 6.08 \\
  NGC~7089 & 0.013 & 0.005 & 0.052 & 0.009 & -1.65 &       &       & 5.70 & 6.38 \\
  NGC~7099 & 0.015 & 0.010 & 0.022 & 0.010 & -2.27 &-0.027 & 0.020 & 5.11 & 5.79 \\
  Pal  12   &     &       &       &       & -0.85 &       &       & 3.76 & 4.33 \\
  Pal  13   &     &       &       &       & -1.88 &       &       & 3.49 & 4.36 \\
  NGC~7492 &     &       &       &       & -1.78 &       &       & 4.51 & 5.25 \\
\hline
\end{tabular}
\end{table}

\begin{table}
\caption{Additional parameters for selected GCs}
\begin{tabular}{lrcrrrrrcr}
\hline 
  \multicolumn{1}{c}{Name} &
  \multicolumn{1}{c}{IQR} &
  \multicolumn{1}{c}{IQR} &
  \multicolumn{1}{c}{IQR} &
  \multicolumn{1}{c}{dRGB} &
  \multicolumn{1}{c}{err} &
  \multicolumn{1}{c}{f(FG)} &
  \multicolumn{1}{c}{err} &
  \multicolumn{1}{c}{GC} &
  \multicolumn{1}{c}{$\delta$}\\
  \multicolumn{1}{c}{} &
  \multicolumn{1}{c}{[O/Na]} &
  \multicolumn{1}{c}{source} &
  \multicolumn{1}{c}{[Al/Mg]} &
  \multicolumn{1}{c}{}&
  \multicolumn{1}{c}{} &
  \multicolumn{1}{c}{} &
  \multicolumn{1}{c}{} &
  \multicolumn{1}{c}{type} &
  \multicolumn{1}{c}{[Al/Mg]}\\  \hline
  NGC~104  &  0.472 & 1 & 0.091 & 0.369 & 0.009 & 0.175 & 0.009 & 1 & 0.3\\
  NGC~288  &  0.776 & 1 & 0.059 & 0.276 & 0.008 & 0.542 & 0.031 & 1 & 0.2\\
  NGC~362  &  0.644 & 1 & 0.405 & 0.275 & 0.005 & 0.279 & 0.015 & 2 & 0.4\\
  NGC~1261 &        &   &	 & 0.29 & 0.01 & 0.359 & 0.016 & 2 & \\
  Eridanus  &        &   &  &  &  &  &  &  & \\
  Pal  2    &        &   &  &  &  &  &  &  & \\
  NGC~1851 &  0.693 & 1 & 0.45 & 0.342 & 0.005 & 0.264 & 0.015 & 2 & 0.4\\
  NGC~1904 &  0.759 & 1 & 0.438 &  &  &  &  &  & \\
  NGC~2298 &        &   &  & 0.243 & 0.017 & 0.37 & 0.037 & 1 & \\
  NGC~2419 &        &   &  &  & & 0.37 & 0.01    &  & \\
  NGC~2808 &  0.999 & 1 & 0.935 & 0.457 & 0.009 & 0.232 & 0.014 & 1 & 1.25\\
  Pal  3    &        &   &  &  &  &  &  &  & \\
  NGC~3201 &  0.634 & 1 & 0.383 & 0.292 & 0.016 & 0.436 & 0.036 & 1 & 0.5\\
  Pal  4    &        &   &  &  &  &  &  &  & \\
  NGC~4147 &  0.560 & 2 &  &  &  &  &  &  & \\
  NGC~4372 &  0.390 & 3 &  &  &  &  &  &  & \\
  Rup  106  &  0.160 &28 &  &  &  & 1.0 & 0.1 &  & \\
  NGC~4590 &  0.372 & 1 & 0.274 & 0.132 & 0.007 & 0.381 & 0.024 & 1 & 0.3\\
  NGC~4833 &  0.945 & 1 & 0.81 & 0.26 & 0.008 & 0.362 & 0.025 & 1 & 0.65\\
  NGC~5024 &  0.400 & 4 & 0.8 & 0.209 & 0.005 & 0.328 & 0.02 & 1 & 0.5\\
  NGC~5053 &  0.950 & 4 &  & 0.102 & 0.013 & 0.544 & 0.062 & 1 & 1.1\\
  NGC~5139 &  1.200 & 5 &  & 0.39 & 0.01 & 0.086 & 0.01 & 2 & 0.85\\
  NGC~5272 &  0.610 & 6 & 0.9 & 0.279 & 0.007 & 0.305 & 0.014 & 1 & 0.55\\
  NGC~5286 &  0.700 & 7 &  & 0.303 & 0.007 & 0.342 & 0.015 & 2 & \\
  NGC~5466 &        &   & 0.4 & 0.141 & 0.016 & 0.467 & 0.063 & 1 & 0.5\\
  NGC~5634 &  0.756 & 1 & 0.44 &  &  &  &  &  & \\
  NGC~5694 &  0.260 & 9 & 0.2 &  &  &  &  &  & \\
  IC  4499  &        &   &  &  &  &  &  &  & \\
  NGC~5824 &        &30 & 0.92 &  &  &  &  &  & \\
  Pal  5    &        & 8 &  &  &  &  &  &  & \\
  NGC~5897 &  0.620 &11 &  & 0.149 & 0.008 & 0.547 & 0.042 & 1 & \\
  NGC~5904 &  0.741 & 1 & 0.541 & 0.332 & 0.013 & 0.235 & 0.013 & 1 & 0.6\\
  NGC~5927 &  0.600 &10 & 0.1 & 0.422 & 0.02 &  &  &  & 0.1\\
  NGC~5946 &        &   &  &  &  &  &  &  & \\
  NGC~5986 &  0.580 &14 & 0.6 & 0.294 & 0.008 & 0.246 & 0.012 & 1 & 0.65\\
  Lynga  7  &        &   &  &  &  &  &  &  & \\
  Pal  14   &  0.500 &26 &  &  &  &  &  &  & \\
  NGC~6093 &  0.784 & 1 & 0.56 & 0.305 & 0.015 & 0.351 & 0.029 & 1 & 0.5\\
  NGC~6101 &        &   &  & 0.14 & 0.009 & 0.654 & 0.032 & 1 & \\
  NGC~6121 &  0.373 & 1 & 0.18 & 0.27 & 0.012 & 0.285 & 0.037 & 1 & 0.0\\
  NGC~6139 &  0.647 & 1 & 0.436 &  &  &  &  &  & \\
  NGC~6144 &        &   &  & 0.21 & 0.012 & 0.444 & 0.037 & 1 & \\
\hline
\end{tabular}
\label{tab:A2}
\addtocounter{table}{-1}
\end{table}

\begin{table}
\caption{Cont...}
\begin{tabular}{lrcrrrrrcr}
\hline 
  \multicolumn{1}{c}{Name} &
  \multicolumn{1}{c}{IQR} &
  \multicolumn{1}{c}{IQR} &
  \multicolumn{1}{c}{IQR} &
  \multicolumn{1}{c}{dRGB} &
  \multicolumn{1}{c}{err} &
  \multicolumn{1}{c}{f(FG)} &
  \multicolumn{1}{c}{err} &
  \multicolumn{1}{c}{GC} &
  \multicolumn{1}{c}{$\delta$}\\
  \multicolumn{1}{c}{} &
  \multicolumn{1}{c}{[O/Na]} &
  \multicolumn{1}{c}{source} &
  \multicolumn{1}{c}{[Al/Mg]} &
  \multicolumn{1}{c}{}&
  \multicolumn{1}{c}{} &
  \multicolumn{1}{c}{} &
  \multicolumn{1}{c}{} &
  \multicolumn{1}{c}{type} &
  \multicolumn{1}{c}{[Al/Mg]}\\  \hline
  Ter  3    &        &   &  &  &  &  &  & & \\
  NGC~6171 &  0.522 & 1 & 0.1 & 0.351 & 0.017 & 0.397 & 0.031 & 1 & 0.0\\
  1636-283  &        &   &  &  &  &  &  &  & \\
  NGC~6205 &  0.890 & 6 & 1.1 & 0.291 & 0.006 & 0.184 & 0.013 & 1 & 0.9\\
  NGC~6218 &  0.863 & 1 & 0.271 & 0.274 & 0.009 & 0.4 & 0.029 & 1 & 0.3\\
  NGC~6229 &  0.810 &13 &  &  &  &  &  &  & \\
  NGC~6235 &        &   &  &  &  &  &  &  & \\
  NGC~6254 &  0.565 & 1 & 0.75 & 0.31 & 0.007 & 0.364 & 0.028 & 1 & 0.6\\
  Pal  15   &        &   &  &  &  &  &  &  & \\
  NGC~6266 &  1.160 &16 &  &  &  &  &  &  & \\
  NGC~6273 &        &15 &  0.32 &  &  &  &  & & \\
  NGC~6284 &        &   &  &  &  &  &  &  & \\
  NGC~6287 &        &   &  &  &  &  &  &  & \\
  NGC~6304 &        &   &  & 0.32 & 0.024 &  &  &  & \\
  NGC~6316 &        &   &  &  &  &  &  &  & \\
  NGC~6333 &        &   &  &  &  &  &  &  & \\
  NGC~6341 &        &   & 0.9 & 0.177 & 0.005 & 0.304 & 0.015 & 1 & 0.9\\
  NGC~6342 &        &   &  &  &  &  &  &  & \\
  NGC~6352 &        &17 & 0.14 & 0.395 & 0.015 & 0.474 & 0.035 & 0 & \\
  NGC~6356 &        &   &  &  &  &  &  &  & \\
  IC  1257  &        &   &  &  &  &  &  &  & \\
  NGC~6362 &  0.280 &18 & 0.11 & 0.292 & 0.011 & 0.574 & 0.035 & 1 & 0.0\\
  NGC~6366 &  0.280 &12 &  & 0.291 & 0.064 & 0.418 & 0.045 & 1 & 0.15\\
  Ter  4    &        &   &  &  &  &  &  &  & \\
  Liller  1 &        &   &  &  &  &  &  &  & \\
  NGC~6380 &        &   &  &  &  &  &  &  & \\
  Ter  2    &        &   &  &  &  &  &  &  & \\
  NGC~6388 &  0.644 & 1 & 0.529 & 0.494 & 0.01 & 0.245 & 0.01 & 2 & 0.55\\
  NGC~6397 &  0.274 & 1 &  & 0.117 & 0.023 & 0.345 & 0.036 & 1 & 0.1\\
  NGC~6401 &        &   &  &  &  &  &  &  & \\
  NGC~6402 &  0.525 &33 & 0.270 &  &  &  &  &  & \\
  Pal  6    &        &   &  &  &  &  &  &  & \\
  NGC~6426 &        &   &  &  &  &  &  &  & \\
  Ter  5    &        &   &  &  &  &  &  &  & \\
  NGC~6440 &  0.370 & 19& 0.41 &  &  &  &  & &  \\
  NGC~6441 &  0.660 & 1 &  & 0.512 & 0.015 &  &  &  & 0.2\\
  UKS  1    &        &   &  &  &  &  &  &  & \\
  NGC~6496 &        &   &  & 0.331 & 0.038 & 0.674 & 0.035 & 1 & \\
  Djorg  2  &        &   &  &  &  &  &  &  & \\
  NGC~6517 &        &   &  &  &  &  &  &  & \\
  Ter  10   &        &   &  &  &  &  &  &  & \\
  NGC~6528 &  0.650 &29 & 0.14 &  &  &  &  &  & \\
\hline
\end{tabular}
\addtocounter{table}{-1}
\end{table}

\begin{table}
\caption{Cont...}
\begin{tabular}{lrcrrrrrcr}
\hline 
  \multicolumn{1}{c}{Name} &
  \multicolumn{1}{c}{IQR} &
  \multicolumn{1}{c}{IQR} &
  \multicolumn{1}{c}{IQR} &
  \multicolumn{1}{c}{dRGB} &
  \multicolumn{1}{c}{err} &
  \multicolumn{1}{c}{f(FG)} &
  \multicolumn{1}{c}{err} &
  \multicolumn{1}{c}{GC} &
  \multicolumn{1}{c}{$\delta$}\\
  \multicolumn{1}{c}{} &
  \multicolumn{1}{c}{[O/Na]} &
  \multicolumn{1}{c}{source} &
  \multicolumn{1}{c}{[Al/Mg]} &
  \multicolumn{1}{c}{}&
  \multicolumn{1}{c}{} &
  \multicolumn{1}{c}{} &
  \multicolumn{1}{c}{} &
  \multicolumn{1}{c}{type} &
  \multicolumn{1}{c}{[Al/Mg]}\\  \hline
    NGC~6535 &  0.440 & 1 &  & 0.142 & 0.02 & 0.536 & 0.081 & 1 & 0.4\\
  NGC~6541 &        &   &  & 0.275 & 0.007 & 0.396 & 0.02 & 1 & \\
  NGC~6544 &        &   &  &  &  &  &  &  & \\
  NGC~6553 &        &   &  &  &  &  &  &  & \\
  IC   1276 &        &   &  &  &  &  &  &  & \\
  Ter  12   &        &   &  &  &  &  &  &  & \\
  NGC~6569 &  0.930 &31 & 0.22 &  &  &  &  &  & \\
  NGC~6584 &        &   &  & 0.221 & 0.014 & 0.451 & 0.026 & 1 & \\
  NGC~6624 &        &   &  & 0.444 & 0.015 & 0.279 & 0.02 & 1 & \\
  NGC~6626 &  1.260 &20 & 0.79 &  &  &  &  &  & \\
  NGC~6637 &        &   &  & 0.367 & 0.011 & 0.425 & 0.017 & 1 & \\
  NGC~6638 &        &   &  &  &  &  &  &  & \\
  NGC~6642 &        &   &  &  &  &  &  &  & \\
  NGC~6652 &        &   &  & 0.341 & 0.014 & 0.344 & 0.026 & 1 & \\
  NGC~6656 &  0.700 &21 & 0.45 & 0.293 & 0.012 & 0.274 & 0.02 & 2 & 0.45\\
  Pal  8    &        &   &  &  &  &  &  &  & \\
  NGC~6681 &  0.440 &22 & 0.28 & 0.309 & 0.005 & 0.234 & 0.019 & 1 & 0.2\\
  NGC~6712 &        &   &  &  &  &  &  &  & \\
  NGC~6715 &  1.169 & 1 & 1.16 & 0.404 & 0.009 & 0.267 & 0.012 & 2 & 0.6\\
  NGC~6717 &        &   &  & 0.293 & 0.012 & 0.637 & 0.039 &  & \\
  NGC~6723 &        &   &  & 0.352 & 0.006 & 0.363 & 0.017 & 1 & \\
  NGC~6749 &        &   &  &  &  &  &  &  & \\
  NGC~6752 &  0.772 & 1 & 0.56 & 0.32 & 0.015 & 0.294 & 0.023 & 1 & 0.85\\
  NGC~6760 &        &   &  &  &  &  &  &  & \\
  NGC~6779 &        &   &  & 0.256 & 0.007 & 0.469 & 0.041 & 1 & \\
  Ter  7    &  0.20  &32 &  &  &  &  &  &  & \\
  Pal  10   &        &   &  &  &  &  &  &  & \\
  Arp  2    &        &   &  &  &  &  &  &  & \\
  NGC~6809 &  0.725 & 1 & 0.451 & 0.211 & 0.012 & 0.311 & 0.029 & 1 & 0.5\\
  Ter  8    &  0.120 & 1 & 0.38 &  &  & 0.93 & 0.07 &  & \\
  Pal  11   &        &   &  &  &  &  &  &  & \\
  NGC~6838 &  0.257 & 1 & 0.175 & 0.334 & 0.014 & 0.622 & 0.038 & 1 & 0.0\\
  NGC~6864 &  0.730 &23 &  &  &  &  &  &  & \\
  NGC~6934 &        &   &  & 0.312 & 0.015 & 0.326 & 0.02 & 2 & \\
  NGC~6981 &        &   &  & 0.24 & 0.009 & 0.542 & 0.027 & 1 & \\
  NGC~7006 &  0.240 &24 & 0.13 &  &  &  &  &  & \\
  NGC~7078 &  0.501 & 1 & 0.478 & 0.217 & 0.003 & 0.399 & 0.019 & 2 & 0.7\\
  NGC~7089 &  0.700 &25 & 0.9 & 0.302 & 0.009 & 0.224 & 0.014 & 2 & 0.5\\
  NGC~7099 &  0.607 & 1 & 0.497 & 0.14 & 0.009 & 0.38 & 0.028 & 1 & 0.3\\
  Pal  12   &  0.140 &27 &  &  &  &  &  &  & \\
  Pal  13   &        &   &  &  &  &  &  &  & \\
  NGC~7492 &        &   &  &  &  &  &  &  & \\
\hline
\end{tabular}
\end{table}

\section{Appendix 2: Summary of data for MC clusters}

This Appendix collects data for extra-galactic massive clusters used in this review. References for individual columns are as follows:

\begin{itemize}
\item Age: SMC: \citet{Glatt2008, Martocchia2017}
\item LMC: Age/[Fe/H]/Mass:  \citet{Mackey2003a, Ferraro2006b, Niederhofer2016, Martocchia2018a}
\item Fornax: Age/[Fe/H]/Mass:  \citet{Mackey2003b}
\item Mass and Metallicity: \citet{Glatt2011}
\item dCUnBI: \citet{Martocchia2017}
\item Hodge~11: [Fe/H]: \citet{Mateluna2012}
\end{itemize}

\begin{table}
\caption{Data for extra-galactic massive clusters}
\begin{tabular}{lcccccccl}
\hline
Cluster & Age & [Fe/H] & $\log{M/M_\odot}$ & FG/tot & dCUNBI & d(C/N) & IQR(Na/O) & Source \\
        & Gyr &        &                   &        &        &        &           &        \\
\hline
SMC-Kron~3    &  6.5 & -1.08 &  5.17 &   0.68 &         &  0.99 &       & \citet{Hollyhead2018} \\
SMC-NGC~121  & 10.5 & -1.46 &  5.50 &   0.70 &    0.14 &       &       & \citet{Niederhofer2017a} \\
&&&&&&&& \citet{Dalessandro2016}\\
SMC-Lindsay~1 &  7.5 & -1.14 &  5.28 &   0.70 &    0.08 &       &       & \citet{Niederhofer2017b} \\
          &      &       &       &        &         &       &       & \citet{Hollyhead2017} \\
SMC-NGC~339  &  6.0 & -1.12 &  4.96 &   0.59 &    0.08 &       &       & \citet{Niederhofer2017b} \\
SMC-NGC~416  &  6.0 & -1.00 &  5.24 &   0.47 &    0.12 &       &       & \citet{Niederhofer2017b} \\
SMC-NGC~419  &  1.5 & -0.70 &  5.23 &   1.00 &    0.00 &       &       & \citet{Martocchia2017} \\
LMC-NGC~1783  & 1.8 & -0.36 &  5.26 &   1.00 &         &       &  1.20 & \citet{Zhang2018} \\
LMC-NGC~1786  & 13.4 & -1.87 &  5.57 &   0.42 &         &       &  1.20 & \citet{Mucciarelli2009} \\
LMC-NGC~1806  &  1.7 & -0.30 &  5.01 &   1.00 &    0.00 &       &       & \citet{Mucciarelli2014c} \\
LMC-NGC~1978  &  1.9 & -0.38 &  5.33 &   0.80 &    0.06 &       &       & \citet{Martocchia2018b} \\
LMC-NGC~2210  & 13.4 & -1.63 &  5.48 &   0.40 &         &       &  0.78 & \citet{Mucciarelli2009} \\
LMC-NGC~2257  & 13.4 & -1.63 &  5.41 &   0.33 &         &       &  0.51 & \citet{Mucciarelli2009} \\
LMC-Hodge~6   &  2.0 & -0.30 &  4.90 &   0.87 &         &       &       & \citet{Hollyhead2019} \\
LMC-Hodge~11 & 13.4 & -2.06 & 5.63 &&&&& \citet{Mateluna2012} \\
\hline
\end{tabular}
\label{tab:extra}
\end{table}

\begin{table}
\caption{Data for MW open clusters (GES is GAIA-ESO Survey)}
\begin{tabular}{lccrrl}
\hline
Cluster      & M$_J$ & $\log{M}$  &age   &[Fe/H] & Spectr. ref. \\
             &  &(M$_\odot)$ &(Gyr) &       & \\
\hline
Berkeley39   &-5.071  &      &6.5  &-0.20  & \citet{Bragaglia2012}   \\
Berkeley81   &-5.624  &      &1    &+0.23  & \citet{Magrini2015} (GES) \\
Collinder261 &-6.211  &      &6    &-0.03  & \citet{Maclean2015} \\
IC4756       &-4.535  &3.428 &0.8  &-0.02  & \citet{Bagdonas2018} \\
NGC~2360      &-5.475  &2.980 &1.1  &-0.07  & \citet{PenaSuarez2018} \\
NGC~2420      &-5.435  &      &2    &-0.16  & \citet{Souto2016} (APOGEE) \\
NGC~2682      &-5.025  &2.446 &4    &+0.03  & \citet{Maclean2015} \\
NGC~3114      &-5.459  &3.247 &0.16 &-0.01  & \citet{KatimeSantrich2013} \\
NGC~3680      &-4.077  &2.124 &1.8  &-0.06  & \citet{PenaSuarez2018} \\
NGC~5822      &-5.346  &3.519 &0.9  &-0.09  & \citet{PenaSuarez2018} \\
NGC~6134      &-4.986  &      &0.9  &+0.15  & \citet{Mikolaitis2010} \\
NGC~6705      &-5.295  &      &0.3  &+0.10  & \citet{Cantat2014} (GES) \\
NGC~6791      &-7.298  &3.700 &8    &+0.40  & \citet{Bragaglia2014,Villanova2018} \\
NGC~6802      &-5.139  &      &0.9  &+0.10  & \citet{Tang2017} (GES) \\
NGC~6940      &-6.612  &2.770 &1.1  &+0.04  & \citet{BocekTopcu2016} \\
NGC~752       &-4.198  &3.440 &1.6  &-0.02  & \citet{Maclean2015} \\
NGC~7789      &-6.789  &3.922 &1.6  &+0.03  & \citet{Maclean2015} \\
Pismis18      &-7.306  &      &0.7  &+0.23  & \citet{Hatzidimitriou2019} (GES) \\
Praesepe     &-3.317  &3.367 &0.73 &+0.13  & \citet{Maclean2015} \\
Ruprecht147  &-3.489  &2.148 &2.5  &+0.08  & \citet{Bragaglia2018} \\
Trumpler20   &-7.40   &      &1.5  &+0.17  & \citet{Donati2014} (GES) \\
Trumpler23   &-5.129  &3.0   &0.8  &+0.14  & \citet{Overbeek2017} (GES) \\
\hline
\multicolumn{6}{c}{} \\
\multicolumn{6}{l}{Notes: M$_J$\ and $\log{M}$\ from \citet{Piskunov2008}; $\log{M}$\ for NGC~6791 based on 5000~M$_\odot$ in \citet{Platais2011}.}\\
\hline
\end{tabular}\label{tab:oc}
\end{table}

\end{document}